\documentclass[a4paper,11pt]{article}
\pdfoutput=1 

\usepackage{jheppub} 

\usepackage{amsmath}
\usepackage{amssymb}
\usepackage{amsfonts}
\usepackage{mathrsfs}
\usepackage{graphicx}
\usepackage{booktabs,adjustbox}
\usepackage{bbm}
\usepackage{bm}
\usepackage{array,longtable}
\usepackage{multirow}
\usepackage{slashed}
\usepackage[utf8]{inputenc}
\usepackage{verbatim}
\usepackage{float}
\usepackage{url}
\usepackage{color}

\usepackage{diagbox}

\usepackage{hyperref}

\allowdisplaybreaks[4]

\title{\bf \boldmath $B_c^- \to J/\psi (\to \mu^+ \mu^-)\tau^- (\to \pi^- \nu_\tau, \rho^- \nu_\tau, \ell^-\bar{\nu}_\ell\nu_\tau)\bar{\nu}_\tau$ decays with visible final-state kinematics}

\author[a,b,1]{Xin-Qiang Li,\note{Corresponding author.}}
\author[a]{Xin Xu,}
\author[a,c]{Ya-Dong Yang,}
\author[a]{and Dong-Hui Zheng}

\affiliation[a]{ Institute of Particle Physics and Key Laboratory of Quark and Lepton Physics~(MOE), Central China Normal University, Wuhan, Hubei 430079, China}
\affiliation[b]{ Center for High Energy Physics, Peking University, Beijing 100871, China}
\affiliation[c]{ Institute of Particle and Nuclear Physics, Henan Normal University, Xinxiang 453007, China}

\emailAdd{xqli@mail.ccnu.edu.cn}
\emailAdd{xuxin@mails.ccnu.edu.cn}
\emailAdd{zhengdh@mails.ccnu.edu.cn}
\emailAdd{yangyd@mail.ccnu.edu.cn}

\abstract{The semitauonic $B_c^- \to J/\psi \tau^-\bar{\nu}_\tau$ decay provides an ideal and clean mode to scrutinize possible new physics effects in $b \to c \tau^- \bar{\nu}_\tau$ transitions as indicated by the current data on $R(D^{(*)})$ anomalies. In this work, we use the spin density matrix method to obtain the maximum information on the underlying physics of $B_c^- \to J/\psi \tau^-\bar{\nu}_\tau$ decay with both polarized $\tau$ lepton and $J/\psi$ meson. Their subsequent decays, with $J/\psi\to \mu^+ \mu^-$ as well as $\tau^- \to \pi^- \nu_\tau$, $\tau^- \to \rho^- \nu_\tau$ and $\tau^- \to \ell^-\bar{\nu}_\ell\nu_\tau$, are exploited to extract the energy and angular distributions of the charged final-state particles in the processes. Starting with the most general effective Hamiltonian relevant for the $b \to c \tau^- \bar{\nu}_\tau$ transitions, including all possible Lorentz structures of the dimension-six operators with both left- and right-handed neutrinos, we first derive the five-fold differential decay rate in terms of the visible final-state kinematics. From this distribution, we then construct in total 34 normalized observables, among which nine refer to the CP-violating triple product asymmetries that vanish within the Standard Model. We also construct five new observables based on the combinations of these normalized observables that can only be attributed to the right-handed neutrinos. On the other hand, considering the low statistics of the fully differential distribution, we introduce some integrated observables with only one kinematic variable left, which are more promising to be measured due to the largely increased statistics. The sensitivities of all these observables to the different new physics scenarios are investigated in detail. Finally, assuming an ideal circumstance, we give an estimate of the statistical uncertainties of the nine CP-conserving observables at LHCb and found that $\tau^-\to \pi^-\nu_\tau$ has the highest analyzing power among the three $\tau$ decay channels.}

\begin{document} 
\maketitle
\flushbottom

\section{Introduction}
\label{sec:introduction}

The semitauonic $b \to c \tau^- \bar{\nu}_\tau$ decays have raised considerable interest
over the last few years in both testing the Standard Model (SM) and searching for new physics (NP) beyond it. Many processes induced by the $b \to c \tau^- \bar{\nu}_\tau$ transitions at the quark level have been measured by several experiments, showing some degrees of deviations from the SM predictions~\cite{Lees:2012xj,Lees:2013uzd,Aaij:2015yra,Huschle:2015rga,Hirose:2016wfn,Aaij:2017uff,Hirose:2017dxl,Aaij:2017deq,Aaij:2017tyk,Belle:2019ewo,Belle:2019rba,LHCb:2022piu,LHCb:2023zxo}.\footnote{As argued in ref.~\cite{Fedele:2022iib}, once additional constraints are taken into account, the current experimental data on $R(D^{(*)})$ and $R(\Lambda_c)$ can be addressed neither within nor beyond the SM simultaneously. Therefore, further refined measurements are required in order to reach a coherent pattern of all the experimental data.} Particularly, the current world average performed by the Heavy Flavor Averaging Group (HFLAV)~\cite{HFLAV:2022pwe} shows that the measured ratios $ R(D^{(*)})\equiv\mathcal{B}(\bar{B}\to D^{(*)} \tau \bar{\nu}_\tau)/\mathcal{B}(\bar{B}\to D^{(*)} \ell\bar{\nu}_\ell)$ (with $\ell=e,\,\mu$) deviate from the SM predictions by about $3.2\sigma$~\cite{Amhis2022}. These deviations could be explained consistently within both model-independent and model-dependent frameworks (see, \textit{e.g.}, refs.~\cite{Bifani:2018zmi,Gambino:2020jvv,Bernlochner:2021vlv,London:2021lfn} for a recent review). For instance, by performing a global fitting analysis with dimension-six effective operators containing only left-handed neutrino fields~\cite{Alok:2017qsi,Hu:2018veh,Alok:2019uqc,Murgui:2019czp,Blanke:2018yud,Blanke:2019qrx,Shi:2019gxi,Cheung:2020sbq,Kumbhakar:2020jdz,Iguro:2022yzr}, it is found that some different combinations of these effective operators can well explain the observed deviations. The case with right-handed neutrinos has also been suggested as a possible alternative to evade some current phenomenological constraints on the effective operators with left-handed neutrinos~\cite{Greljo:2018ogz,Asadi:2018wea,Robinson:2018gza,Azatov:2018kzb,Mandal:2020htr}. However, the masses of these right-handed neutrinos must be small enough in order to be consistent with the measured $\bar{B}\to D^{(*)} \tau \bar{\nu}_\tau$ invariant-mass distributions~\cite{Lees:2013uzd,Huschle:2015rga}. Therefore, the masses of both left- and right-handed neutrinos will be ignored throughout this work.

The $B_c^- \to J/\psi \tau^- \bar{\nu}_\tau$ decay proceeds through the same quark-level transition as in $\bar{B}\to D^{(*)} \tau \bar{\nu}_\tau$ decays. This implies that the $B_c^- \to J/\psi \tau^- \bar{\nu}_\tau$ decay will also get affected by the possible NP scenarios designed for explaining the $R(D^{(*)})$ anomalies. In fact, a similar observable $R(J/\psi)\equiv\mathcal{B}(B_c^- \to J/\psi \tau^- \bar{\nu}_\tau)/\mathcal{B}(B_c^- \to J/\psi \mu^- \bar{\nu}_\mu)$ has been measured by the LHCb collaboration~\cite{Aaij:2017tyk}, which shows a $1.8\sigma$ discrepancy with the latest SM prediction made by the HPQCD collaboration~\cite{Harrison:2020nrv}. However, contrary to the $R(D^{(*)})$ measurements, the background contributions from partially reconstructed $B_c$ decays are significantly reduced thanks to the strongly peaking $\mu^+\mu^-$ spectrum and the clean muon final state in $B_c^- \to J/\psi \tau^- \bar{\nu}_\tau$ decay~\cite{Aaij:2017tyk,Bernlochner:2021vlv}. Furthermore, the $B_c$ lifetime is almost three times shorter than that of the $B_{u,d,s}$ mesons~\cite{ParticleDataGroup:2022pth}, which can be used to improve the separation of the $B_c$ decay from the $B_{u,d,s}$ decays, providing therefore an extra handle to discriminate
against the large background that originates from the $B_{u,d,s}$ decays~\cite{Aaij:2017tyk,Bernlochner:2021vlv}. All these features make the $B_c^- \to J/\psi \tau^- \bar{\nu}_\tau$ decay an ideal and clean mode to scrutinize the possible NP effects as indicated by the $R(D^{(*)})$ anomalies.

However, as pointed out in refs.~\cite{Hagiwara:2014tsa,Bordone:2016tex,Alonso:2016gym,Ligeti:2016npd,Asadi:2018sym,Alonso:2018vwa,Bhattacharya:2020lfm,Hu:2020axt,Nierste:2008qe,Tanaka:2010se,Alonso:2017ktd,Asadi:2020fdo,Penalva:2021gef,Hu:2021emb,Penalva:2021wye}, the $\tau$ three-momentum in these semitauonic decays cannot be determined precisely since the $\tau$ lepton is very short-lived and its decay products contain at least one undetected neutrino. One way out here is to consider only the visible final-state kinematics in the subsequent $\tau$ decays, while integrating out all the variables that cannot be directly measured. To this end, we shall make use of the three subsequent decays of the $\tau$ lepton, $\tau^- \to \pi^- \nu_\tau$, $\tau^- \to \rho^- \nu_\tau$ and $\tau^- \to \ell^-\bar{\nu}_\ell\nu_\tau$, to construct all the {\it measurable} distributions. These three channels account for more than $70\%$ of the total $\tau$ decay width~\cite{ParticleDataGroup:2022pth}. Moreover, the energy of the visible decay product (\textit{i.e.}, $\pi$, $\rho$ and $\ell$ for the three different decay channels, respectively) can serve as a $\tau$ polarimeter~\cite{Kiers:1997zt,Nierste:2008qe,Tanaka:2010se,Sakaki:2012ft,Ivanov:2017mrj,Alonso:2017ktd,Asadi:2020fdo,Penalva:2021gef,Hu:2021emb,Penalva:2021wye}. In addition, by considering the subsequent decay $J/\psi\to \mu^+ \mu^-$, we can further extract the spin asymmetries of the $J/\psi$ meson along with that of the $\tau$ lepton. Therefore, the full cascade decay we are considering is $B_c^- \to J/\psi (\to \mu^+ \mu^-) \tau^- (\to \pi^- \nu_\tau, \rho^- \nu_\tau, \ell^-\bar{\nu}_\ell\nu_\tau)\bar{\nu}_\tau$, which includes three visible final states $\mu^+$, $\mu^-$ and $\{\pi^-, \rho^-, \ell^-\}$, with their three-momenta all being able to be measured.

From the theoretical point of view, the main obstacle for studying the semileptonic $B_c^- \to J/\psi l^- \bar{\nu}_l$ (with $l=e,\,\mu,\,\tau$) decays is the precise determination of the $B_c\to J/\psi$ transition form factors. In the literature, a wide range of different approaches has been used to evaluate these form factors, such as the quark models~\cite{Ebert:2003cn,Ivanov:2005fd,Hernandez:2006gt,Ke:2013yka}, the QCD sum rules~\cite{Azizi:2009ny,Leljak:2019eyw,Azizi:2019aaf,Bordone:2022drp}, the Bethe-Salpeter equation~\cite{Liu:1997hr,AbdEl-Hady:1999jux,Yao:2021pyf}, the 
relativistic constituent quark model on the light front~\cite{Anisimov:1998xv,Wang:2008xt,Tran:2018kuv,Tang:2020org}, the perturbative QCD calculations~\cite{Wang:2012lrc,Shen:2014msa,Rui:2016opu,Hu:2019qcn}, as well as the nonrelativistic QCD (NRQCD) approach~\cite{Qiao:2011yz,Qiao:2012vt,Qiao:2012hp,Zhu:2017lqu,Zhu:2017lwi,Shen:2021dat,Tao:2022yur,Colangelo:2022lpy}. Particularly, there exist high-precision lattice QCD determinations of the $B_c\to J/\psi$ vector and axial-vector form factors~\cite{Harrison:2020gvo}, which will be adopted by us. The scalar and pseudo-scalar form factors, which are also needed for a full description of possible NP effects in a most general model-independent framework, can be directly related to the vector and axial-vector ones through the equations of motion~\cite{Colangelo:2022lpy}. However, we are still missing a direct determination of the tensor form factors from lattice QCD, which precludes an accurate NP analysis. To this end, we shall follow the results obtained in ref.~\cite{Tang:2022nqm}, where the tensor form
factors are related to the (axial-)vector ones by using the NRQCD relations including the next-to-leading-order relativistic corrections, and then determined in terms of the lattice QCD results for the latter. These form factors are all parameterized in a $z$ expansion to cover the full kinematic range of the dilepton invariant mass squared $q^2$, with $0\leq q^2\leq (m_{B_c}-m_{J/\psi})^2$~\cite{Colangelo:2022lpy,Harrison:2020gvo,Tang:2022nqm}.

This paper is organized as follows. In section~\ref{sec:angular}, we first define the various observables in terms of the asymmetries of spins and/or angles with the help of spin density matrix method~\cite{Fano:1957zz,Bourrely:1980mr,Leader:2011vwq}, and then give the analytic results for the five-fold differential decay rate of $B_c^- \to J/\psi (\to \mu^+ \mu^-) \tau^- (\to \pi^- \nu_\tau, \rho^- \nu_\tau, \ell^-\bar{\nu}_\ell\nu_\tau)\bar{\nu}_\tau$ decays in terms of the visible final-state kinematics. In section~\ref{sec:Num}, we present our numerical results for the normalized observables and discuss their sensitivities to the different NP scenarios. Some combinations of these observables that can only be attributed to the right-handed neutrinos are also discussed. Due to the limited experimental statistics, we also present the integrated observables with only one kinematic variable left. In section~\ref{sec:sensitivity}, taking the LHCb experiment as an example, we estimate the statistical uncertainty in extracting the spin and spin-angular asymmetries from the full five-fold differential decay rate. Our conclusions are finally made in section~\ref{sec:conclusions}. For convenience, details of the calculation procedures of the three spin density matrices along with the explicit expressions of the observables as well as the phase-space integrations are presented in  appendices~\ref{app:Spin density matrices} and \ref{app:phase space}, respectively.

\section{Energy and angular distributions}
\label{sec:angular}

In this section, we begin by describing our method for calculating the full energy and angular distributions of $B_c^- \to J/\psi (\to \mu^+ \mu^-)\tau^- (\to \pi^- \nu_\tau, \rho^- \nu_\tau, \ell^-\bar{\nu}_\ell\nu_\tau)\bar{\nu}_\tau$ decays.

\subsection{Effective Hamiltonian}
\label{subsec:EH}

Assuming that the NP scale is much higher than the electroweak scale, we can integrate out all the heavy degrees of freedom, and thus both the SM and NP contributions can be described by a low-energy effective Hamiltonian. With both the left- and right-handed neutrinos as well as all the possible Lorentz structures of the dimension-six four-fermion operators taken into account, the most general effective Hamiltonian relevant for the $b \to c \tau^- \bar{\nu}_\tau$ transitions can be written as~\cite{Mandal:2020htr}
\begin{equation} \label{eq:Hamiltonian}
 \mathcal{H}_{\mathrm{eff}} = \frac{4G_{F} V_{cb}}{\sqrt{2}} \left[\mathcal{O}^V_{LL} + \sum_{\tiny \substack{X=V,S,T \\ A,B=L,R}} C^X_{AB}\,\mathcal{O}^X_{AB}\right],
\end{equation}
where $G_F$ is the Fermi constant and $V_{cb}$ is the Cabibbo–Kobayashi–Maskawa matrix element involved. The ten four-fermion operators are defined, respectively, by\footnote{Note that the tensor operators with different quark and lepton chiralities vanish identically, which can be derived from the Dirac-algebra identity $\sigma^{\mu\nu}\gamma_5 = -\frac{i}{2}\epsilon^{\mu\nu\alpha\beta} \sigma_{\alpha\beta}$. We use the convention $\epsilon_{0123} = - \epsilon^{0123} = 1$.}
\begin{equation}
	\begin{aligned}
     	\mathcal{O}^V_{AB}&\equiv(\bar{c}\gamma^\mu P_Ab)(\bar{\tau}\gamma_\mu P_B\nu_\tau),\\[2mm]
		\mathcal{O}^S_{AB}&\equiv(\bar{c}P_Ab)(\bar{\tau}P_B\nu_\tau),\\[2mm]
		\mathcal{O}^T_{AB}&\equiv\delta_{AB}(\bar{c}\sigma^{\mu\nu}P_Ab)(\bar{\tau}\sigma_{\mu\nu}P_B\nu_\tau),
	\end{aligned}
\end{equation}
with $P_{L(R)} = (1\mp\gamma_5)/2$ and $\sigma^{\mu\nu} = \frac{i}{2}[\gamma^\mu,\,\gamma^\nu]$. All the NP effects are encoded in the short-distance Wilson coefficients $C^X_{AB}$, which are defined at the characteristic energy scale $\mu = m_b$, with $m_b$ being the bottom-quark mass. Within the SM, all $C^X_{AB}= 0$ and the only non-zero operator comes from $\mathcal{O}^V_{LL}$ (\textit{i.e.}, the first term in eq.~\eqref{eq:Hamiltonian}).

\subsection{Spin density matrices}
\label{sec:Spin density matrix}

The cascade processes $B_c^- \to J/\psi (\to \mu^+ \mu^-)\tau^- (\to \pi^- \nu_\tau, \rho^- \nu_\tau, \ell^-\bar{\nu}_\ell\nu_\tau)\bar{\nu}_\tau$ can be broken down into four successive decays of the $B_c$ meson and the three intermediate states. Explicitly, the fully differential decay width can be written as
\begin{equation} \label{eq:decay rate}
	\begin{aligned}
		d\Gamma  =&\frac{1}{2m_{B_c}}\,d\Pi_2(p_{B_c};p_{J/\psi},q)\,\frac{dq^2}{2\pi\,}d\Pi_2(q;p_{\tau},p_{\bar{\nu}_{\tau}})\\
		&\times\frac{dp_\tau^2}{2\pi}\,\frac{1}{(p_\tau^2-m_\tau^2)^2+m_\tau^2\Gamma_\tau^2}\,d\Pi_{2(3)}(p_\tau;p_d(,p_{\bar{\nu}_\ell}),p_{\nu_\tau})\\
		&\times\frac{dp_{J/\psi}^2}{2\pi}\,\frac{1}{(p_{J/\psi}^2-m_{J/\psi}^2)^2+m_{J/\psi}^2\Gamma_{J/\psi}^2}\,d\Pi_2(p_{J/\psi};p_{\mu^-},p_{\mu^+})\\
		&\times \mathrm{Tr}\left[\rho^{B_c}(s,s^\prime,\lambda,\lambda^\prime)\left(\rho^\tau(s,s^\prime)\otimes\rho^{J/\psi}(\lambda,\lambda^\prime)\right)^T\right],
	\end{aligned}
\end{equation}
where $\rho^{B_c}(s,s^\prime,\lambda,\lambda^\prime)$ denotes the spin density matrix~\cite{Bourrely:1980mr} of the $B_c^- \to J/\psi\tau^- \bar{\nu}_\tau$ process, while the decay density matrices for the $\tau$ lepton and the $J/\psi$ meson are denoted by $\rho^\tau(s,s^\prime)$ and $\rho^{J/\psi}(\lambda,\lambda^\prime)$, respectively. The indices $s,s^\prime=\{1/2,-1/2\}$ and $\lambda,\lambda^\prime=\{1,0,-1\}$ characterize the helicities of the particles $\tau$ and $J/\psi$, respectively. Here, $q=p_{B_c}-p_{J/\psi}=p_{\tau}+p_{\bar{\nu}_{\tau}}$ denotes the momentum transfer to the lepton pair, and $p_{d}$ refers to the momentum of the visible $\tau$ decay product, with $d=\pi,\rho,\ell$ corresponding to the three different channels of the $\tau$ lepton. Since the decay widths of both $\tau$ and $J/\psi$ are much smaller than their respective masses~\cite{ParticleDataGroup:2022pth}, we can apply to eq.~\eqref{eq:decay rate} the narrow-width approximation, 
\begin{equation}
	\lim_{\Gamma\to0^+}\frac{1}{\pi}\frac{m\Gamma}{(p^2-m^2)^2+m^2\Gamma^2}=\delta(p^2-m^2).
\end{equation}
This will put $\tau$ and $J/\psi$ on their mass-shell, respectively. The two- and three-body phase spaces in eq.~\eqref{eq:decay rate} are all Lorentz invariant, and their integrations can be therefore performed in any frame of reference without loss of generality. For convenience, we present the details of these phase-space integrations in appendix~\ref{app:phase space}.

Since the polarizations of both $\tau$ and $J/\psi$ in the decays are considered, the spin density matrix $\rho^{B_c}(s,s^\prime,\lambda,\lambda^\prime)$ is now a $6\times6$ Hermitian matrix and can be parameterized as~\cite{Bourrely:1980mr,Leader:2011vwq,Boudjema:2009fz}
\begin{equation}
	\label{eq:rhoBc}
	\begin{aligned}
		\frac{\rho^{B_c}(s,s^\prime,\lambda,\lambda^\prime)}{\mathrm{Tr}\left[\rho^{B_c}\right]}=\frac{1}{6}\bigg[&I\otimes I+\frac{3}{2}P^U_iI\otimes t_i+\sqrt{\frac{3}{2}}T^U_{ij}I\otimes \left(t_it_j+t_jt_i\right)\\
		&+P_U^{i^\prime}\sigma^{i^\prime}\otimes I+\frac{3}{2}P_i^{i^\prime}\sigma^{i^\prime}\otimes t_i+\sqrt{\frac{3}{2}}T^{i^\prime}_{ij}\sigma^{i^\prime}\otimes \left(t_it_j+t_jt_i\right)\bigg],
	\end{aligned}
\end{equation}
where $\sigma^{i^\prime}$ are the Pauli matrices, with
\begin{equation}
	\sigma^{x}=\begin{pmatrix}
		0&1\\1&0
	\end{pmatrix},\quad 
	\sigma^{y}=\begin{pmatrix}
		0&-i\\i&0
	\end{pmatrix},\quad 
	\sigma^{z}=\begin{pmatrix}
		1&0\\0&-1
	\end{pmatrix},
\end{equation}
and $t_i$ the three-dimensional traceless spin operators for a spin-1 particle, with
\begin{equation}
	t_{x}=\frac{1}{\sqrt{2}}\begin{pmatrix}
		0&1&0\\1&0&1\\0&1&0
	\end{pmatrix},\quad 
	t_{y}=\frac{i}{\sqrt{2}}\begin{pmatrix}
		0&-1&0\\1&0&-1\\0&1&0
	\end{pmatrix},\quad 
	t_{z}=\begin{pmatrix}
		1&0&0\\0&0&0\\0&0&-1
	\end{pmatrix}.
\end{equation}
For a normalized $6\times6$ Hermitian matrix, there are in total 35 independent real parameters which, in our definition, refer to the three components of the $\tau$ vector polarization $P_U^{i^\prime}$, the three components of the $J/\psi$ vector polarization $P^U_{i}$, the five components of the $J/\psi$ tensor polarization $T^U_{ij}$ that is a symmetric traceless rank–2 tensor, as well as the $(3\times3+3\times5)=24$ components of the mixed $\tau$-$J/\psi$ polarizations $P_i^{i^\prime}$ and $T^{i^\prime}_{ij}$.

Note that all the four sequential decays in eq~\eqref{eq:decay rate} are given in the rest frames of the corresponding decaying particles. Further details of the calculation procedures as well as the explicit expressions of the decay density matrices $\rho^\tau(s,s^\prime)$ and $\rho^{J/\psi}(\lambda,\lambda^\prime)$ can be found in appendices~\ref{app:rho_tau} and \ref{app:rho_J}, respectively.

\subsection{Observables}
\label{sec:Observables}

Let us now introduce the spin basis to discuss the polarizations defined in the last subsection. For the polarizations of the $J/\psi$ meson, we choose
\begin{equation}
	\vec{n}_L=\frac{\vec{p}_{J/\psi}}{|\vec{p}_{J/\psi}|},\quad \vec{n}_\perp=\frac{(\vec{p}_{J/\psi}\times\vec{p}_\tau)\times\vec{p}_{J/\psi}}{|(\vec{p}_{J/\psi}\times\vec{p}_\tau)\times\vec{p}_{J/\psi}|},\quad \vec{n}_T=\frac{\vec{p}_{J/\psi}\times\vec{p}_\tau}{|\vec{p}_{J/\psi}\times\vec{p}_\tau|},
\end{equation}
where the subscripts $L$, $\perp$ and $T$ denote the polarization components that are longitudinal, perpendicular and transverse to the $J/\psi$ momentum, respectively. Particularly, in the $J/\psi$ rest frame, $\vec{n}_L$, $\vec{n}_\perp$ and $\vec{n}_T$ correspond to the Cartesian basis $\vec{n}_z$, $\vec{n}_x$ and $\vec{n}_y$, respectively. For the polarizations of the $\tau$ lepton, we choose
\begin{equation}
	N^L=\left(\frac{|\vec{p}_\tau|}{m_\tau},\frac{E_\tau\vec{p}_\tau}{m_\tau|\vec{p}_\tau|}\right),\quad N^\perp=\left(0,\frac{(\vec{p}_{J/\psi}\times\vec{p}_\tau)\times\vec{p}_\tau}{|(\vec{p}_{J/\psi}\times\vec{p}_\tau)\times\vec{p}_\tau|}\right),\quad N^T=\left(0,\frac{\vec{p}_{J/\psi}\times\vec{p}_\tau}{|\vec{p}_{J/\psi}\times\vec{p}_\tau|}\right),
\end{equation}
where the basis vectors are defined in a Lorentz covariant form for our later convenience. 

Combining the parametrization of eq.\eqref{eq:rhoBc} with the explicit expressions of $\rho^\tau(s,s^\prime)$ and $\rho^{J/\psi}(\lambda,\lambda^\prime)$ as detailed in appendices~\ref{app:rho_tau} and \ref{app:rho_J}, we can obtain
\begin{equation} \label{eq:Tr rho h}
	\begin{aligned}
		& \frac{\mathrm{Tr}\left[\rho^{B_c}(s,s^\prime,\lambda,\lambda^\prime)\left(\rho^\tau(s,s^\prime)\otimes\rho^{J/\psi}(\lambda,\lambda^\prime)\right)^T\right]}{\mathrm{Tr}\left[\rho^{B_c}\right]\mathrm{Tr}\left[\rho^\tau\right]\mathrm{Tr}\left[\rho^{J/\psi}\right]}=\frac{1}{6}\left[1+\sqrt{\frac{3}{2}}T^U_{ij}(\vec{n}_i\cdot\hat{p}_{\mu^-})(\vec{n}_j\cdot\hat{p}_{\mu^-})\right.\\[0.15cm]
		&\hspace{1.5cm}\left.-\frac{2\alpha_\tau m_\tau}{m_\tau^2-m_d^2} P_U^{i^\prime}(N^{i^\prime}\cdot p_{d})-\frac{2\alpha_\tau m_\tau}{m_\tau^2-m_d^2}\sqrt{\frac{3}{2}} T^{i^\prime}_{ij}(N^{i^\prime}\cdot p_{d})(\vec{n}_i\cdot\hat{p}_{\mu^-})(\vec{n}_j\cdot\hat{p}_{\mu^-})\right],
	\end{aligned}
\end{equation}
for the hadronic $\tau^- \to \pi^-\nu_\tau$ and $\tau^- \to \rho^-\nu_\tau$ decays, while
\begin{equation} \label{eq:Tr rho l}
	\begin{aligned}
		& \frac{\mathrm{Tr}\left[\rho^{B_c}(s,s^\prime,\lambda,\lambda^\prime)\left(\rho^\tau(s,s^\prime)\otimes\rho^{J/\psi}(\lambda,\lambda^\prime)\right)^T\right]}{\mathrm{Tr}\left[\rho^{B_c}\right]\mathrm{Tr}\left[\rho^\tau\right]\mathrm{Tr}\left[\rho^{J/\psi}\right]}=\frac{1}{6}\left[1+\sqrt{\frac{3}{2}}T^U_{ij}(\vec{n}_i\cdot\hat{p}_{\mu^-})(\vec{n}_j\cdot\hat{p}_{\mu^-})\right.\\[0.15cm]
		&\hspace{1.8cm}\left.-\frac{m_\tau}{p_{\bar{\nu}_\ell}\cdot p_\tau} P_U^{i^\prime}(N^{i^\prime}\cdot p_{\bar{\nu}_\ell})-\frac{m_\tau}{p_{\bar{\nu}_\ell}\cdot p_\tau}\sqrt{\frac{3}{2}} T^{i^\prime}_{ij}(N^{i^\prime}\cdot p_{\bar{\nu}_\ell})(\vec{n}_i\cdot\hat{p}_{\mu^-})(\vec{n}_j\cdot\hat{p}_{\mu^-})\right],
	\end{aligned}
\end{equation}
for the leptonic $\tau^- \to \ell^- \bar{\nu}_\ell\nu_\tau$ decay. Here we have already used the spin bases introduced above, and the indices $i,j,i^\prime$ should be summed over the components $\{L,\perp,T\}$. As the decay $J/\psi \to \mu^+ \mu^-$ is an electromagnetic process, where parity is conserved, we are unable to extract the vector polarizations of $J/\psi$ through this channel~\cite{Boudjema:2009fz}. Furthermore, since $\sum_i(\vec{n}_i\cdot\hat{p}_{\mu^-})(\vec{n}_i\cdot\hat{p}_{\mu^-})=1$, the three vector polarizations are not linearly independent. These facts imply that we cannot extract all of these polarization coefficients from the decays. For convenience, we can redefine
\begin{equation} \label{eq:redefine}
	\tilde{T}^U_{ii}=\sqrt{\frac{2}{3}}+T^U_{ii},\qquad \tilde{T}^{i^\prime}_{ii}=\sqrt{\frac{2}{3}}P^{i^\prime}_U+T^{i^\prime}_{ii}.
\end{equation}
Note that there is no information lost in this redefinition, since the tensor polarizations $T^U_{ii}$ and $T^{i^\prime}_{ii}$ are all traceless, and we can regain the $\tau$ vector polarizations through the relation $P^{i^\prime}_U=\sum_i\tilde{T}^{i^\prime}_{ii}/\sqrt{6}$.

For the $B_c^- \to J/\psi\tau^- \bar{\nu}_\tau$ process, which is a three-body decay of a massive spinless particle, there are only two independent kinematic parameters that can be chosen as the dilepton invariant mass squared $q^2$ and the helicity angle $\theta_\tau$. Here $\theta_\tau$ is the angle between the flight direction of the $\tau$ and $J/\psi$ in the centre-of-mass frame of the $\tau \bar{\nu}_\tau$ pair. Since the spin density matrix $\rho^{B_c}(s,s^\prime,\lambda,\lambda^\prime)$ depends on $\theta_\tau$, we can further consider asymmetries with respect to this angle together with the spin asymmetries. To this end, we can define the following spin and spin-angular asymmetries:
\begin{equation}
	\begin{aligned}
		\frac{d\Gamma}{dq^2}\left\langle \tilde{T}^{i^\prime}_{ij}\right\rangle=&\int_{-1}^{1}d\cos\theta_\tau\frac{d^2\Gamma}{dq^2d\cos\theta_\tau}\tilde{T}^{i^\prime}_{ij},\\[3mm]
		\frac{d\Gamma}{dq^2}\tilde{Z}^{i^\prime}_{ij}=&\left(\int_{0}^{1}-\int_{-1}^{0}\right)d\cos\theta_\tau\frac{d^2\Gamma}{dq^2d\cos\theta_\tau}\tilde{T}^{i^\prime}_{ij},\\[3mm]
		\frac{d\Gamma}{dq^2}\tilde{A}^{\{U,L\}}_{ii}\left(\tilde{A}^{\{\perp,T\}}_{\{\perp L,TL\}}\right)=&\frac{5}{2}\int_{-1}^{1}d\cos\theta_\tau P^0_2(\cos\theta_\tau)\frac{d^2\Gamma}{dq^2d\cos\theta_\tau}\tilde{T}^{\{U,L\}}_{ii}\left(\tilde{T}^{\{\perp,T\}}_{\{\perp L,TL\}}\right),
	\end{aligned}
\end{equation}
where $\frac{d\Gamma}{dq^2}$ is the unpolarized differential decay rate. Explicit expressions of these observables expressed in terms of the transversity amplitudes can be found in appendix~\ref{app:rho_Bc}.

\subsection{Visible final-state kinematics}
\label{sec:Visible kinematics}

\begin{figure}[t]
	\centering
	\includegraphics[width=0.58\textwidth]{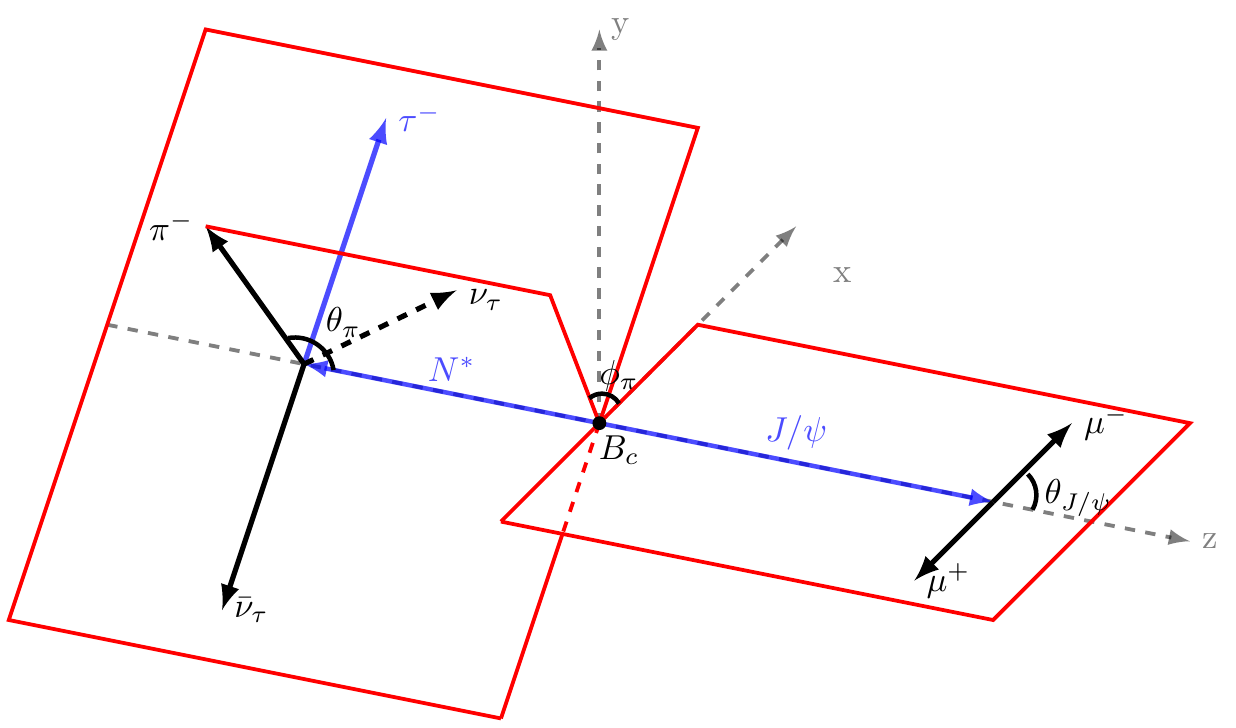}
	\caption{Definitions of the visible angles in the $B_c^- \to J/\psi (\to \mu^+ \mu^-)\tau^- (\to \pi^- \nu_\tau)\bar{\nu}_\tau$ decay.} \label{fig:angle} 
\end{figure}

So far, we have been discussing all the successive decays in the rest frames of their parent particles. However, as mentioned already in the introduction, it is generally not possible to fully reconstruct the $\tau$ momentum experimentally~\cite{Hagiwara:2014tsa,Bordone:2016tex,Alonso:2016gym,Ligeti:2016npd,Asadi:2018sym,Alonso:2018vwa,Bhattacharya:2020lfm,Hu:2020axt,Nierste:2008qe,Tanaka:2010se,Alonso:2017ktd,Asadi:2020fdo,Penalva:2021gef,Hu:2021emb,Penalva:2021wye}. Therefore, the kinematic variables that are defined in the $\tau$ rest frame or referred to the $\tau$ momentum direction are hard to be reconstructed in practice. This requires us to build the angular distributions based on a set of new frames of reference, which are illustrated in figure~\ref{fig:angle} by considering the channel $\tau^-\to \pi^- \nu_\tau$ as an example. Here, $\theta_{J/\psi}$ denotes the polar angle of $\mu^-$ in the $J/\psi$ rest frame, while $E_d$, $\theta_d$ and $\phi_d$ represent the energy, the polar and the azimuthal angle of the visible product $d$, with $d=\pi, \rho, \ell$, as viewed in the $\tau \bar{\nu}_\tau$ centre-of-mass frame. 

With the above setup of the kinematics, we can then write the scalar products appearing in eqs.~\eqref{eq:Tr rho h} and \eqref{eq:Tr rho l} explicitly as
\begin{equation}
	\begin{aligned}
		N^L\cdot p_d=&\frac{|\vec{p}_\tau|E_d}{m_\tau}-\frac{|\vec{p}_d|E_\tau}{m_\tau}\cos\theta_{\tau d},\qquad N^\perp\cdot p_d=|\vec{p}_d|\frac{\cos\theta_d-\cos\theta_\tau\cos\theta_{\tau d}}{\sin\theta_\tau},\\[2mm]
		N^T\cdot p_d=&-|\vec{p}_d|\sin\theta_d\sin(\phi_d-\phi_\tau),
	\end{aligned}
\end{equation}
for the leptonic, and
\begin{equation}
	\vec{n}_L\cdot\hat{p}_{\mu^-}=\cos\theta_{J/\psi},\quad\vec{n}_\perp\cdot\hat{p}_{\mu^-}=\sin\theta_{J/\psi}\cos\phi_\tau,\quad\vec{n}_T\cdot\hat{p}_{\mu^-}=-\sin\theta_{J/\psi}\sin\phi_\tau,
\end{equation}
for the hadronic side. Here $\theta_\tau$ and $\phi_\tau$ denote the polar and the azimuthal angle of the $\tau$ lepton relative to the z-axis, while $\theta_{\tau d}$ and $\phi_{\tau d}$  characterize the direction of the charged particle $d$ produced from the $\tau$ decay. They are related to each other through 
\begin{equation}
	\begin{aligned}
		\cos\theta_\tau&=\cos\theta_d\cos\theta_{\tau d}-\cos\phi_{\tau d}\sin\theta_d\sin\theta_{\tau d},\\[3mm]
		\sin\phi_\tau&=\frac{\cos\theta_{\tau d}\sin\phi_d\sin\theta_d+(\cos\phi_{\tau d}\cos\theta_d\sin\phi_d+\sin\phi_{\tau d}\cos\phi_d)\sin\theta_{\tau d}}{\sqrt{1-\cos^2\theta_\tau}}.
	\end{aligned}
\end{equation}
For the $\tau^-\to \pi^-(\rho^-)\nu_\tau$ channels, the angle $\theta_{\tau d}$ can be expressed in terms of other visible variables as
\begin{equation}
	\cos\theta_{\tau d}=\frac{2 E_{\tau} E_{d}-m_{\tau}^{2}-m_{d}^{2}}{2\left|\vec{p}_{\tau}\right|\left|\vec{p}_{d}\right|},
\end{equation}
while for the $\tau^-\to \ell^-\bar{\nu}_\ell\nu_\tau$ channel, it is an independent variable that cannot be measured. For our later convenience, we introduce the variable
\begin{equation} \label{eq:x-costaud}
	x=\frac{2(p_d\cdot p_\tau)}{m_\tau^2}=\frac{2}{m_\tau^2}\left(E_dE_\tau-|\vec{p}_d||\vec{p}_\tau|\cos\theta_{\tau d}\right),
\end{equation}
which is the energy of the charged particle $d$ in the $\tau$ rest frame up to a constant, and will be integrated out in our final result. In addition, for all the three $\tau$ decay channels, the angle $\phi_{\tau d}$ is also an unmeasurable variable, and hence will be integrated out too. After performing the phase-space integrations, details of which could be found in appendix~\ref{app:phase space}, we can get the normalized five-fold differential decay rate of the process $B_c^- \to J/\psi (\to \mu^+ \mu^-)\tau^- (\to d^- (\bar{\nu}_\ell)\nu_\tau)\bar{\nu}_\tau$, which reads
\begin{align} \label{eq:differential decay rate}
		&\frac{d^5\Gamma}{dq^2dE_d d\cos\theta_d d\phi_dd\cos\theta_{J/\psi}}/\frac{d\Gamma}{dq^2}\nonumber\\[2mm]
		&=\mathcal{N}^U\bigg[\Big(\left\langle \tilde{T}^{U}_{\perp \perp}\right\rangle+\tilde{A}^{U}_{\perp\perp}-\tilde{A}^{U}_{TT}+2\cos\theta_{\tau d}P_1(\cos\theta_d)\tilde{Z}^{U}_{\perp\perp}\nonumber\\[1mm]
		&\qquad\quad+\frac{1}{2}(3\cos^2\theta_{\tau d}-1)P_2(\cos\theta_d)\big(\tilde{A}^{U}_{\perp\perp}+\tilde{A}^{U}_{TT}\big)\nonumber\\[1mm]
		&\qquad\quad-\frac{1}{4}(3\cos^2\theta_{\tau d}-1)P^2_2(\cos\theta_d)\big(\cos2\phi_d(\tilde{A}^{U}_{\perp\perp}-\tilde{A}^{U}_{TT})+\sin2\phi_d\left\langle \tilde{T}^{U}_{\perp T}\right\rangle\big)\Big)\sin^2\theta_{J/\psi}\nonumber\\[1mm]
		&\qquad-\Big(\frac{4}{\pi}\cos\theta_{\tau d}P^1_1(\cos\theta_d)\big(\cos\phi_d\left\langle \tilde{T}^{U}_{\perp L}\right\rangle-\sin\phi_d\left\langle \tilde{T}^{U}_{T L}\right\rangle\big)\nonumber\\[1mm]
		&\qquad\qquad+\frac{1}{2}(3\cos^2\theta_{\tau d}-1)P^1_2(\cos\theta_d)\big(\cos\phi_d\tilde{Z}^{U}_{\perp L}-\sin\phi_d\tilde{Z}^{U}_{T L}\big)\Big)\sin2\theta_{J/\psi}\nonumber\\[1mm]
		&\qquad+\Big(\left\langle \tilde{T}^{U}_{L L}\right\rangle+2\cos\theta_{\tau d}P_1(\cos\theta_d)\tilde{Z}^{U}_{L L}+(3\cos^2\theta_{\tau d}-1)P_2(\cos\theta_d)\tilde{A}^{U}_{L L}\Big)\cos^2\theta_{J/\psi}\bigg]\nonumber\\[2mm]
		&-\mathcal{N}^L\bigg[\Big(\left\langle \tilde{T}^{L}_{\perp\perp}\right\rangle+\tilde{A}^{L}_{\perp\perp}-\tilde{A}^{L}_{TT}+2\cos\theta_{\tau d}P_1(\cos\theta_d)\tilde{Z}^{L}_{\perp\perp}\nonumber\\[1mm]
		&\qquad\qquad+\frac{1}{2}(3\cos^2\theta_{\tau d}-1)P_2(\cos\theta_d)\big(\tilde{A}^{L}_{\perp\perp}+\tilde{A}^{L}_{TT}\big)\nonumber\\[1mm]
		&\qquad\qquad-\frac{1}{4}(3\cos^2\theta_{\tau d}-1)P^2_2(\cos\theta_d)\big(\cos2\phi_d(\tilde{A}^{L}_{\perp\perp}-\tilde{A}^{L}_{TT})+\sin2\phi_d\left\langle \tilde{T}^{L}_{\perp T}\right\rangle\big)\Big)\sin^2\theta_{J/\psi}\nonumber\\[1mm]
		&\qquad\quad-\Big(\frac{4}{\pi}\cos\theta_{\tau d}P^1_1(\cos\theta_d)\big(\cos\phi_d\left\langle \tilde{T}^{L}_{\perp L}\right\rangle-\sin\phi_d\left\langle \tilde{T}^{L}_{T L}\right\rangle\big)\nonumber\\[1mm]
		&\qquad\qquad\quad+\frac{1}{2}(3\cos^2\theta_{\tau d}-1)P^1_2(\cos\theta_d)\big(\cos\phi_d\tilde{Z}^{L}_{\perp L}-\sin\phi_d\tilde{Z}^{L}_{T L}\big)\Big)\sin2\theta_{J/\psi}\nonumber\\[1mm]
		&\qquad\quad+\Big(\left\langle \tilde{T}^{L}_{L L}\right\rangle+2\cos\theta_{\tau d}P_1(\cos\theta_d)\tilde{Z}^{L}_{L L}+(3\cos^2\theta_{\tau d}-1)P_2(\cos\theta_d)\tilde{A}^{L}_{L L}\Big)\cos^2\theta_{J/\psi}\bigg]\nonumber\\[2mm]
		&+\mathcal{N}^\perp\bigg[\Big(-\frac{2}{\pi}\sin^2\theta_{\tau d}P_1(\cos\theta_{d})\big(\left\langle \tilde{T}^{\perp}_{\perp\perp}\right\rangle+\left\langle \tilde{T}^{\perp}_{TT}\right\rangle\big)\nonumber\\[1mm]
		&\qquad\qquad-\frac{3}{2}\cos\theta_{\tau d}\sin^2\theta_{\tau d}P_2(\cos\theta_d)\big(\tilde{Z}^\perp_{\perp\perp}+\tilde{Z}^\perp_{TT}\big)\nonumber\\[1mm]
		&\qquad\qquad+\cos\theta_{\tau d}\sin^2\theta_{\tau d}P^2_2(\cos\theta_d)\big(\frac{2}{\pi}\cos2\phi_d\left\langle \tilde{T}^T_{\perp T}\right\rangle-\frac{3}{2}\sin2\phi_d\tilde{Z}^\perp_{\perp T}\big)\Big)\sin^2\theta_{J/\psi}\nonumber\\[1mm]
		&\qquad\quad-\Big(\sin^2\theta_{\tau d}P^1_1(\cos\theta_d)\big(\cos\phi_d(\left\langle \tilde{T}^T_{T L}\right\rangle+\frac{1}{2}\tilde{A}^T_{T L})+\sin\phi_d(\left\langle \tilde{T}^T_{\perp L}\right\rangle+\frac{1}{2}\tilde{A}^T_{\perp L})\big)\nonumber\\[1mm]
		&\qquad\qquad\quad+\frac{3}{2}\cos\theta_{\tau d}\sin^2\theta_{\tau d}P^1_2(\cos\theta_d)\big(\cos\phi_d\tilde{A}^\perp_{\perp L}-\sin\phi_d\tilde{A}^\perp_{T L}\big)\Big)\sin2\theta_{J/\psi}\nonumber\\[1mm]
		&\qquad\quad-\Big(\frac{4}{\pi}\sin^2\theta_{\tau d}P_1(\cos\theta_d)\left\langle \tilde{T}^\perp_{L L}\right\rangle+3\cos\theta_{\tau d}\sin^2\theta_{\tau d}P_2(\cos\theta_d)\tilde{Z}^\perp_{L L}\Big)\cos^2\theta_{J/\psi}\bigg],
\end{align}
where we have introduced the abbreviations, 
\begin{equation}
	\begin{aligned}
		\mathcal{N}^U&=\frac{\sqrt{6}m^2_\tau}{16\pi|\vec{p}_\tau|(m_\tau^2-m_d^2)},\qquad \mathcal{N}^L=\frac{\sqrt{6}\alpha_\tau m^2_\tau E_\tau|\vec{p}_d|}{8\pi|\vec{p}_\tau|(m_\tau^2-m_d^2)^2}\left(\frac{|\vec{p}_\tau|E_d}{|\vec{p}_d|E_\tau}-\cos\theta_{\tau d}\right),\\[2mm]
	\mathcal{N}^\perp&=\frac{\sqrt{6}\alpha_\tau m^3_\tau|\vec{p}_d|}{8\pi|\vec{p}_\tau|(m_\tau^2-m_d^2)^2},
	\end{aligned}
\end{equation}
for the hadronic $\tau^-\to \pi^-(\rho^-)\nu_\tau$ channels, and
\begin{equation}
	\begin{aligned}
		\mathcal{N}^U&=\int dx\frac{\sqrt{6}}{8\pi|\vec{p}_\tau|}\frac{\theta(1+y^2-x)\left[x\left(3-2x\right)-y^2\left(4-3x\right)\right]}{1-8y^2+8y^6-y^8-24y^4\ln y},\\[2mm] 
		\mathcal{N}^L&=\int dx\frac{\sqrt{6}|\vec{p}_d|E_\tau}{4\pi m^2_\tau|\vec{p}_\tau|}\frac{\theta(1+y^2-x)\left(1+3y^2-2x\right)}{1-8y^2+8y^6-y^8-24y^4\ln y}\left(\frac{|\vec{p}_\tau|E_d}{|\vec{p}_d|E_\tau}-\cos\theta_{\tau d}\right),\\[2mm]
		\mathcal{N}^\perp&=\int dx\frac{\sqrt{6}|\vec{p}_d|}{4\pi m_\tau|\vec{p}_\tau|}\frac{\theta(1+y^2-x)\left(1+3y^2-2x\right)}{1-8y^2+8y^6-y^8-24y^4\ln y},
	\end{aligned}
\end{equation}
for the leptonic $\tau^-\to \ell^-\bar{\nu}_\ell\nu_\tau$ channel, where $\theta(x)$ denotes the step function and the integration over the variable $x$ is implicit. Explicit expressions of the observables that can be extracted from the differential distribution given by eq.~\eqref{eq:differential decay rate} are listed in appendix~\ref{app:rho_Bc}. It should be noted that some observables can only be extracted from eq.~\eqref{eq:differential decay rate} in a combination way. We also find the following interesting relations among the observables:
\begin{equation}
	\begin{aligned}
		&\frac{1}{2}\left\langle \tilde{T}^{\{U,L\}}_{\perp\perp}\right\rangle+\tilde{A}^{\{U,L\}}_{\perp\perp}=\frac{1}{2}\left\langle \tilde{T}^{\{U,L\}}_{TT}\right\rangle+\tilde{A}^{\{U,L\}}_{TT},\qquad&& \tilde{Z}^{\{U,L\}}_{\perp\perp}=\tilde{Z}^{\{U,L\}}_{TT},\\[2mm] &\left\langle \tilde{T}^{\{\perp,T\}}_{\perp\perp}\right\rangle-\left\langle \tilde{T}^{\{\perp,T\}}_{TT}\right\rangle=\pm\frac{3\pi}{2}\tilde{Z}^{\{T,\perp\}}_{\perp T}, &&\tilde{Z}^{\{\perp,T\}}_{\perp\perp}-\tilde{Z}^{\{\perp,T\}}_{TT}=\pm\frac{8}{3\pi}\left\langle \tilde{T}^{\{T,\perp\}}_{\perp T}\right\rangle,\\[2mm]
		&\frac{1}{2}\left\langle \tilde{T}^{\{\perp,T\}}_{\perp L}\right\rangle+\tilde{A}^{\{\perp,T\}}_{\perp L}=\pm \tilde{Z}^{\{T,\perp\}}_{T L},&&\frac{1}{2}\left\langle \tilde{T}^{\{T,\perp\}}_{T L}\right\rangle+\tilde{A}^{\{T,\perp\}}_{T L}=\pm \tilde{Z}^{\{\perp,T\}}_{\perp L},
	\end{aligned}
\end{equation}
which hold in the presence of any of the ten NP operators in eq.~\eqref{eq:Hamiltonian}. Therefore, these relations can provide no extra information for searching and distinguishing the different NP effects, and have already been used in eq.~\eqref{eq:differential decay rate} to get a more compact result.

\section{Numerical results}
\label{sec:Num}

In the previous sections, we have defined various asymmetries of the spins and/or angles that can be extracted from the fully differential distribution of the visible final-state kinematics in the $B_c^- \to J/\psi (\to \mu^+ \mu^-)\tau^- (\to \pi^- \nu_\tau, \rho^- \nu_\tau, \ell^-\bar{\nu}_\ell\nu_\tau)\bar{\nu}_\tau$ decays. In order to get a general idea about the sensitivities of these asymmetries to the different Wilson coefficients $C^X_{AB}$ in eq.~\eqref{eq:Hamiltonian}, we shall select in this section some best-fit values inferred from the $R(D^{(*)})$ resolutions as the NP benchmark points, and study how these observables will be affected by these NP scenarios. 

\subsection{$B_c\to J/\psi$ transition form factors}
\label{subsec:FFs}

For the $B_c\to J/\psi$ (axial-)vector form factors, $V(q^2)$ and $A_{0,1,2}(q^2)$, we shall use the high-precision lattice QCD results obtained in ref.~\cite{Harrison:2020gvo}. As a direct determination of the $B_c\to J/\psi$ tensor form factors, $T_{1,2,3}(q^2)$, from lattice QCD is still missing so far, we shall adopt the results presented in ref.~\cite{Tang:2022nqm}, where the tensor form factors are related to the (axial-)vector ones by using the NRQCD relations including the next-to-leading-order relativistic corrections, and then determined in terms of the lattice QCD results for the latter. In this way, we can parametrize all the $B_c\to J/\psi$ transition form factors in a $z$ expansion to cover the full $q^2$ range in the decays~\cite{Colangelo:2022lpy,Harrison:2020gvo,Tang:2022nqm}.

\subsection{NP benchmark points}
\label{subsec:NPBPs}

Since we include both the left- and right-handed neutrinos, the most general effective Hamiltonian given by eq.~\ref{eq:Hamiltonian} contains in total ten four-fermion operators. The large number of free parameters makes it difficult to perform a global fit to the full basis of these operators, and it means little to have a complete discussion about all the possible NP scenarios. Therefore, for the NP scenarios with purely left-handed neutrinos, we choose only the following four benchmark points as discussed in ref.~\cite{Iguro:2022yzr}:
\begin{equation} \label{eq:BP1-4}
	\begin{aligned}
		&\mathrm{BP1:} && C^V_{RL}=0.02\pm i 0.43,\\[2mm]
		&\mathrm{BP2:} && C^S_{LL}=-0.58\pm i 0.88,\\[2mm]
		&\mathrm{BP3:} && C^T_{LL}=0.06\pm i 0.16,\\[2mm]
		&\mathrm{BP4:} && C^V_{RL}=\pm i 0.68, C^S_{LL}=+8.4C^T_{LL}=0.04\mp i 0.65~\mbox{for the $R_2$ leptoquark}.
	\end{aligned}
\end{equation}
Let us first consider the NP scenarios where only a single Wilson coefficient $C^X_{AB}$ is present at a time. The case with a SM-like $C^V_{LL}$ equals to a global modification of the SM prediction by the factor $1+C^V_{LL}$ at the amplitude level, and thus its effect on the normalized observables is completely cancelled. Therefore, we are not going to discuss this scenario even though it can well resolve the $R(D^{(*)})$ anomalies~\cite{Hu:2018veh,Murgui:2019czp}. The BP1 contains only $C^V_{RL}$ that is naively suppressed by the small factor $v^4/\Lambda^4$, with $v$ and $\Lambda$ denoting the electroweak and the NP scale respectively, because the corresponding operator can only be generated at tree level starting from a dimension-eight operator in the standard model effective field theory~(SMEFT) formalism~\cite{Hu:2018veh,Alonso:2014csa,Aebischer:2015fzz} (for a recent review, see ref.~\cite{Brivio:2017vri} and references therein). This means that a sizable contribution from $C^V_{RL}$ would indicate an effective field theory with non-linear realization of the SM gauge group $SU(2)_L\times U(1)_Y$~\cite{Burgess:2021ylu,London:2022rjt}. The scenarios with a single scalar operator $\mathcal{O}^S_{LL}$ or $\mathcal{O}^S_{RL}$, like BP2, are already ruled out by the LEP data on the leptonic $B_c^- \to \tau^- \bar{\nu}_\tau$ decay rate~\cite{Li:2016vvp,Celis:2016azn,Alonso:2016oyd,Akeroyd:2017mhr}. However, as pointed out in refs.~\cite{Blanke:2018yud,Blanke:2019qrx}, there is a sizable charm-mass dependence of the $B_c$ lifetime (which is also confirmed later in ref.~\cite{Aebischer:2021ilm}) and the  transverse momentum $p_T$ dependence of the fragmentation function $f(b\to B_c)$ extracted from the Tevatron and LHC data has been entirely overlooked when applied to the LEP $Z$-peak analyses~\cite{Akeroyd:2017mhr}. Therefore, a more conservative bound, with ${\cal B}(B_c^- \to \tau^- \bar{\nu}_\tau)\lesssim60\%$, is obtained and the scenarios with a single scalar operator are still revived at present~\cite{Blanke:2018yud,Blanke:2019qrx}. Finally, the scenario with a single $C^T_{LL}$ gives unique predictions for some observables like the $D^\ast$
longitudinal polarization fraction~\cite{Belle:2019ewo}, which can be used to distinguish it from the other NP scenarios~\cite{Hu:2018veh,Murgui:2019czp,Iguro:2022yzr}.

For the NP scenarios with purely right-handed neutrinos, on the other hand, we choose only the following two benchmark points as obtained in ref.~\cite{Mandal:2020htr}:
\begin{equation} \label{eq:BP5-6}
	\begin{aligned}
		&\mathrm{BP5:} &&C^T_{RR}=\frac{1}{8}C^S_{RR}=0.054 &&\mbox{for the $\tilde{R}_2$ leptoquark},\\
		&\mathrm{BP6:} &&C^V_{RR}=0.422,C^T_{RR}=-\frac{1}{8}C^S_{RR}=0.022 &&\mbox{for the $S_1$ leptoquark}.\\
	\end{aligned}
\end{equation}
Here the benchmark points BP4--BP6 correspond to the scenarios with different leptoquarks, and the values of these Wilson coefficients have already run from the NP scale down to the $m_b$ scale. Following the same treatments as in refs.~\cite{Becirevic:2019tpx,Boer:2019zmp,Asadi:2020fdo,Harrison:2020nrv,Alguero:2020ukk}, we consider only the central values of these best-fit results to qualitatively discuss the influence of these different NP scenarios on the observables. 

\subsection{CP-violating observables}
\label{sec:CP-violating observables}

Among all the observables that can be extracted from the fully differential decay rate given by eq.~\eqref{eq:differential decay rate}, some are related to the triple products~(TP) of the kinematic variables involved or, to be more specific, to the sine of the azimuthal angle $\sin\phi_d$. Since the TP get a minus sign under time reversal, the corresponding observables may serve as a powerful tool for displaying the CP-violating effects~\cite{Kayser:1989vw,Datta:2003mj,Gronau:2011cf} according to the CPT theorem. 

From the explicit expressions listed in appendix~\ref{app:rho_Bc}, we can see that these kinds of observables are all proportional to ${\rm Im}[\mathcal{A}_i \mathcal{A}_j^{*}]$, where $\mathcal{A}_i$ and $\mathcal{A}_j$ are two different transversity amplitudes that can be written in general as
\begin{equation}
	\mathcal{A}_i=|\mathcal{A}_i|e^{i\phi_i}e^{i\delta_i}, \qquad \mathcal{A}_j=|\mathcal{A}_j|e^{i\phi_j}e^{i\delta_j},
\end{equation}
with $\phi_{i(j)}$ and $\delta_{i(j)}$ denoting the weak and strong phases respectively. With these definitions, we can get
\begin{equation}
	\mathrm{Im}[\mathcal{A}_i\mathcal{A}_j^*]=|\mathcal{A}_i||\mathcal{A}_j|\left[\underbrace{\sin(\phi_i-\phi_j)\cos(\delta_i-\delta_j)}_{\mbox{CP-odd}}+\underbrace{\cos(\phi_i-\phi_j)\sin(\delta_i-\delta_j)}_{\mbox{CP-even}}\right],
\end{equation}
where the first term in the bracket is non-zero and hence a clear signal of CP violation only when the weak-phase difference between $\mathcal{A}_i$ and $\mathcal{A}_j$ is non-negligible. The second term is, on the other hand, non-zero only in the presence of a strong-phase difference between $\mathcal{A}_i$ and $\mathcal{A}_j$, regardless of whether there exists a weak-phase difference or not. Therefore, it is in fact not CP-violating and usually dubbed as the ``fake TP"~\cite{Kayser:1989vw,Datta:2003mj,Gronau:2011cf}. Strictly speaking, we should compare these observables with the corresponding ones of the CP-conjugated process to get a true TP. However, we shall assume here the strong-phase difference to be zero, because $B_c\to J/\psi$ is the only hadronic transition in the decays considered and all the transversity amplitudes will have approximately the same strong phases~\cite{Datta:2004re,Datta:2004jm,Bhattacharya:2019olg}. For simplicity, all these observables will be simply called the CP-violating ones from now on.

\begin{figure}[t]
	\centering
	\includegraphics[width=0.31\textwidth]{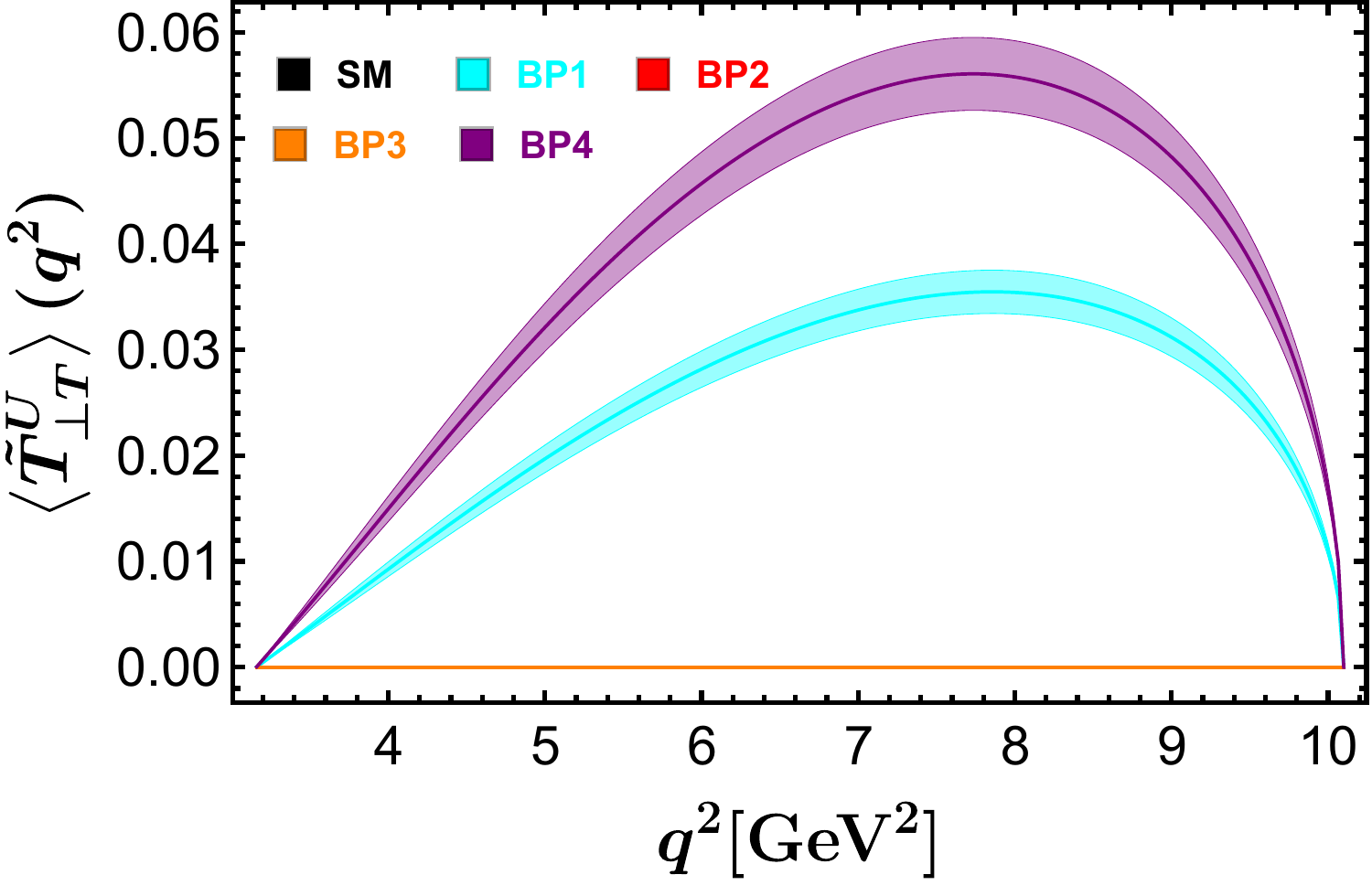}\quad
	\includegraphics[width=0.31\textwidth]{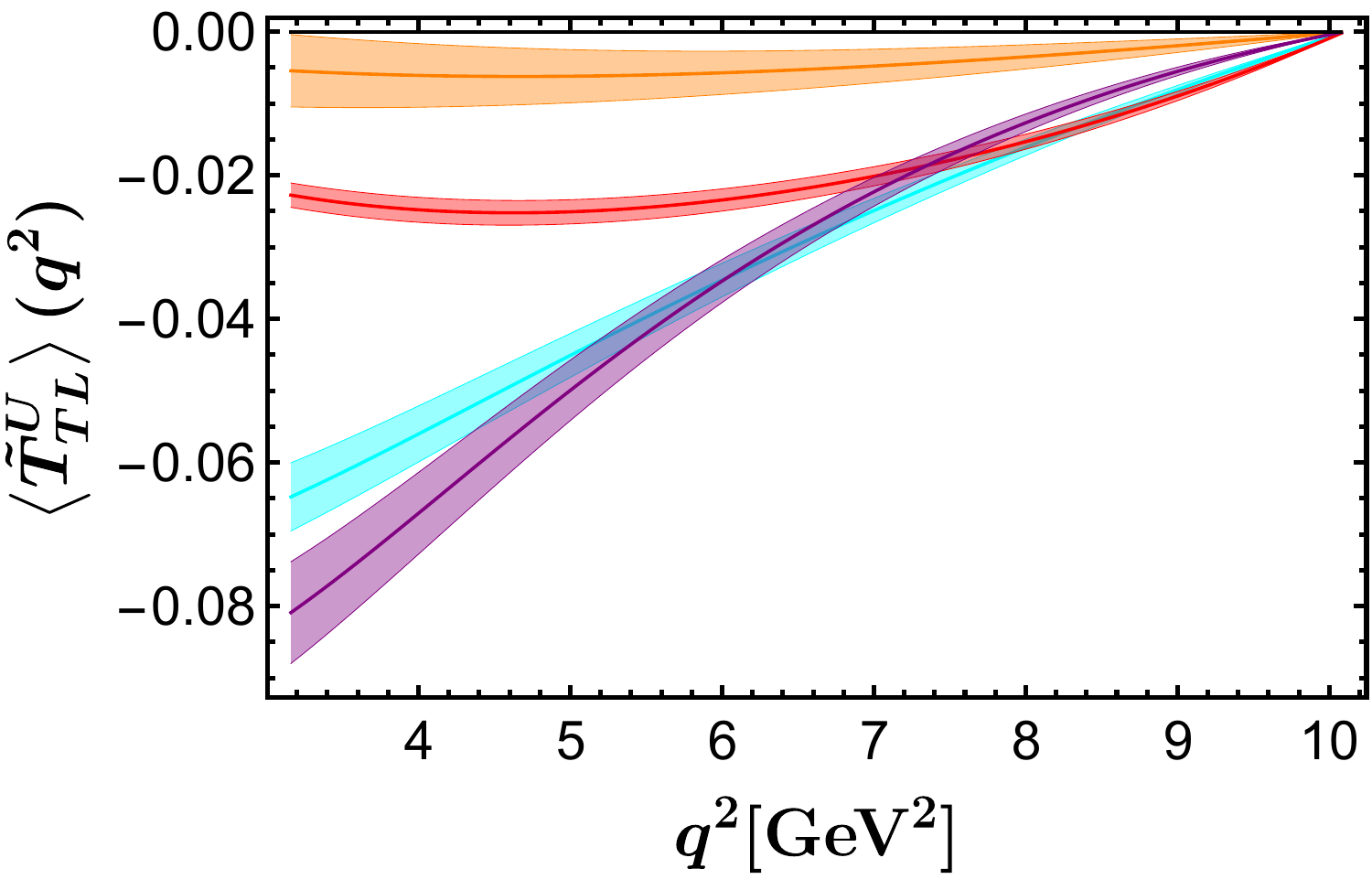}\quad
	\includegraphics[width=0.31\textwidth]{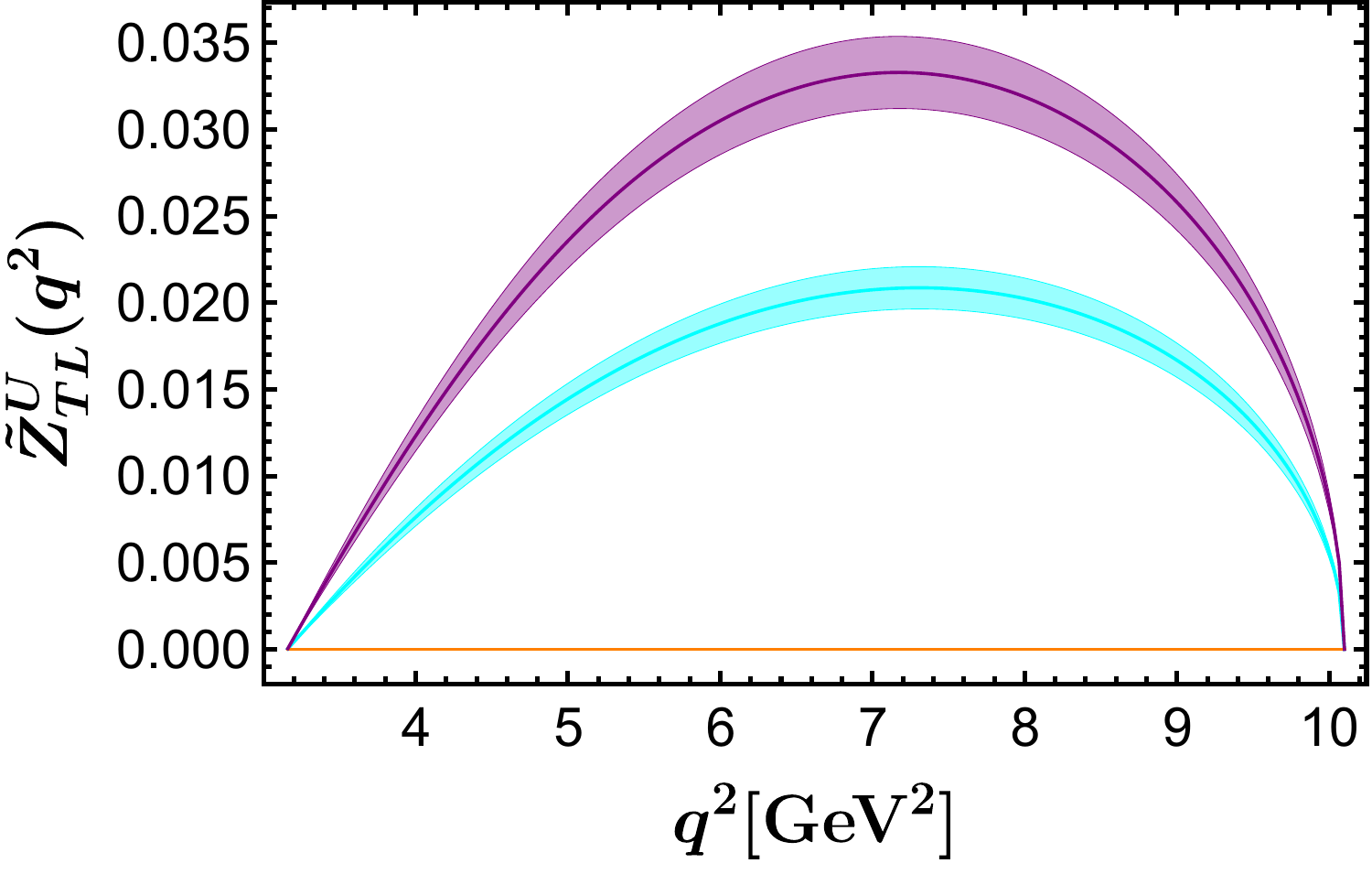}
	\\[4mm]
	\includegraphics[width=0.31\textwidth]{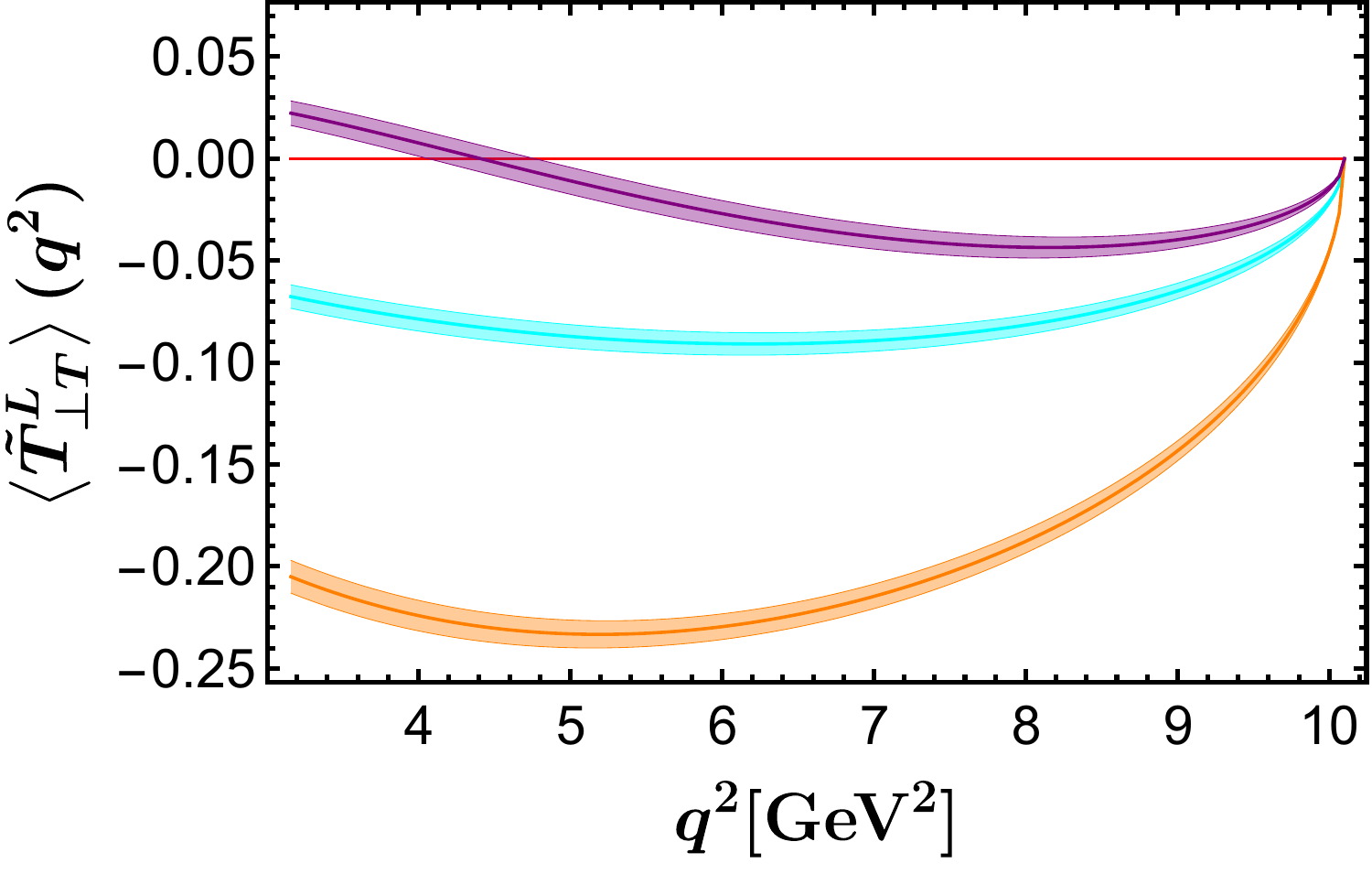}\quad
	\includegraphics[width=0.31\textwidth]{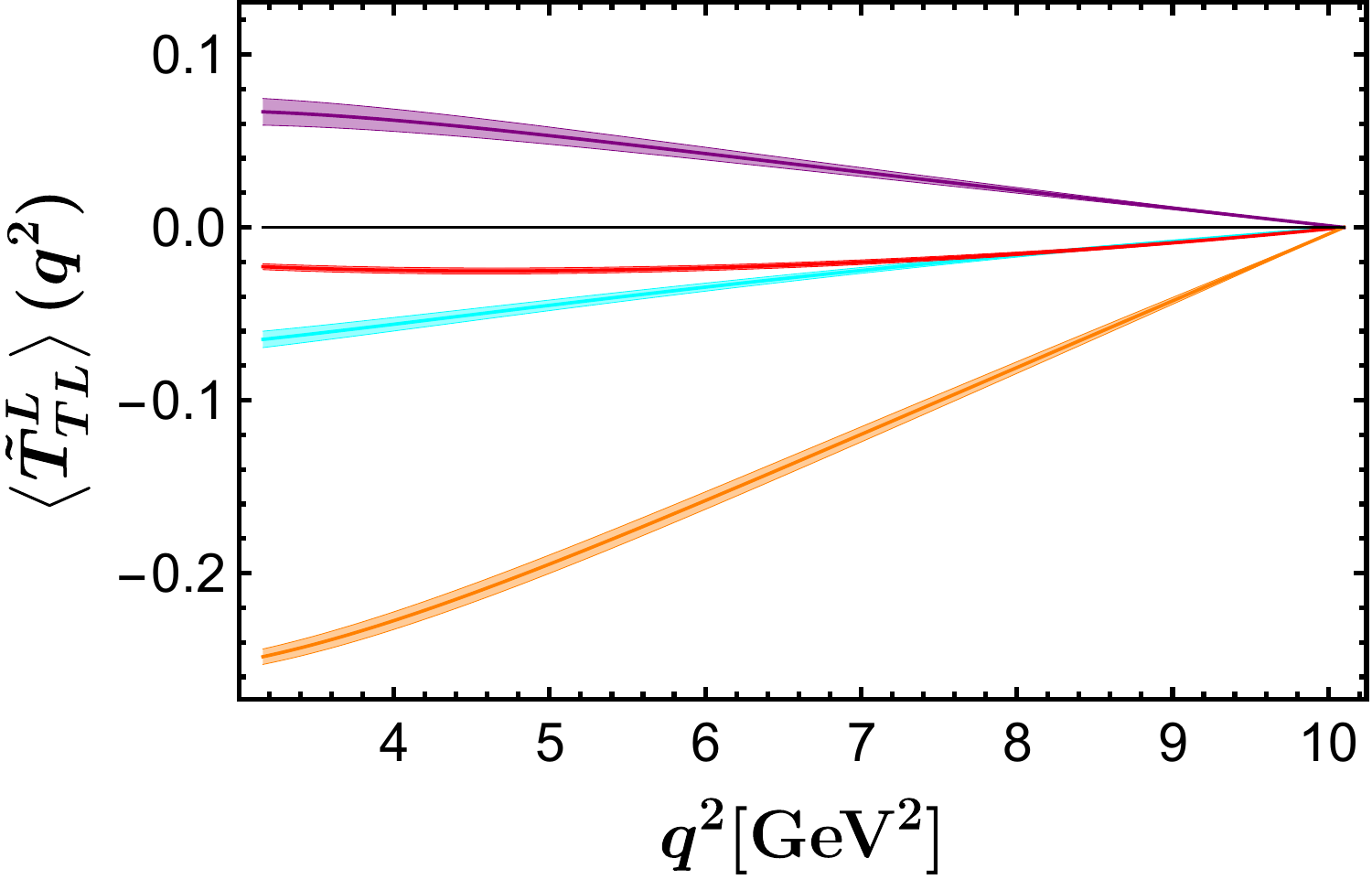}\quad
	\includegraphics[width=0.31\textwidth]{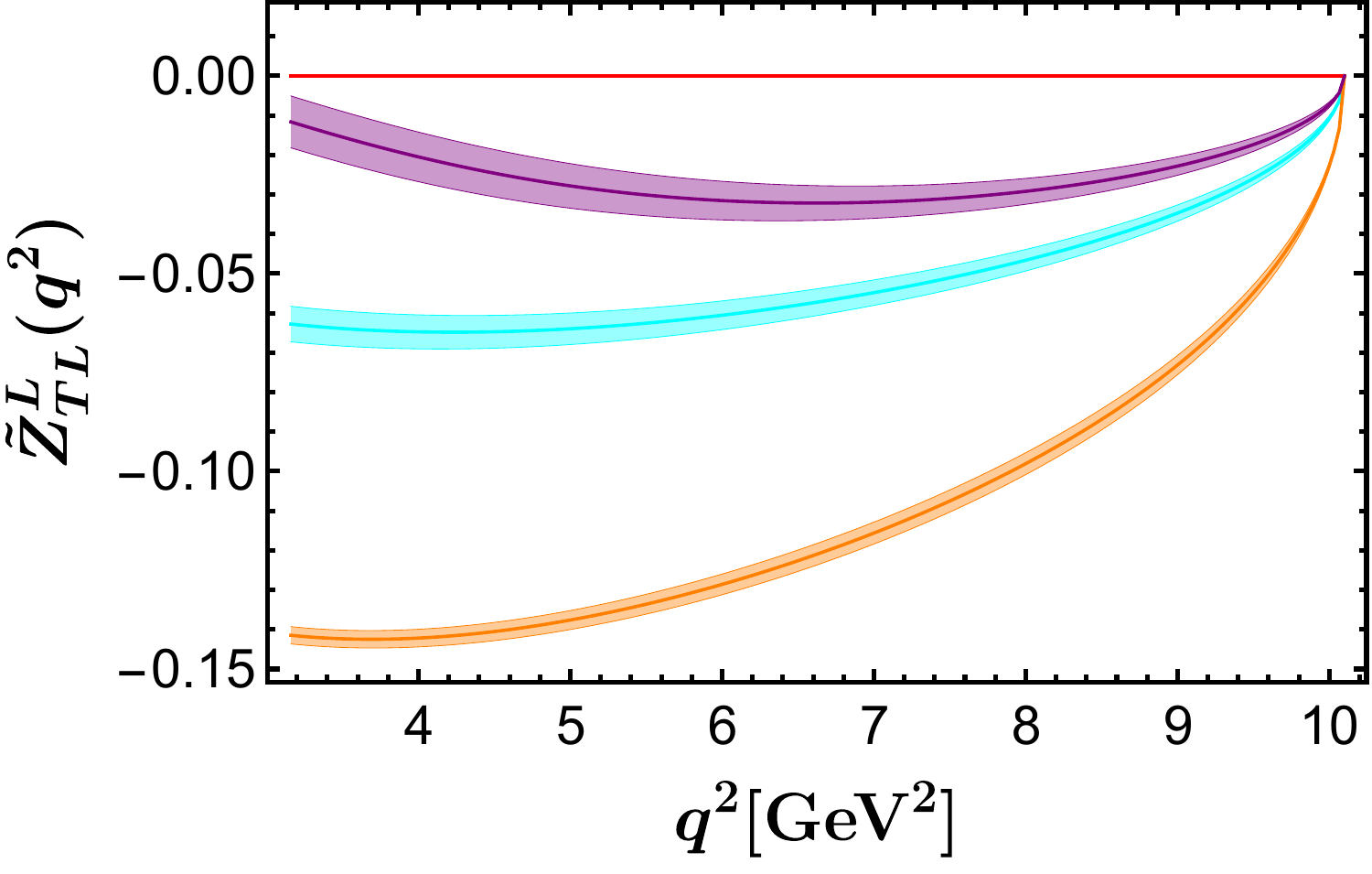}
	\\[4mm]
	\includegraphics[width=0.31\textwidth]{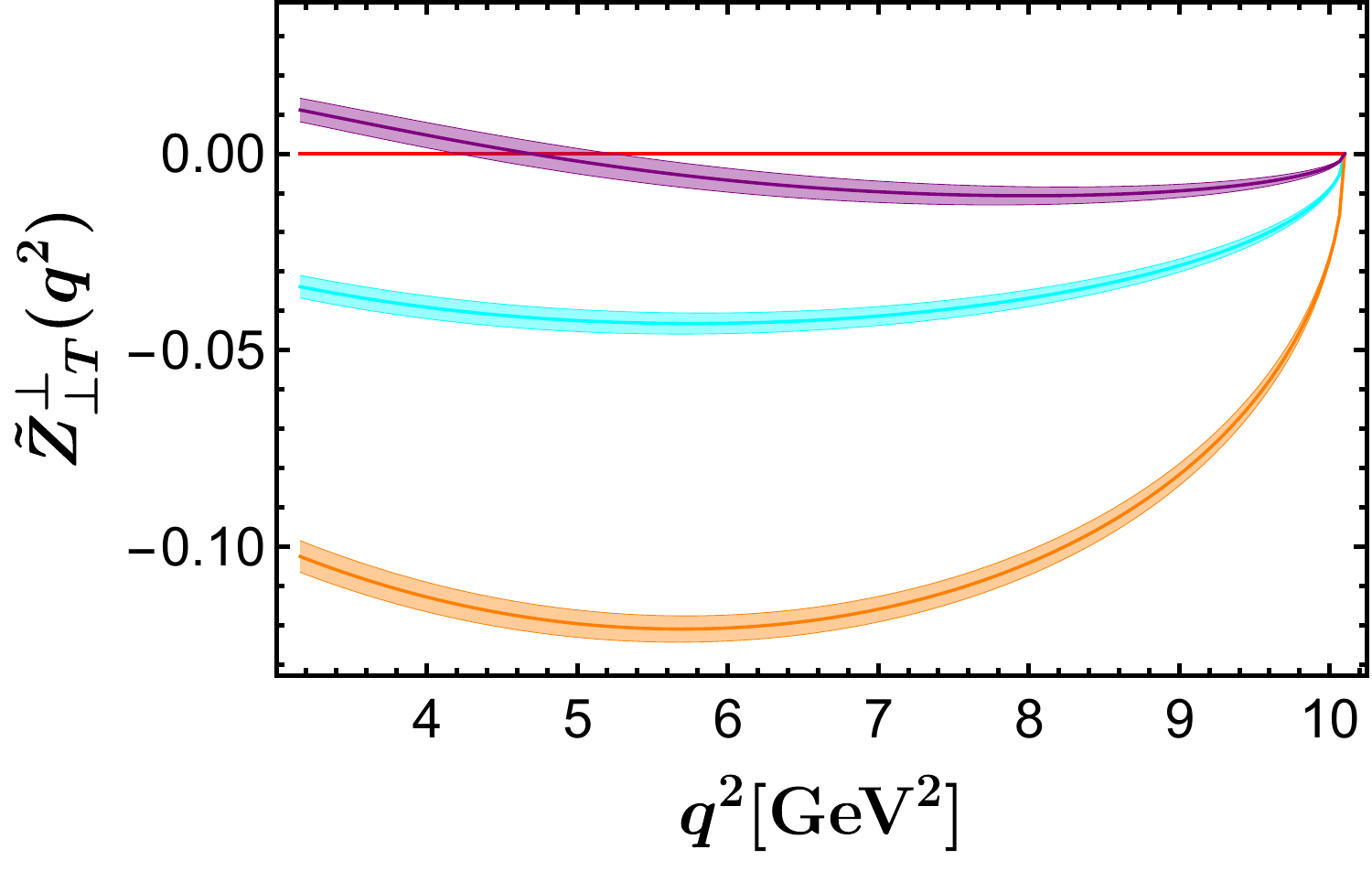}\quad
	\includegraphics[width=0.31\textwidth]{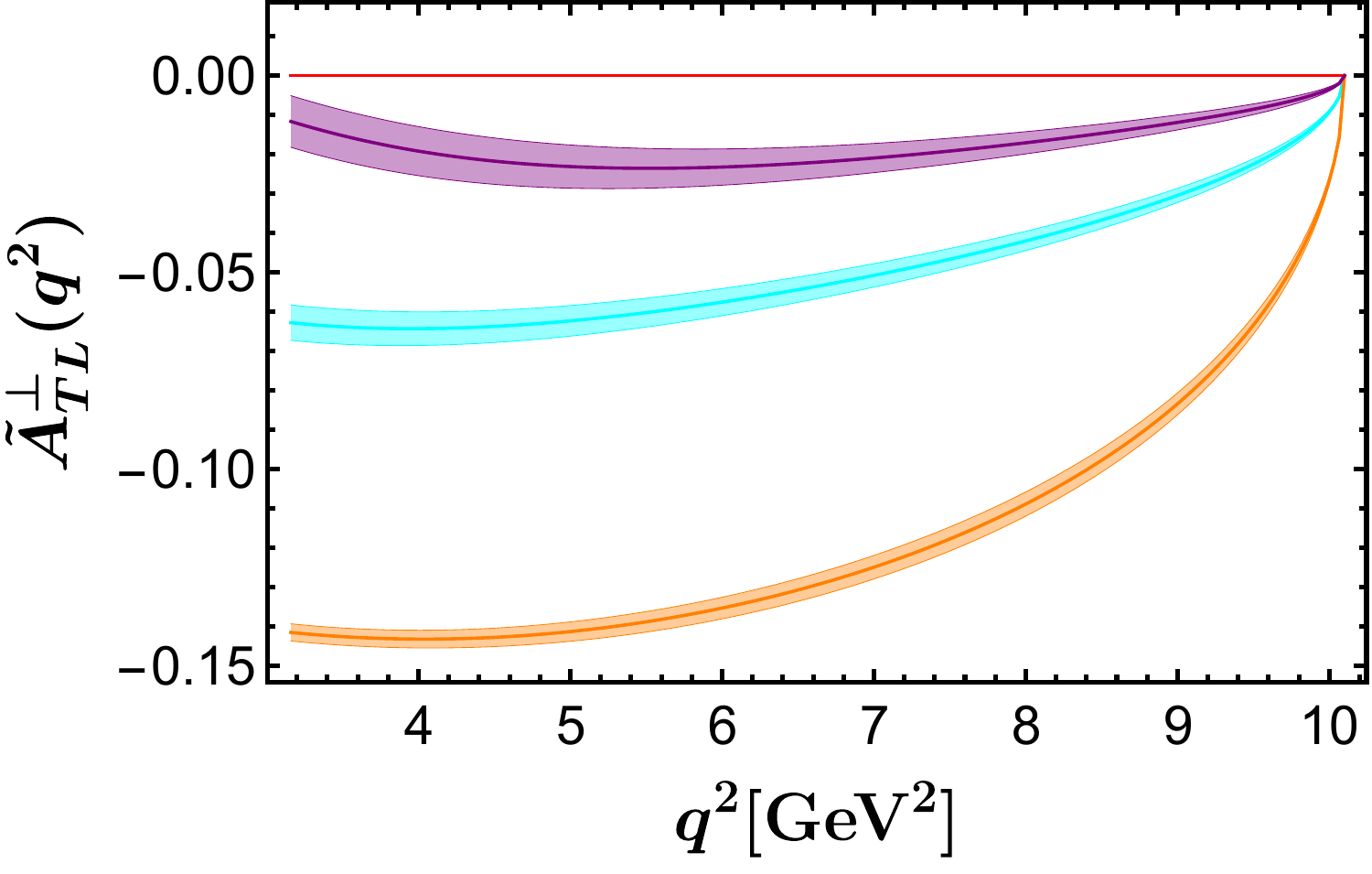}\quad
	\includegraphics[width=0.31\textwidth]{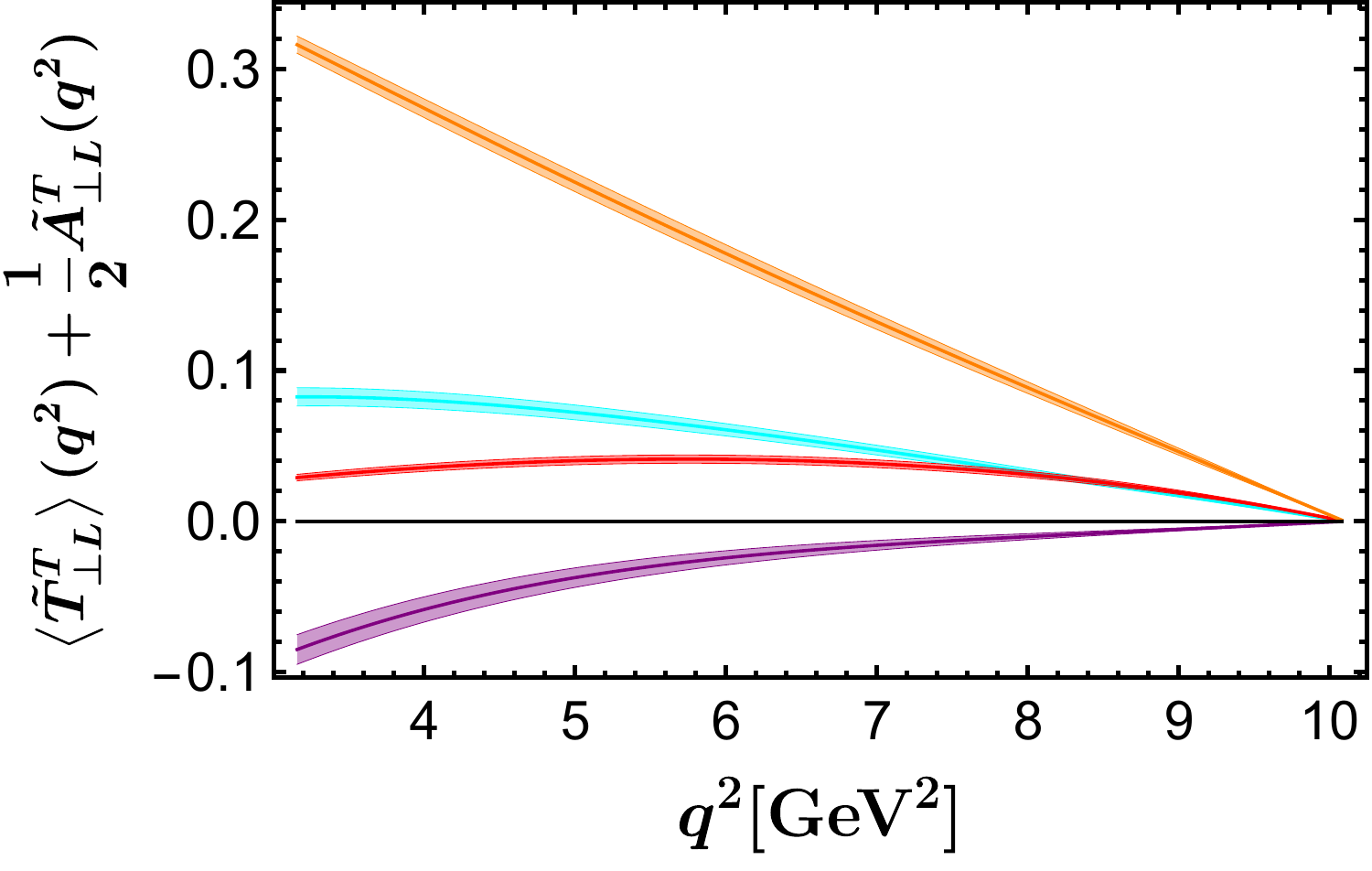}
	\caption{The CP-violating observables as a function of $q^2$, predicted both within the SM and in the four complex NP benchmark points. The black, cyan, red, orange and purple curves represent the results of the SM, BP1, BP2, BP3 and BP4, respectively. The band of each curve is induced by the theoretical uncertainties of $B_c\to J/\psi$ transition form factors~\cite{Harrison:2020gvo,Tang:2022nqm}.} \label{fig:CP-odd}
\end{figure}

Since there exist no direct constraints on the imaginary parts of the NP Wilson coefficients $C^X_{A,B}$ from the current experimental data, the complex NP benchmark points can only be fitted up to a two-fold ambiguity, as indicated by eq.~\eqref{eq:BP1-4}. Here, for simplicity, we choose the imaginary parts to be positive for BP1--BP3, whereas for BP4 a positive $C^V_{RL}$ and a negative $C^S_{LL}$ will be assumed. The numerical results of all the CP-violating observables as a function of $q^2$ are illustrated in figure~\ref{fig:CP-odd}. Since the weak-phase difference of the processes considered is zero within the SM, any observation of these observables being different from zero will be a definite signal of CP-violating NP. It can also be seen that these different NP benchmark points can be distinguished from each other through these CP-violating observables. This makes the measurements of them very promising at the future experiments like the LHCb~\cite{LHCb:2018roe}. 

\subsection{CP-conserving observables}
\label{sec:CP-conserving observables}

\begin{figure}[t]
	\centering
	\includegraphics[width=0.31\textwidth]{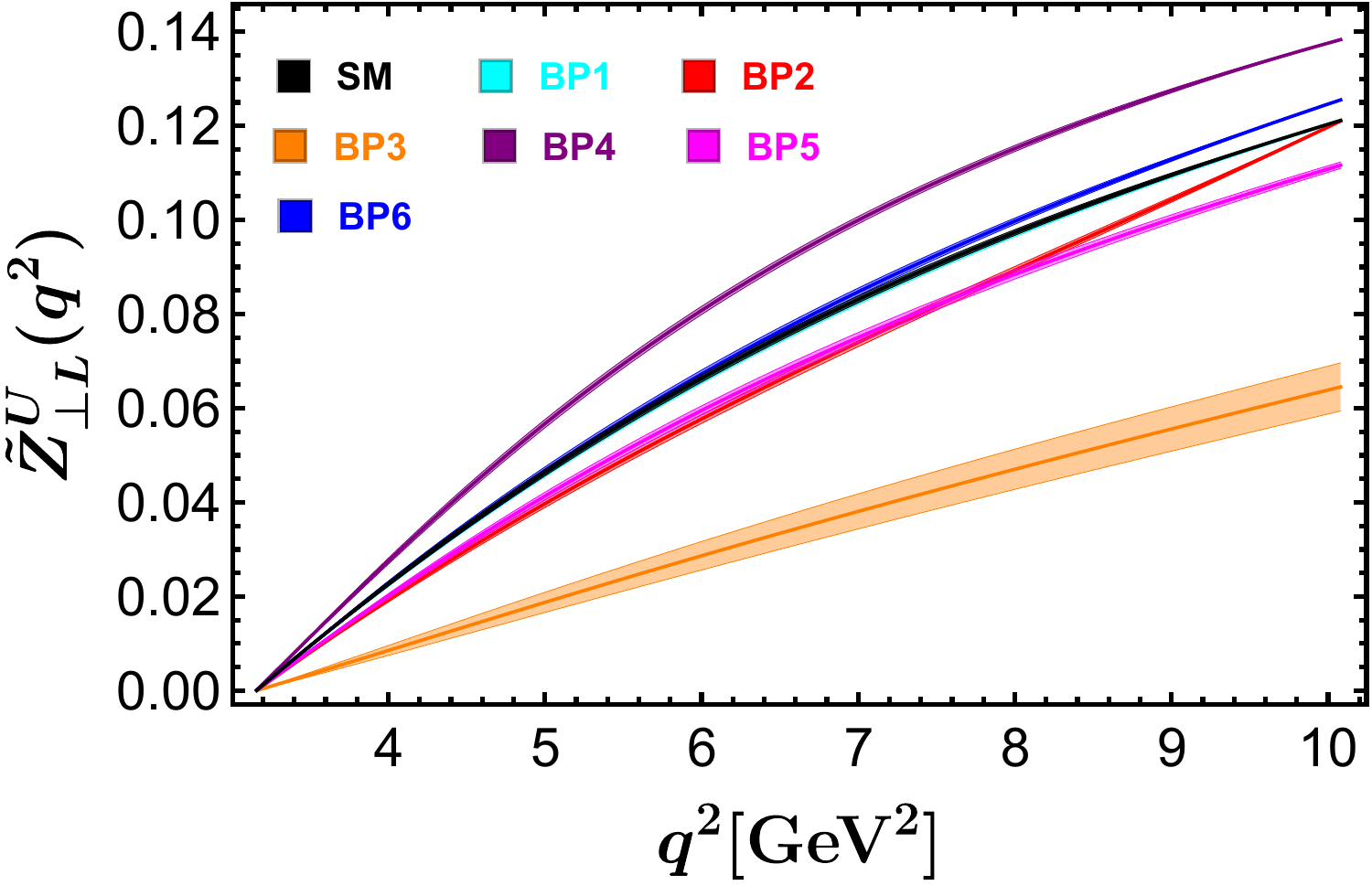}\quad
	\includegraphics[width=0.31\textwidth]{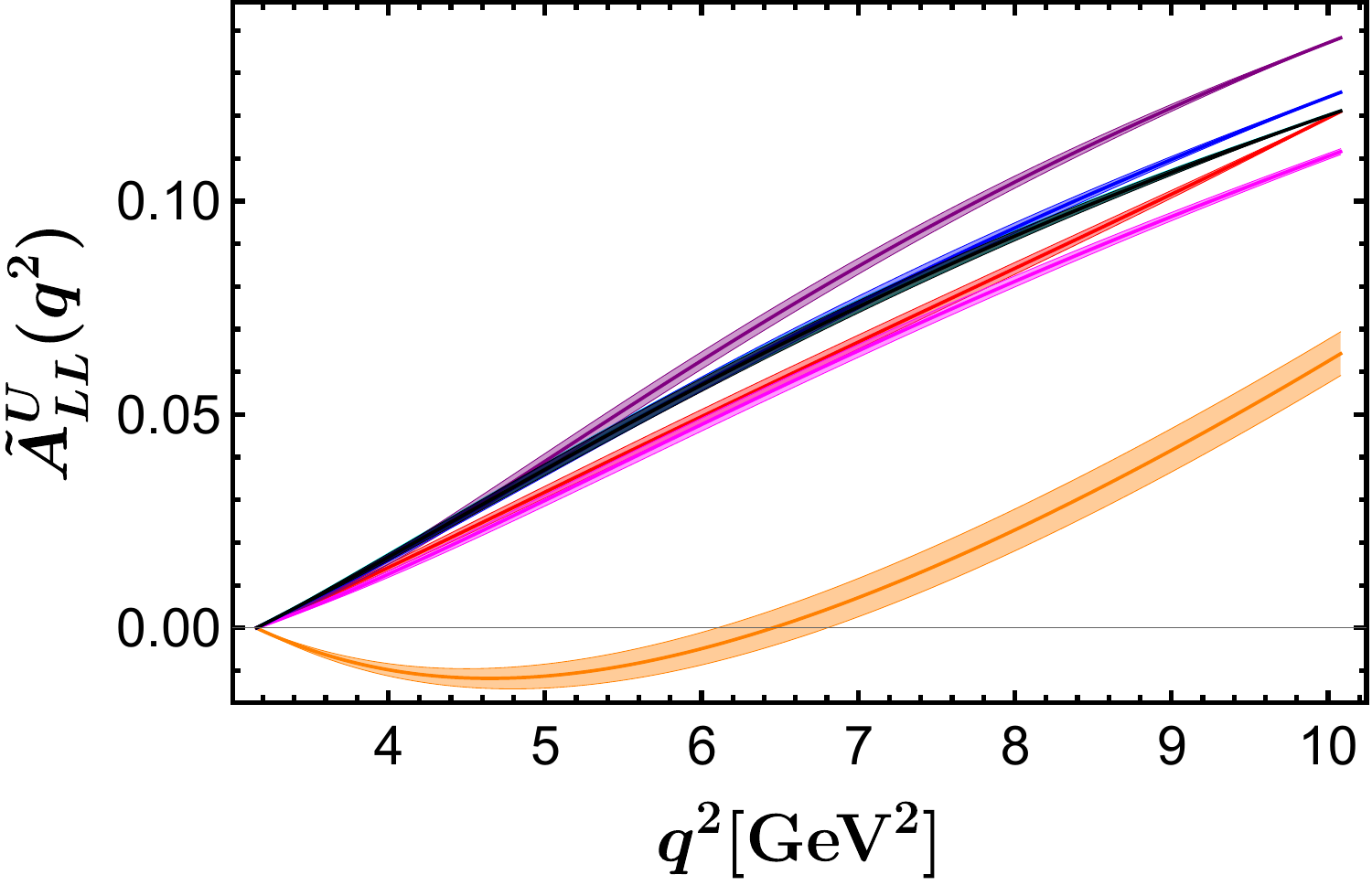}\quad
	\includegraphics[width=0.31\textwidth]{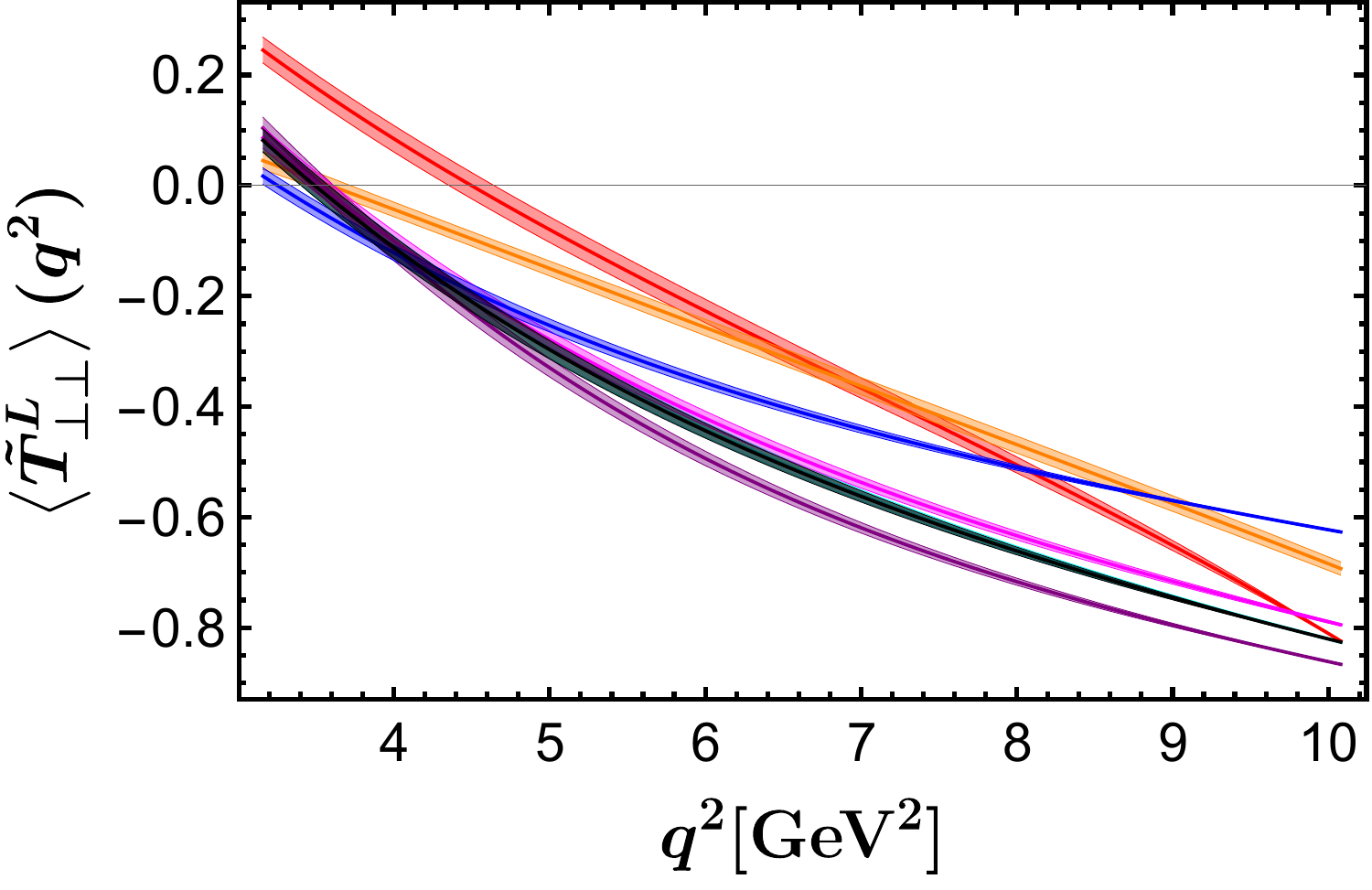}
	\\[4mm]
	\includegraphics[width=0.31\textwidth]{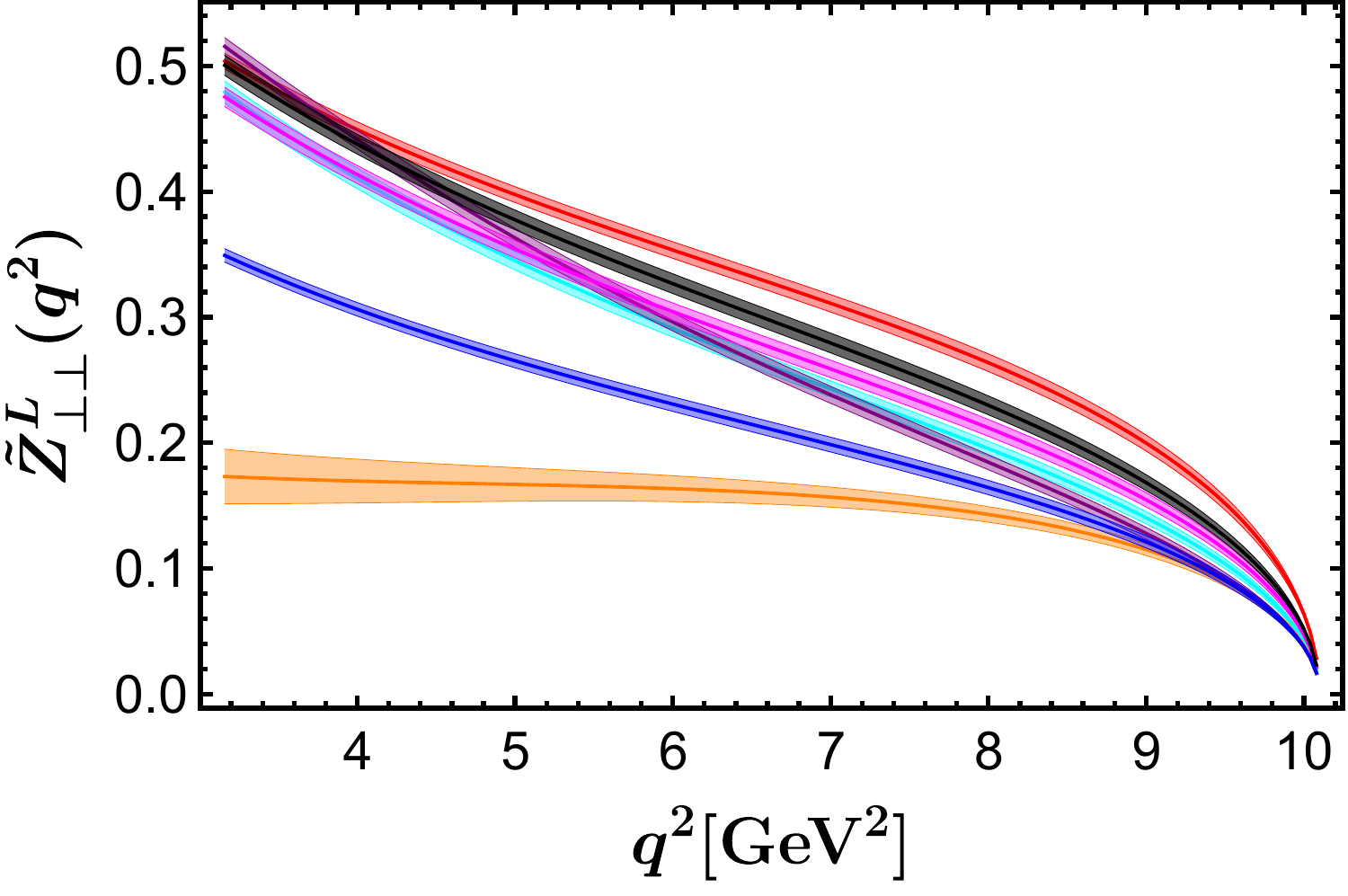}\quad
	\includegraphics[width=0.31\textwidth]{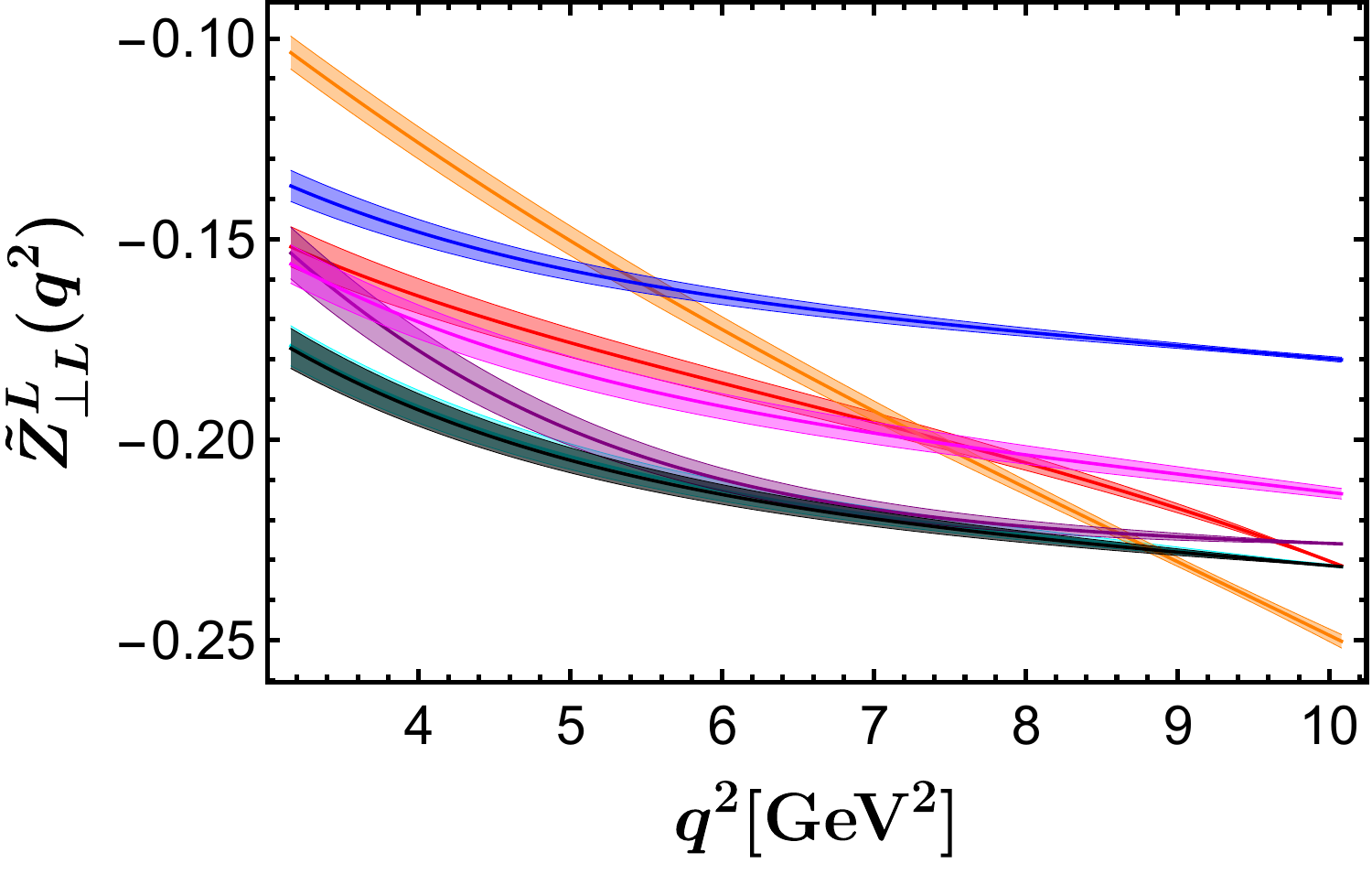}\quad
	\includegraphics[width=0.31\textwidth]{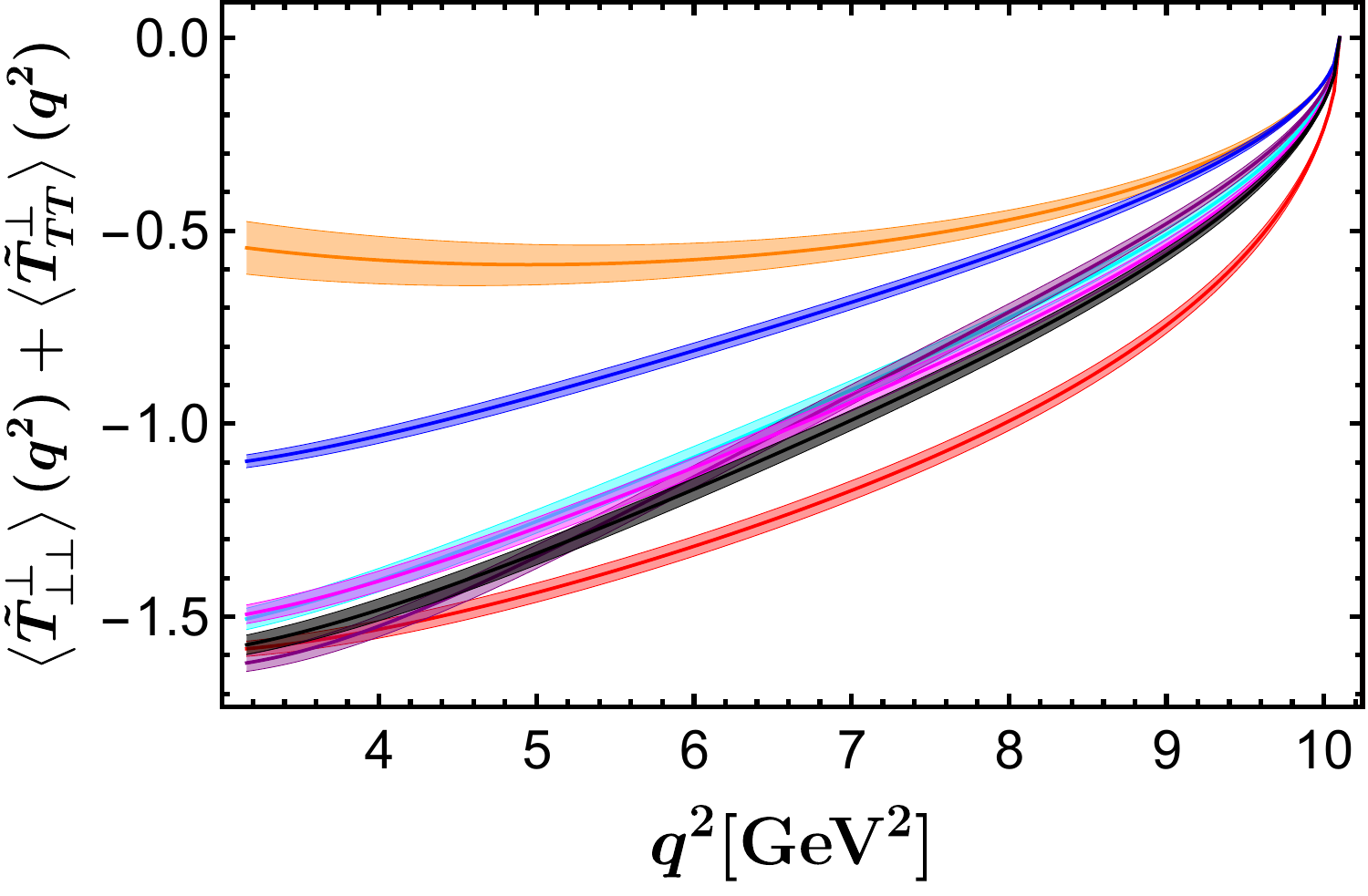}
	\\[4mm]
	\includegraphics[width=0.31\textwidth]{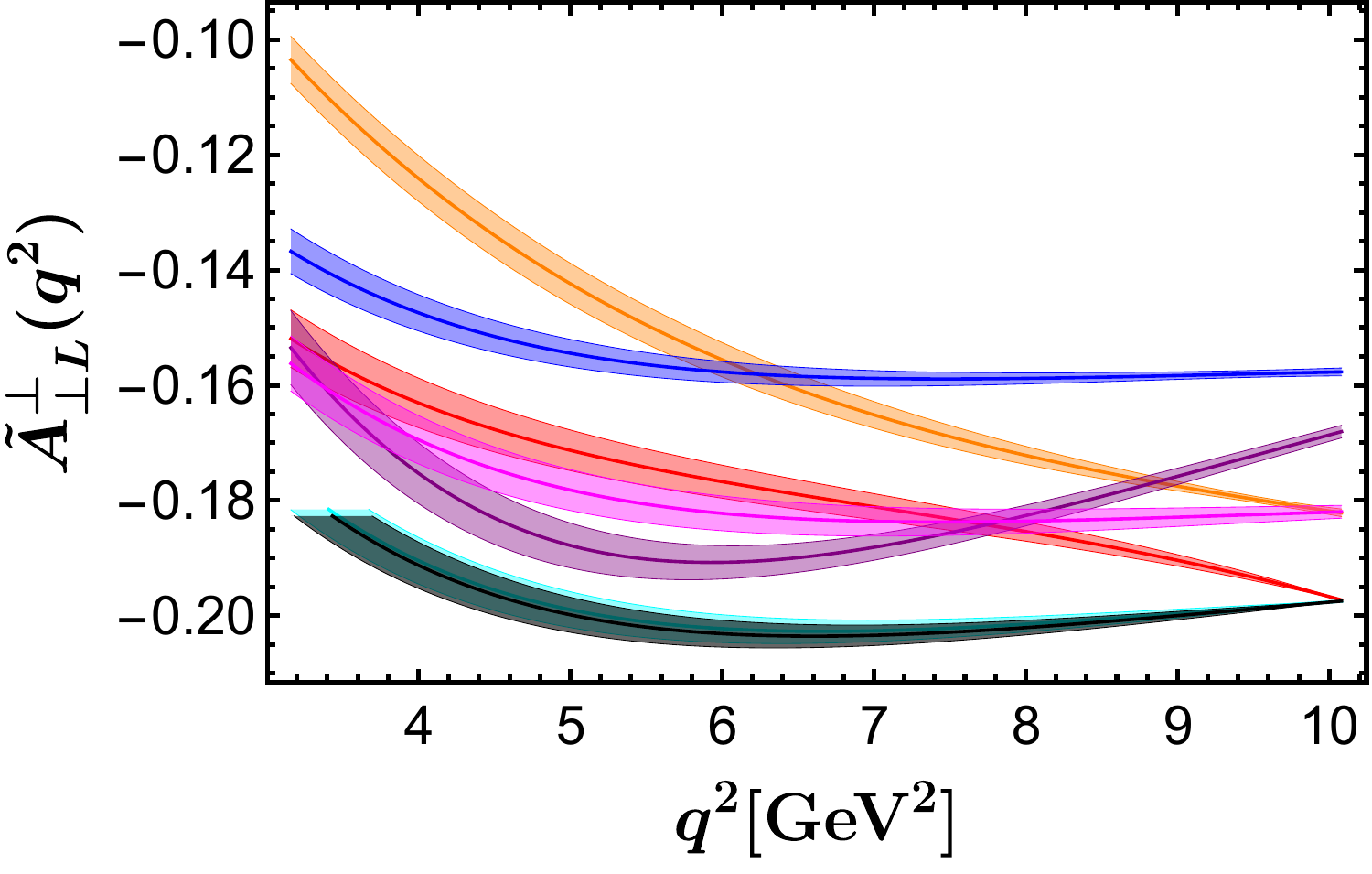}\quad
	\includegraphics[width=0.31\textwidth]{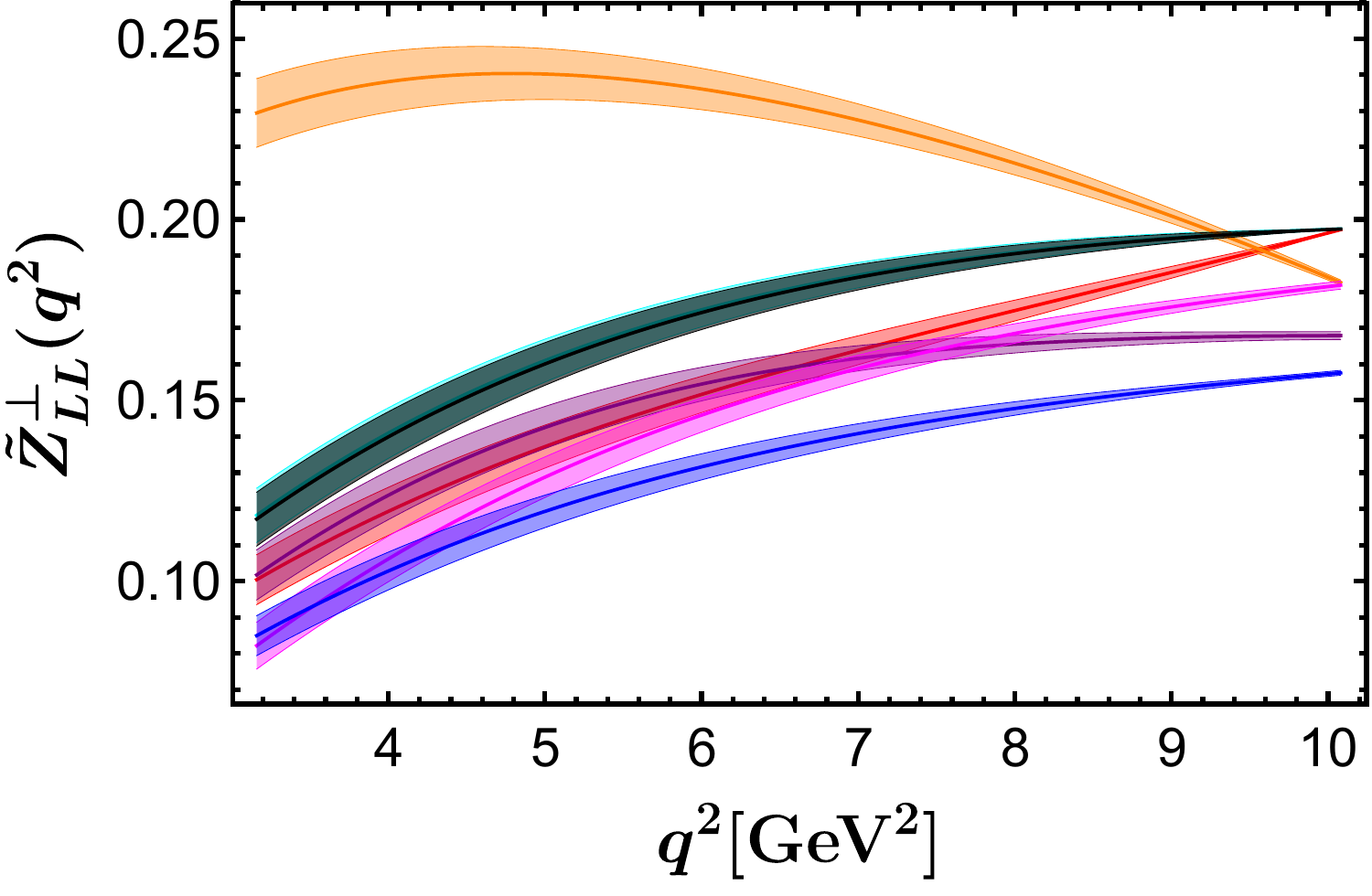}\quad
	\includegraphics[width=0.31\textwidth]{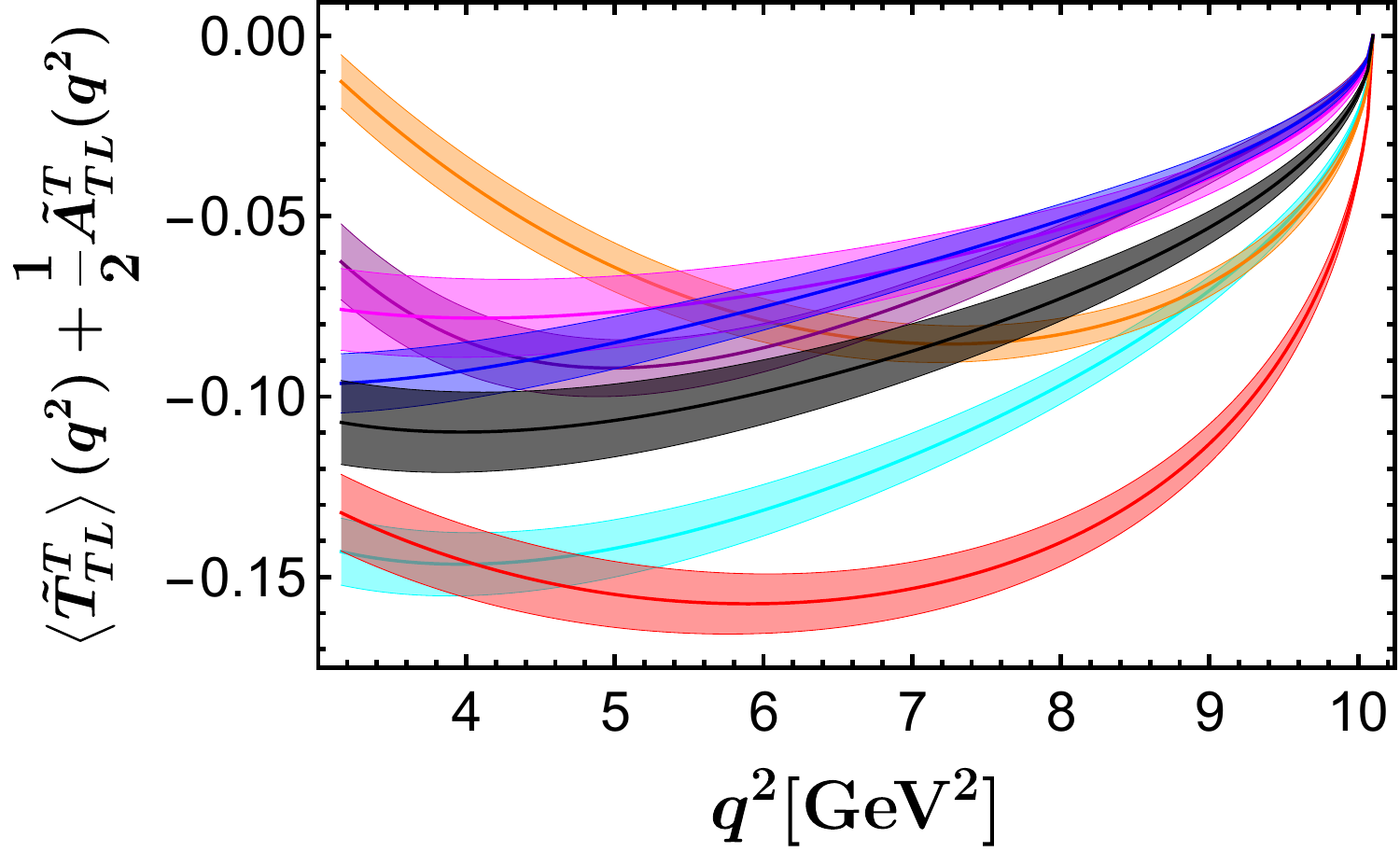}
	\caption{The nine CP-conserving observables as a function of $q^2$, predicted both within the SM and in the six NP benchmark points. Here, compared with figure~\ref{fig:CP-odd}, two more NP benchmark points, BP5 and BP6, are considered, which are represented by the magenta and blue curves, respectively. The other captions are the same as in figure~\ref{fig:CP-odd}.} \label{fig:CP-even}
\end{figure}

The CP-conserving observables can be constructed even in the absence of any NP contribution. In order to demonstrate the NP contributions to these observables, we should compare the experimental measurements of these observables with the corresponding SM predictions. There are in total 25 normalized CP-conserving observables that can be extracted from the fully differential decay rate given by eq.~\eqref{eq:differential decay rate}. Here, as an illustration, we only pick up the nine most interesting ones and show in figure~\ref{fig:CP-even} their sensitivities to the different NP scenarios.

From figure~\ref{fig:CP-even}, we can see that all these nine observables can serve to distinguish the different NP scenarios. Specifically, among all the six NP benchmark points given by eqs.~\eqref{eq:BP1-4} and \eqref{eq:BP5-6}, the BP3 has the largest effect on the selected observables except for $\left\langle \tilde{T}^T_{TL}\right\rangle+\frac{1}{2}\tilde{A}^T_{TL}$, which is found to be mainly affected by the BP2. Furthermore, the observable $\left\langle \tilde{T}^L_{\perp\perp}\right\rangle$ can receive a comparable  BP2 contribution with respect to that of BP3, but is more sensitive to BP2 at low $q^2$ region. On the other hand, the observables $\tilde{Z}^L_{\perp L}$ and $\tilde{A}^\perp_{\perp L}$ can receive a large contribution from BP6 that is comparable to that of BP3, but are more sensitive to BP6 at large $q^2$ region. In addition, the observable $\tilde{A}^\perp_{\perp L}$ is sensitive to, except for the BP1, all the remaining five NP benchmark points. However, the BP1 can be well distinguished from the SM by the observable $\left\langle \tilde{T}^T_{TL}\right\rangle+\frac{1}{2}\tilde{A}^T_{TL}$. Therefore, this CP-conserving observable can be used to search for the non-SMEFT realization of the SM gauge group $SU(2)_L\times U(1)_Y$~\cite{Burgess:2021ylu,London:2022rjt}.

Besides the $q^2$ distributions presented in figures~\ref{fig:CP-odd} and \ref{fig:CP-even}, it is also interesting to consider the integrated values of these observables in different $q^2$ intervals from an experimental perspective. To this end, let us define
\begin{equation}\label{eq:binned observable}
	\overline{O_i}^j = \frac{1}{\Gamma}\int_{(q^2_-)^j}^{(q^2_+)^j} dq^2\frac{d\Gamma}{dq^2}O_i,
\end{equation}
where $O_i$ are the different normalized observables and $\{(q^2_-)^j,(q^2_+)^j\}$ refer to the different $q^2$ intervals. As an illustration, we divide equally the full available $q^2$ range, $m_\tau^2\leq q^2\leq (m_{B_c}-m_{J/\psi})^2$, into three bins, and present in table~\ref{tab:binned observable} the SM predictions of the nine CP-conserving observables in these three different $q^2$ bins. This information may serve as a reference point for the future LHCb measurements.

\begin{table}[t]
	\begin{center}	
		\tabcolsep 0.06in
		\renewcommand\arraystretch{1.5}
		\begin{tabular}{c|ccc}
			\hline \hline 
			\diagbox{Observables}{$\{(q^2_-)^j,(q^2_+)^j\}[\rm{GeV^2}]$} & $\{3.16,5.47\}$ & $\{5.47,7.79\}$ & $\{7.79,10.10\}$ \\
			\hline
			$\overline{\tilde {Z}^U_{\perp L}}^j$ &$0.00529(10)$&$0.03336(29)$&$0.04770(58)$ \\[2mm]
			$\overline{\tilde{A}^U_{L L}}^j$ &$0.00421(13)$&$0.02988(60)$&$0.04619(71)$ \\[2mm]
			$\overline{\left\langle \tilde {T}^L_{\perp \perp}\right\rangle}^j$ &$-0.0329(19)$&$-0.2250(43)$&$-0.3245(48)$ \\[2mm]
			$\overline{\tilde {Z}^L_{\perp \perp}}^j$ &$0.0501(18)$&$0.1244(36)$&$0.0782(25)$ \\[2mm]
			$\overline{\tilde {Z}^L_{\perp L}}^j$ &$-0.02596(51)$&$-0.09282(84)$&$-0.1013(12)$ \\[2mm]
			$\overline{\left\langle \tilde{T}^\perp_{\perp\perp}\right\rangle}^j+\overline{\left\langle \tilde{T}^\perp_{TT}\right\rangle}^j$ &$-0.1752(66)$&$-0.442(12)$&$-0.2641(65)$ \\[2mm]
			$\overline{\tilde{A}^\perp_{\perp L}}^j$ &$-0.02538(50)$&$-0.08651(78)$&$-0.0893(11)$ \\[2mm]
			$\overline{\tilde{Z}^\perp_{LL}}^j$ &$0.01997(65)$&$0.0771(16)$&$0.0864(13)$ \\[2mm]
			$\overline{\left\langle \tilde{T}^T_{TL}\right\rangle}^j+\frac{1}{2}\overline{\tilde{A}^T_{TL}}^j$ 
			&$-0.0137(14)$&$-0.0383(34)$&$-0.0246(21)$ \\
			\hline \hline
		\end{tabular}
		\caption{The SM predictions of the nine CP-conserving observables in three different $q^2$ bins. The uncertainties come from the $B_c\to J/\psi$ transition form factors~\cite{Harrison:2020gvo}.} \label{tab:binned observable}
	\end{center}
\end{table}	

\subsection{Observables specific to the right-handed neutrinos}
\label{sec:R_n}

We also find that some combinations of the observables can only be attributed to the right-handed neutrinos. To this end, let us define the following observables $R_n$:
\begin{align} \label{eq:Rn}
		\frac{d\Gamma}{dq^2}R_1&=\frac{d\Gamma}{dq^2}\left(\tilde{Z}^U_{LL}+\tilde{Z}^L_{LL}\right)=-2\sqrt{6}\mathcal{N}q^2{\rm Re}\left[\mathcal{A}_{R,\parallel}^- \mathcal{A}_{R,\perp}^{-*} \right],\nonumber\\[2mm]
		\frac{d\Gamma}{dq^2}R_2&=\frac{d\Gamma}{dq^2}\left(\tilde{Z}^U_{\perp \perp}-\tilde{Z}^U_{LL}-\tilde{Z}^L_{\perp \perp}\right)=\sqrt{6}\mathcal{N}\left\{2m_\tau^2{\rm Re}\left[\mathcal{A}_{R,t} \mathcal{A}_{R,0}^{+*} \right]+q^2{\rm Re}\left[\mathcal{A}_{R,\parallel}^- \mathcal{A}_{R,\perp}^{-*} \right]\right\},\nonumber\\[2mm]
		\frac{d\Gamma}{dq^2}R_3&=\frac{d\Gamma}{dq^2}\left(\frac{1}{8}\left\langle \tilde{T}^U_{LL}\right\rangle+\tilde{A}^U_{LL}+\frac{3}{8}\left\langle \tilde{T}^L_{LL}\right\rangle\right)\nonumber\\[1.5mm]
		&=-\frac{\sqrt{6}}{4}\mathcal{N}\left\{m_\tau^2\left(\left| \mathcal{A}_{R,\parallel}^+\right|
		^2+\left| \mathcal{A}_{R,\perp}^+\right|
		^2\right)-2q^2\left(\left| \mathcal{A}_{R,\parallel}^-\right|
		^2+\left| \mathcal{A}_{R,\perp}^-\right|
		^2\right)\right\},\nonumber\\[2mm]
		\frac{d\Gamma}{dq^2}R_4&=\frac{d\Gamma}{dq^2}\left(\frac{3}{8}\left\langle \tilde{T}^U_{LL}\right\rangle+\frac{1}{8}\left\langle \tilde{T}^L_{LL}\right\rangle+\tilde{A}^L_{LL}\right)\nonumber\\[1.5mm]
		&=\frac{\sqrt{6}}{4}\mathcal{N}\left\{m_\tau^2\left(\left| \mathcal{A}_{R,\parallel}^+\right|
		^2+\left| \mathcal{A}_{R,\perp}^+\right|
		^2\right)+2q^2\left(\left| \mathcal{A}_{R,\parallel}^-\right|
		^2+\left| \mathcal{A}_{R,\perp}^-\right|
		^2\right)\right\},\nonumber\\[2mm]
		\frac{d\Gamma}{dq^2}R_5&=\frac{d\Gamma}{dq^2}\left(\frac{1}{2}\left\langle \tilde{T}^U_{\perp\perp}\right\rangle+\tilde{A}^U_{\perp\perp}-\frac{3}{8}\left\langle \tilde{T}^U_{LL}\right\rangle-\frac{1}{2}\left\langle \tilde{T}^L_{\perp\perp}\right\rangle-\tilde{A}^L_{\perp\perp}+\frac{3}{8}\left\langle \tilde{T}^L_{LL}\right\rangle\right)\nonumber\\[1.5mm]
		&=\frac{\sqrt{6}}{4}\mathcal{N}m_\tau^2\left\{4\left| \mathcal{A}_{R,t}\right|
		^2+4\left| \mathcal{A}_{R,0}^+\right|
		^2-\left| \mathcal{A}_{R,\parallel}^+\right|
		^2-\left| \mathcal{A}_{R,\perp}^+\right|
		^2\right\},
\end{align}
where explicit expressions of the abbreviation $\mathcal{N}$ and the transversity amplitudes $\mathcal{A}_i$ can be found in appendix~\ref{app:rho_Bc}. It is interesting to note that any observation of these observables will be a clear signal of NP with right-handed neutrinos. Notice that the observable $R_2$ can only get a non-zero contribution from a scalar coupling $C^S_{AR}$ through the interference with the vector coupling $C^V_{AR}$ or the tensor coupling $C^T_{RR}$, whereas the observable $R_5$ can be non-zero even with the pure $C^S_{AR}$ NP scenarios. On the other hand, the remaining observables $R_{1,3,4}$ cannot get any contribution from these couplings. 

\begin{figure}[t]
	\centering
	\includegraphics[width=0.31\textwidth]{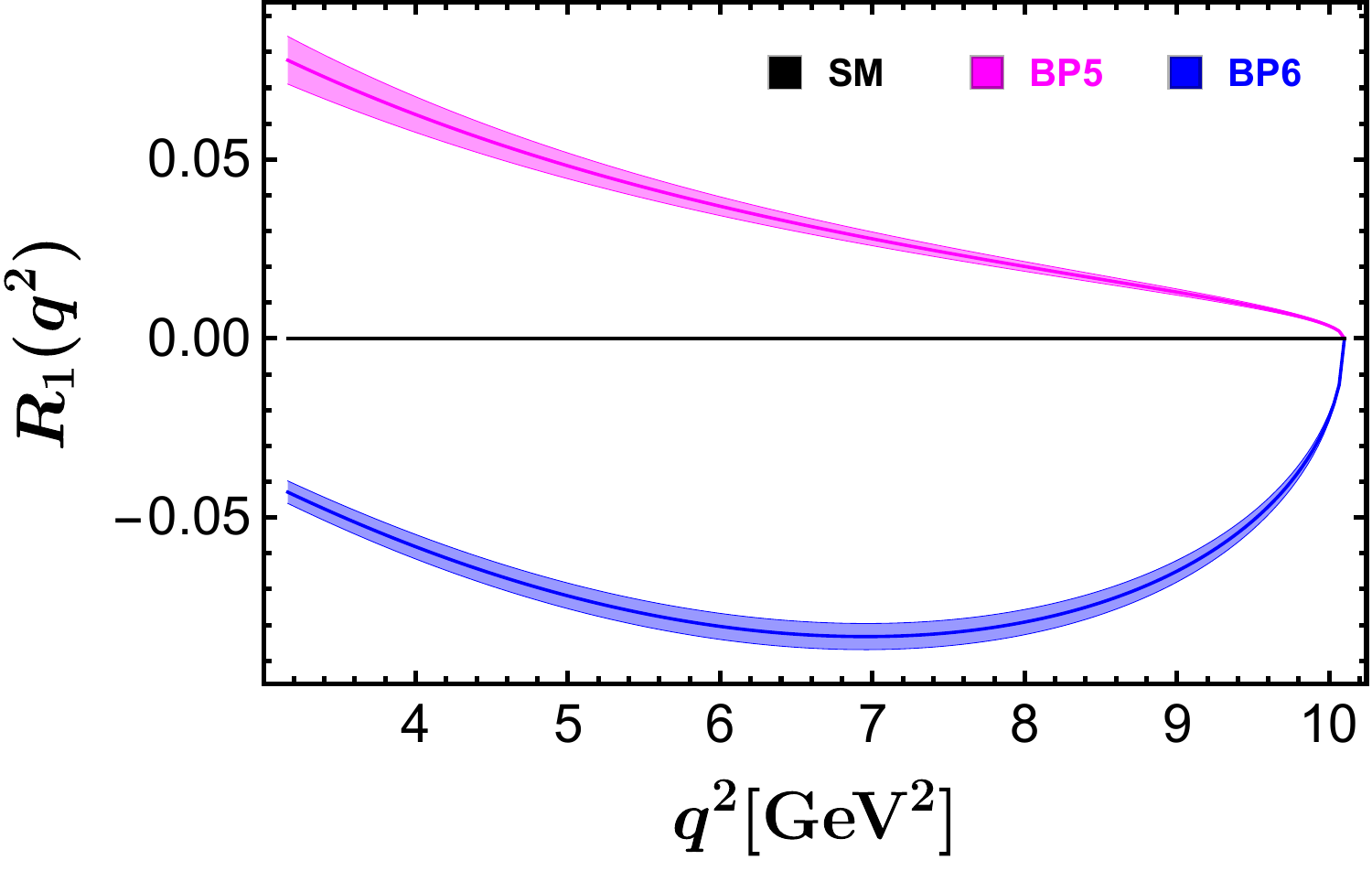}\quad
	\includegraphics[width=0.31\textwidth]{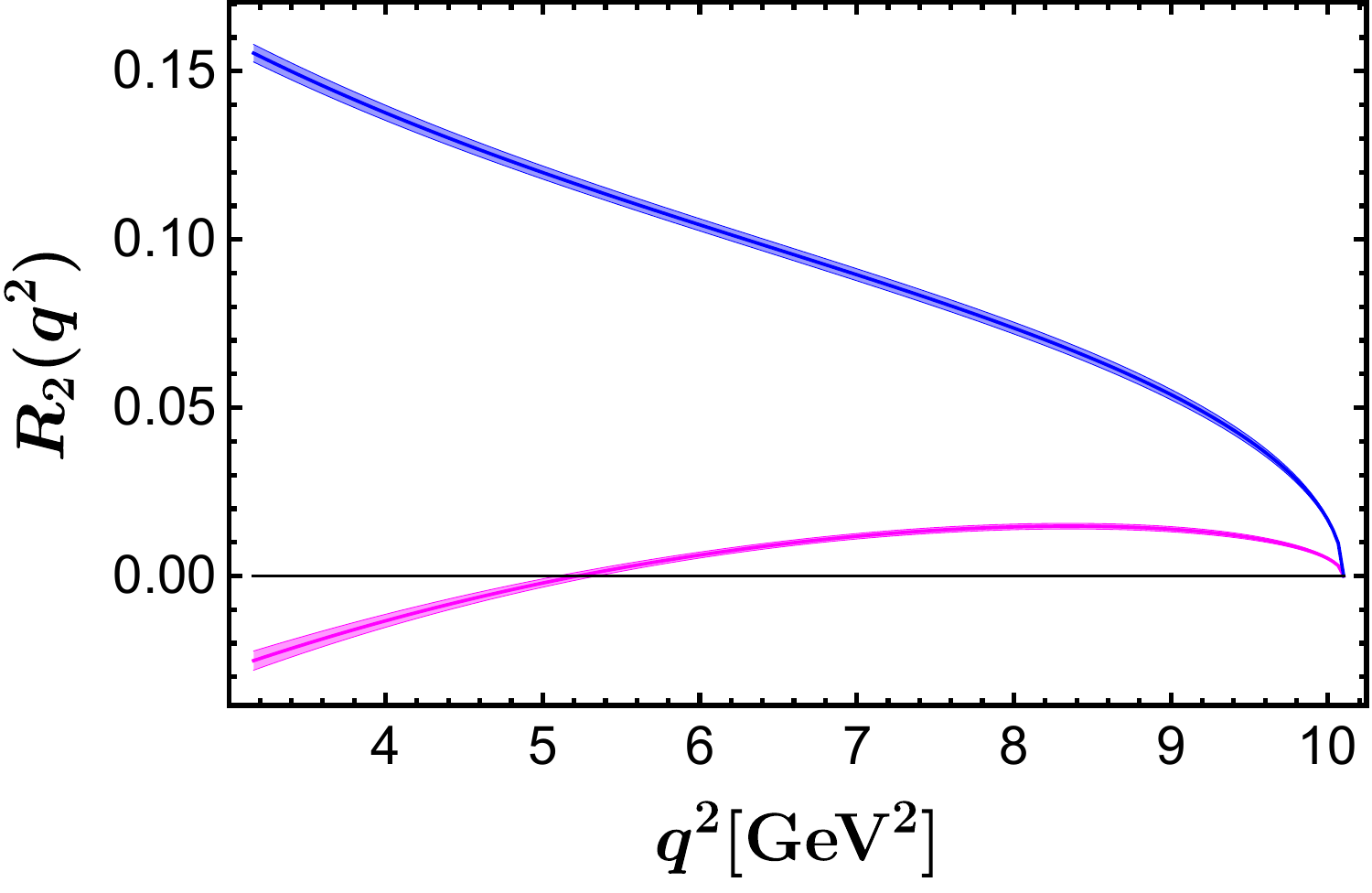}\quad
	\includegraphics[width=0.31\textwidth]{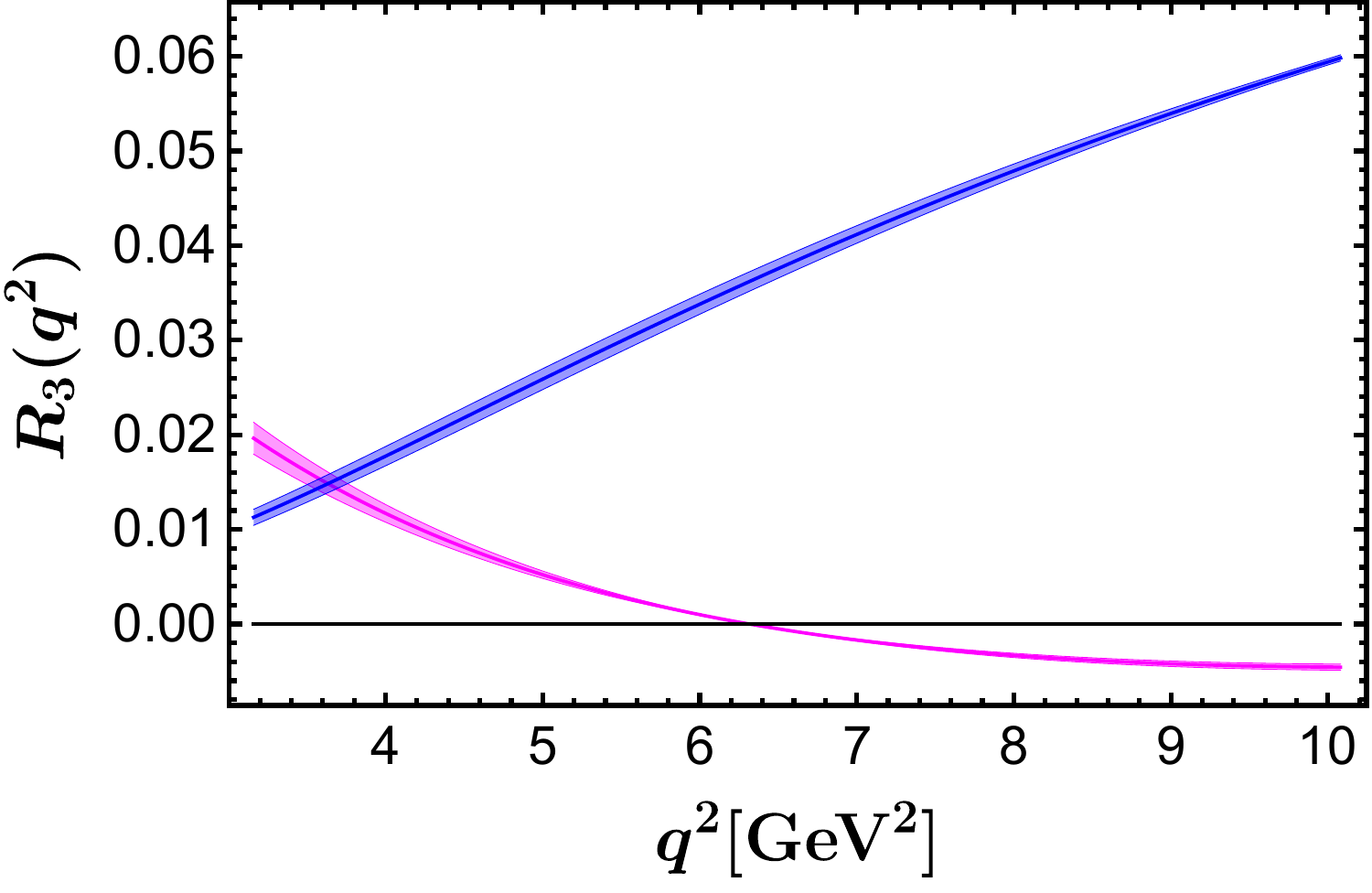}
	\\[4mm]
	\includegraphics[width=0.31\textwidth]{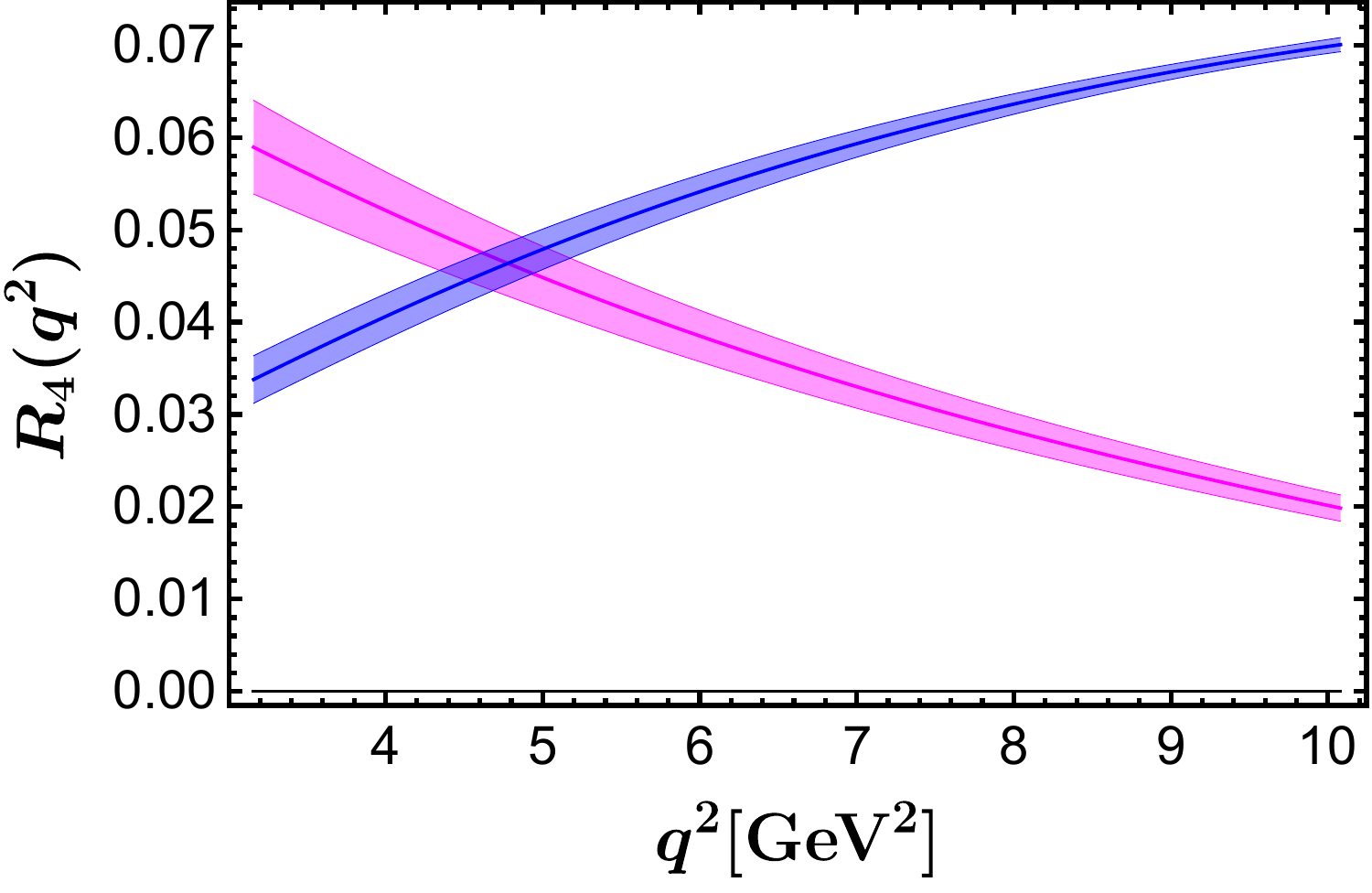}\qquad
	\includegraphics[width=0.31\textwidth]{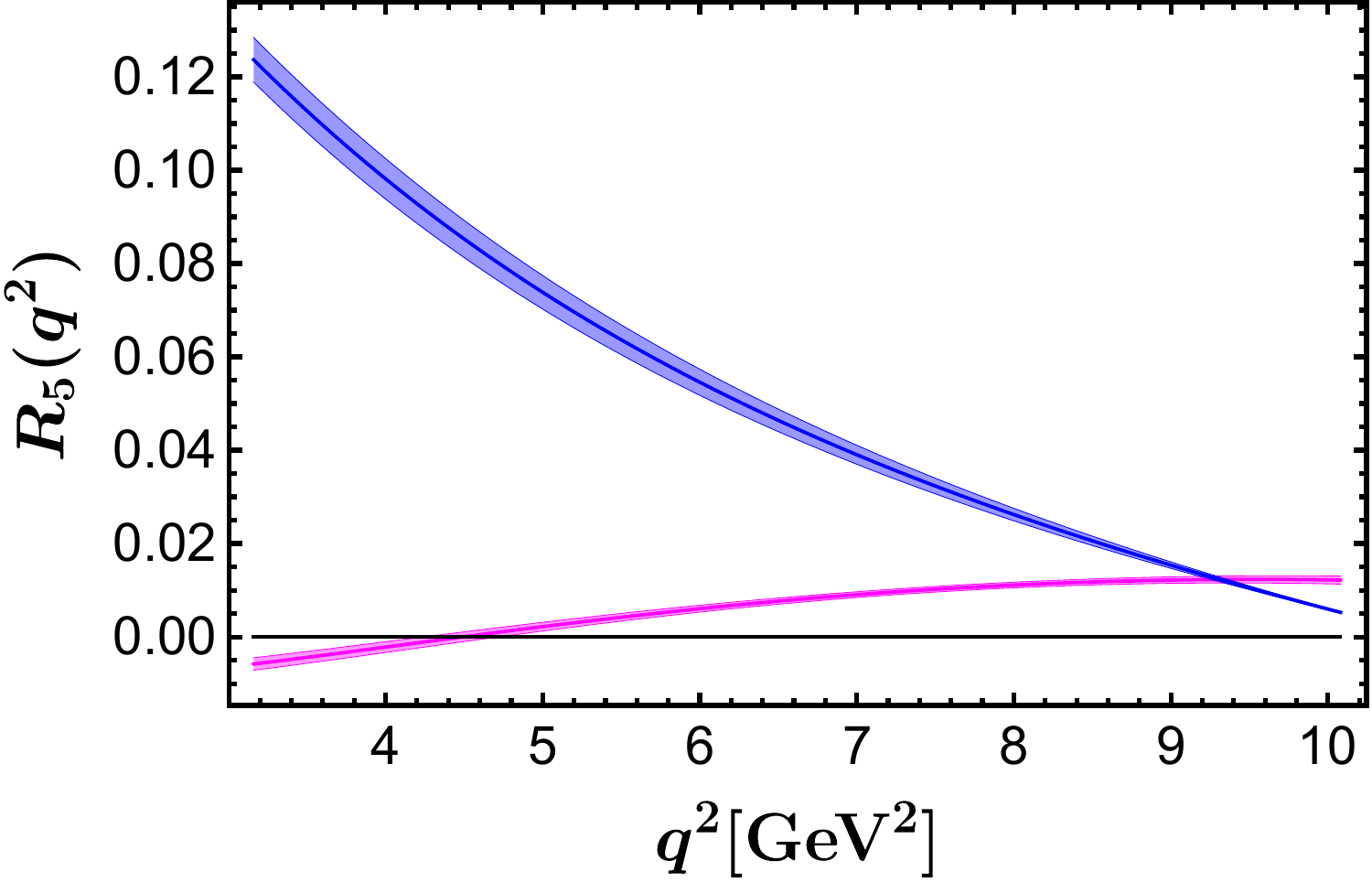}
	\caption{The observables $R_n$ defined by eq.~\eqref{eq:Rn} as a function of $q^2$, predicted both within the SM and in the two NP benchmark points with right-handed neutrinos. The other captions are the same as in figures~\ref{fig:CP-odd} and \ref{fig:CP-even}.} \label{fig:R}
\end{figure}

The sensitivities of these observables to the NP benchmark points BP5 and BP6 are shown in figure~\ref{fig:R}, from which we can see that these two benchmark points can be well distinguished from each other through these observables.

\subsection{Integrated observables}
\label{sec:Integrated observables}

Although the observables discussed above can be used to discern the different NP scenarios, their extractions from the fully differential distribution given by eq.~\eqref{eq:differential decay rate} may suffer from low experimental statistics. A way out is then to consider possible statistically enhanced distributions, which can be obtained by integrating eq.~\eqref{eq:differential decay rate} over one or more of the related kinematic variables. The resulting integrated observables can still be used to study possible NP contributions, as discussed already in refs.~\cite{Hu:2021emb,Penalva:2022vxy}. In this work, we only discuss the distributions with only one visible kinematic variable left.

We start by integrating eq.~\eqref{eq:differential decay rate} over the variables $\phi_d$ and $\theta_{J/\psi}$ to get
\begin{equation} \label{eq:threedistribution}
	\begin{aligned}
		&\frac{d^3\Gamma}{dq^2dE_d d\cos\theta_d}/\frac{d\Gamma}{dq^2}=\frac{4\sqrt{6}}{3}\pi \bigg\{\left(\mathcal{N}^U -\mathcal{N}^L\left\langle P_U^L\right\rangle\right)\\[1.5mm]
		&\quad\quad+\left[2\cos\theta_{\tau  d}\left(\mathcal{N}^UA_{FB}-\mathcal{N}^LZ_U^L\right)-\frac{4}{\pi}\sin^2\theta_{\tau d}\mathcal{N}^\perp \left\langle P_U^\perp\right\rangle\right]P_1\left(\cos\theta_d\right)\\[1.5mm]
		&\quad\quad+\left[\left(3\cos^2\theta_{\tau  d}-1\right)\left(\mathcal{N}^UA_Q-\mathcal{N}^LA_U^L\right)-3\cos\theta_{\tau d}\sin^2\theta_{\tau d}\mathcal{N}^\perp Z^\perp_U\right]P_2\left(\cos\theta_d\right)\bigg\}.
	\end{aligned}
\end{equation}
Since the spin of the $J/\psi$ meson has already been summed over in this distribution, we can use the relations $\sum_i\tilde{T}^{i^\prime}_{ii}=\sqrt{6}P^{i^\prime}_U$ and $\sum_i\tilde{T}^{U}_{ii}=\sqrt{6}$ to regain the $\tau$ vector polarizations. This allows us to define the following observables:
 \begin{equation}
 	\begin{aligned}
 		\frac{d\Gamma}{dq^2}A_{FB}=&\left(\int_{0}^{1}-\int_{-1}^{0}\right)d\cos\theta_\tau\frac{d^2\Gamma}{dq^2d\cos\theta_\tau},\\[2mm]
 		\frac{d\Gamma}{dq^2}A_Q=&\frac{5}{2}\int_{-1}^{1}d\cos\theta_\tau P^0_2(\cos\theta_\tau)\frac{d^2\Gamma}{dq^2d\cos\theta_\tau},\\[2mm]
 		\frac{d\Gamma}{dq^2}\left\langle P^{i^\prime}_U\right\rangle=&\int_{-1}^{1}d\cos\theta_\tau\frac{d^2\Gamma}{dq^2d\cos\theta_\tau}P^{i^\prime}_U,\\[2mm]
 		\frac{d\Gamma}{dq^2}Z^{i^\prime}_U=&\left(\int_{0}^{1}-\int_{-1}^{0}\right)d\cos\theta_\tau\frac{d^2\Gamma}{dq^2d\cos\theta_\tau}P^{i^\prime}_U,\\[2mm]
 		\frac{d\Gamma}{dq^2}A^L_U=&\frac{5}{2}\int_{-1}^{1}d\cos\theta_\tau P^0_2(\cos\theta_\tau)\frac{d^2\Gamma}{dq^2d\cos\theta_\tau}P^L_U,
 	\end{aligned}
 \end{equation}
which are consistent with that derived in refs.~\cite{Asadi:2020fdo,Penalva:2021wye}, but given with different notations. Then, the distributions $d\Gamma/d\cos\theta_d$ and $d\Gamma/dq^2$ can be easily derived after performing further the integration of eq.~\eqref{eq:threedistribution} over the other kinematic variables with the ranges specified in appendix~\ref{app:phase space}. On the other hand, to obtain the $E_d$ distribution, one should firstly perform the $q^2$ integration for a given $E_d$. This requires to invert the limits of $E_d$ presented in appendix~\ref{app:phase space}, and the resulting ranges of $q^2$ for the $\tau^-\to\pi^-(\rho^-)\nu_\tau$ channels are given by
\begin{equation}
	\begin{aligned}
		&\mbox{Part I:} && \frac{m_{\tau}^{4}+m_d^{2} q_{u}^{2}}{2 m_{\tau}^{2} \sqrt{q_{u}^{2}}}\leq E_d \leq \frac{m_{\tau}^{2}+m_d^{2}}{2 m_{\tau}}, \quad\frac{m_\tau^4}{m_d^4}\left(E_d-\left|\vec{p}_d\right|\right)^2\leq q^2\leq q_{u}^{2},\\[2mm]
		&\mbox{Part II:} && \frac{m_{\tau}^{2}+m_d^{2}}{2 m_{\tau}} \leq E_d \leq \frac{m_d^{2}+q_{u}^{2}}{2 \sqrt{q_{u}^{2}}}, \quad\left(E_d+\left|\vec{p}_d\right|\right)^2\leq q^2\leq q_{u}^{2},
	\end{aligned}
\end{equation}
with $q_{u}^{2}=\left(m_{B_c}-m_{J/\psi}\right)^2$ being the upper limit of $q^2$. Similarly, we can get the allowed ranges of $q^2$ for the $\tau^-\to \ell^-\bar{\nu}_\ell\nu_\tau$ channel as
\begin{equation}
	\begin{aligned}
		&\mbox{Part I:} && m_d\leq E_d \leq \frac{m_{\tau}^{4}+m_d^{2} q_{u}^{2}}{2 m_{\tau}^{2} \sqrt{q_{u}^{2}}}, \quad m_\tau^2\leq q^2\leq q_{u}^{2},\quad  x_-\leq x\leq x_+,\\[2mm]
		&\mbox{Part II:} && \frac{m_{\tau}^{4}+m_d^{2} q_{u}^{2}}{2 m_{\tau}^{2} \sqrt{q_{u}^{2}}}\leq E_d \leq \frac{m_{\tau}^{2}+m_d^{2}}{2 m_{\tau}}, \quad m_\tau^2\leq q^2\leq \frac{m_\tau^4}{m_d^4}\left(E_d-\left|\vec{p}_d\right|\right)^2,\quad  x_-\leq x\leq x_+,\\[2mm]
		&\mbox{Part III:} && \frac{m_{\tau}^{4}+m_d^{2} q_{u}^{2}}{2 m_{\tau}^{2} \sqrt{q_{u}^{2}}}\leq E_d \leq \frac{m_{\tau}^{2}+m_d^{2}}{2 m_{\tau}}, \quad \frac{m_\tau^4}{m_d^4}\left(E_d-\left|\vec{p}_d\right|\right)^2\leq q^2\leq q_{u}^{2},\quad x_-\leq x\leq1+y^2,\\[2mm]
		&\mbox{Part IV:} && \frac{m_{\tau}^{2}+m_d^{2}}{2 m_{\tau}} \leq E_d \leq \frac{m_d^{2}+q_{u}^{2}}{2 \sqrt{q_{u}^{2}}}, \quad \left(E_d+\left|\vec{p}_d\right|\right)^2\leq q^2\leq q_{u}^{2},\quad  x_-\leq x\leq1+y^2.
	\end{aligned}
\end{equation}

We then study the distribution of the decay rate with respect to the azimuthal angle $\phi_d$, which can be written as
	\begin{align}
		\frac{d\Gamma}{d\phi_d}&=\int dq^2\int dE_d\, \frac{4}{3}\frac{d\Gamma}{dq^2}\,\bigg\{\sqrt{6}\left(\mathcal{N}^U-\mathcal{N}^L\left\langle P_U^L\right\rangle\right)\nonumber\\[1.5mm]
		&\hspace{1.5cm}-\Big[\left(3\cos^2\theta_{\tau  d}-1\right)\left(\mathcal{N}^U\left(\tilde{A}^U_{\perp\perp}-\tilde{A}^U_{TT}\right)-\mathcal{N}^L\left(\tilde{A}^L_{\perp\perp}-\tilde{A}^L_{TT}\right)\right)\nonumber\\[1.5mm]
		&\hspace{6.0cm}-\frac{8}{\pi}\cos\theta_{\tau d}\sin^2\theta_{\tau d}\mathcal{N}^\perp \left\langle \tilde{T}^T_{\perp T}\right\rangle\Big]\cos2\phi_d\nonumber\\[1.5mm]
		&\hspace{1.5cm}-\Big[\left(3\cos^2\theta_{\tau  d}-1\right)\left(\mathcal{N}^U\left\langle \tilde{T}^U_{\perp T}\right\rangle-\mathcal{N}^L\left\langle \tilde{T}^L_{\perp T}\right\rangle\right)\nonumber\\[1.5mm]
		&\hspace{6.0cm}+6\cos\theta_{\tau d}\sin^2\theta_{\tau d}\mathcal{N}^\perp \tilde{Z}^\perp_{\perp T}\Big]\sin2\phi_d\bigg\}.
	\end{align}
Since $\phi_d$ is defined as the azimuthal angle between the decay planes of $\tau$ and $J/\psi$, this distribution contains the spin information of both $\tau$ and $J/\psi$. As discussed already in section~\ref{sec:CP-violating observables}, the coefficients of $\sin2\phi_d$, $\left\langle \tilde{T}^U_{\perp T}\right\rangle$, $\left\langle \tilde{T}^L_{\perp T}\right\rangle$ and $\tilde{Z}^\perp_{\perp T}$, are all referred to the CP-violating observables, which can be non-zero only under the NP scenarios with weak phases being different from the SM.

Finally, the differential distribution $d\Gamma/d\cos\theta_{J/\psi}$ is given by
\begin{equation}
	\frac{d\Gamma}{d\cos\theta_{J/\psi}}=\int dq^2\frac{\sqrt{6}}{8}\frac{d\Gamma}{dq^2}\left[2\cos^2\theta_{J/\psi}\left\langle \tilde{T}^U_{LL}\right\rangle+\sin^2\theta_{J/\psi}\left(\left\langle \tilde{T}^U_{\perp\perp}\right\rangle+\left\langle \tilde{T}^U_{TT}\right\rangle\right)\right],
\end{equation}
where the spin of the $\tau$ lepton has already been summed over. The $J/\psi$ longitudinal polarization fraction $F^{J/\psi}_L$, which is the analogue of the usually discussed $F^{D^\ast}_L$~\cite{Tanaka:2012nw,Belle:2019ewo}, is given with our notations by
\begin{equation}
	F^{J/\psi}_L\left(q^2\right)=\frac{d\Gamma^{\lambda_{J/\psi}=0}/dq^2}{d\Gamma/dq^2}=\frac{1}{\sqrt{6}}\left(\left\langle \tilde{T}^U_{\perp\perp}\right\rangle+\left\langle \tilde{T}^U_{TT}\right\rangle-\left\langle \tilde{T}^U_{LL}\right\rangle\right).
\end{equation}

\begin{figure}[t]
	\centering
	\includegraphics[width=0.31\textwidth]{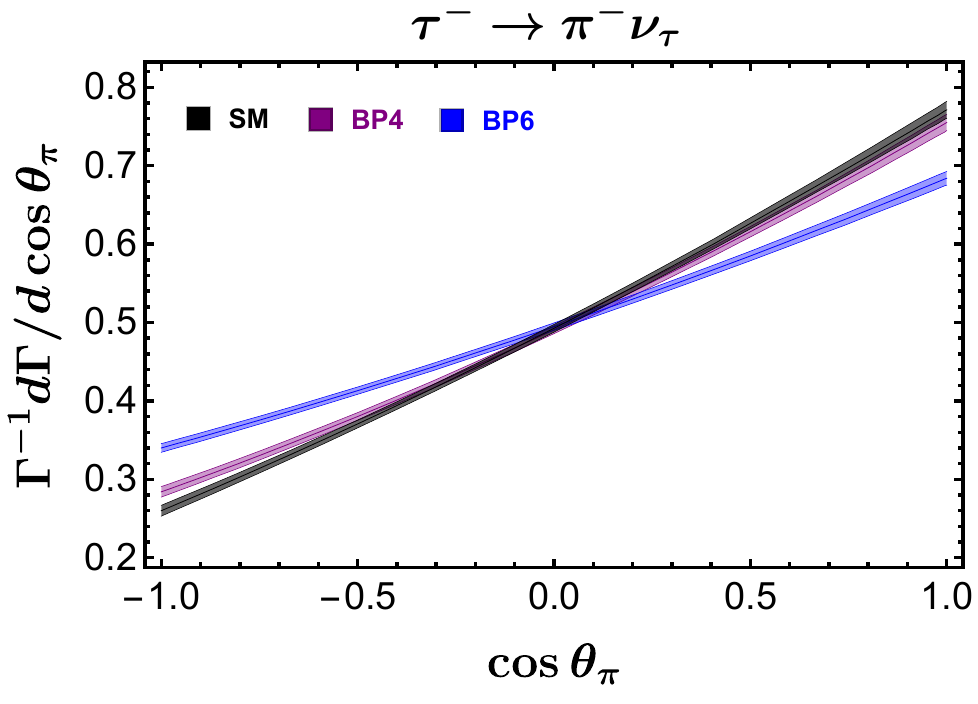}\quad
	\includegraphics[width=0.31\textwidth]{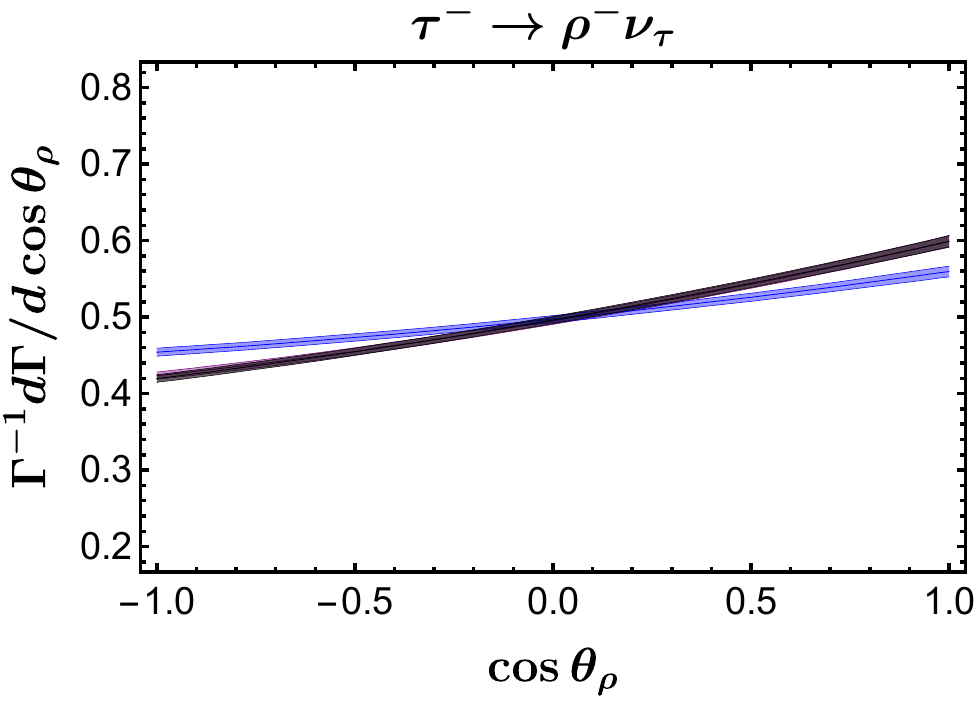}\quad
	\includegraphics[width=0.31\textwidth]{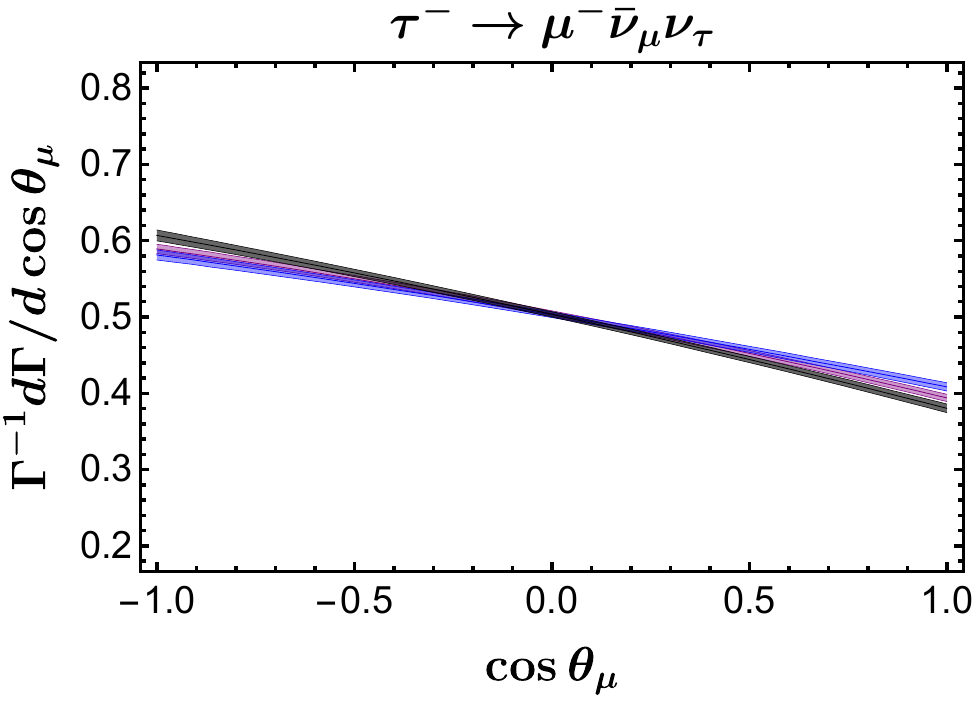}
	\\[4mm]
	\includegraphics[width=0.31\textwidth]{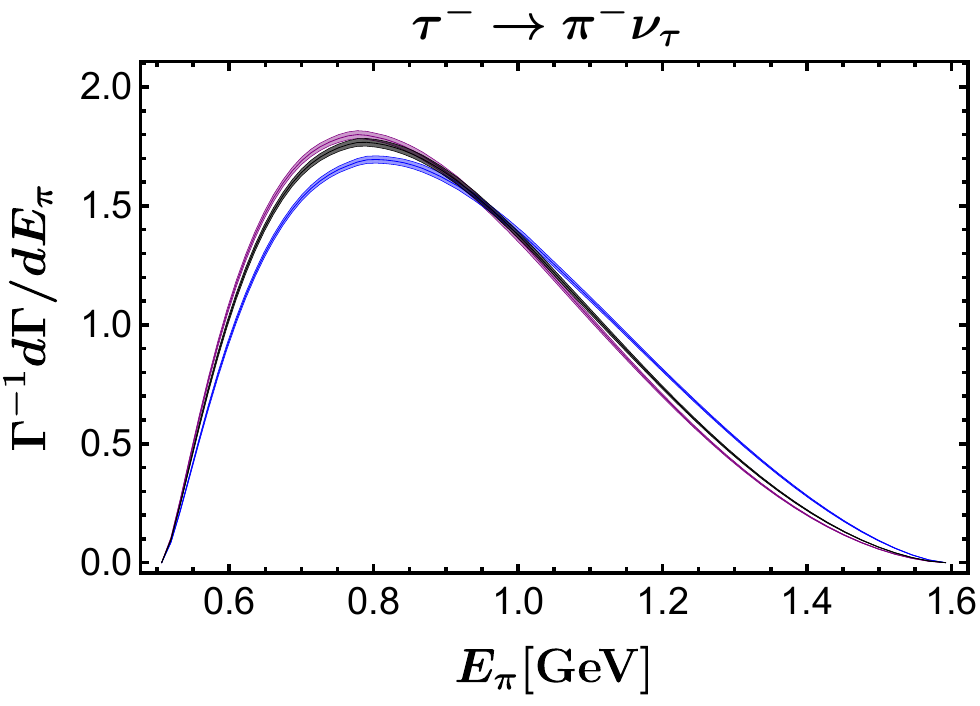}\quad
	\includegraphics[width=0.31\textwidth]{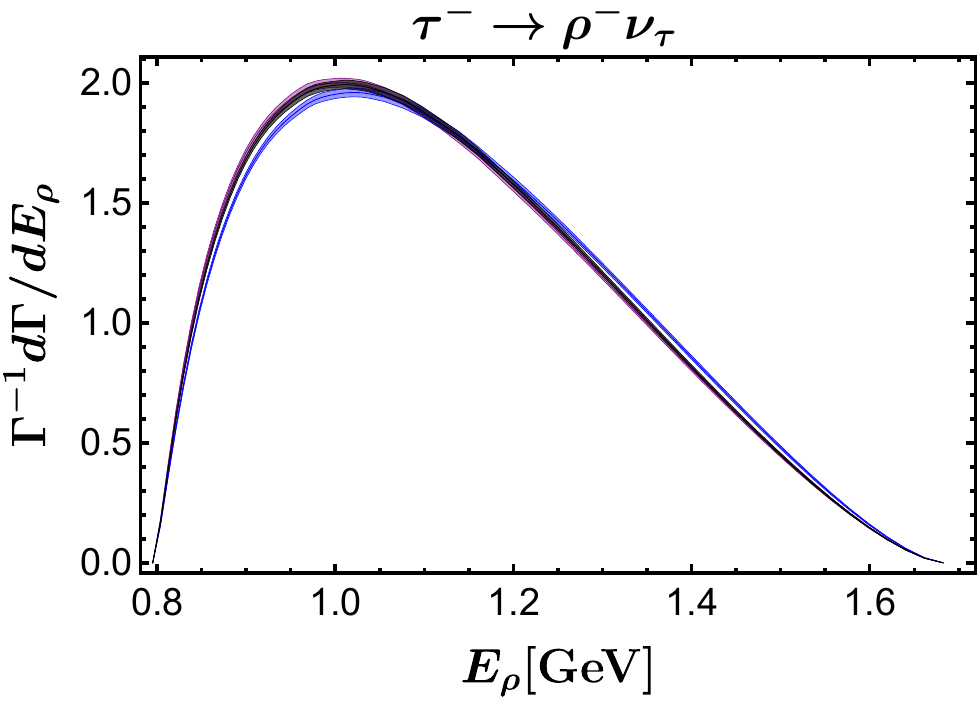}\quad
	\includegraphics[width=0.31\textwidth]{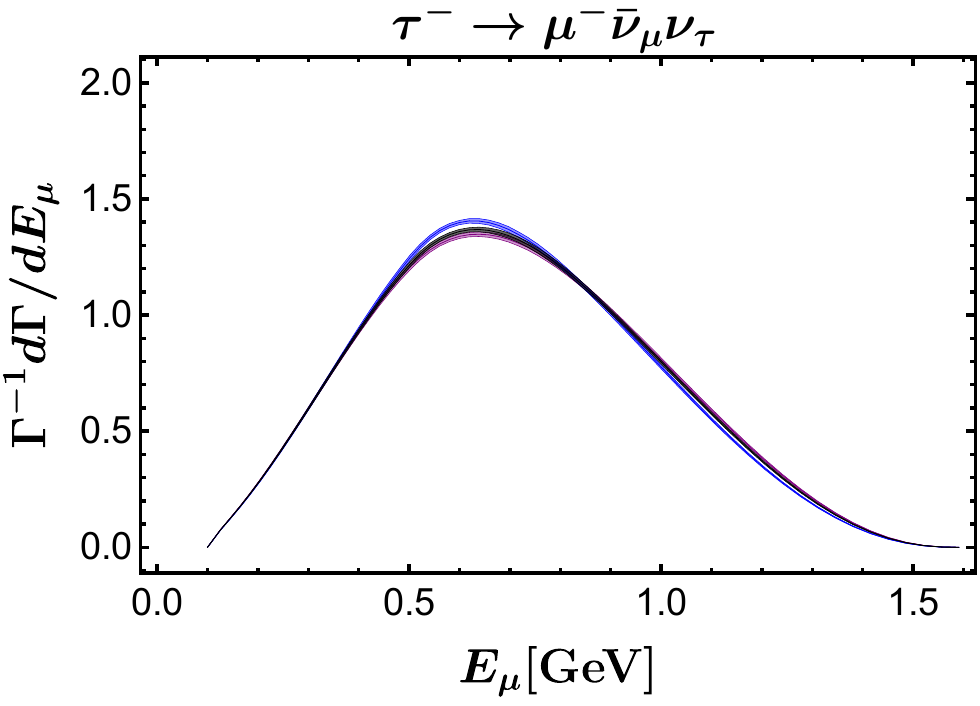}
	\\[4mm]
	\includegraphics[width=0.31\textwidth]{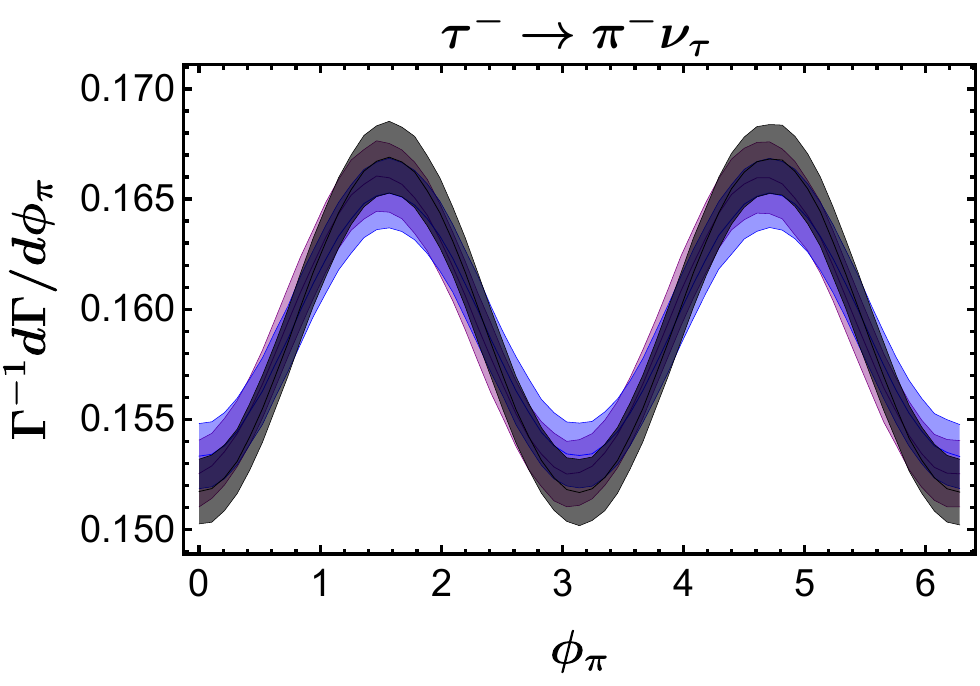}\quad
	\includegraphics[width=0.31\textwidth]{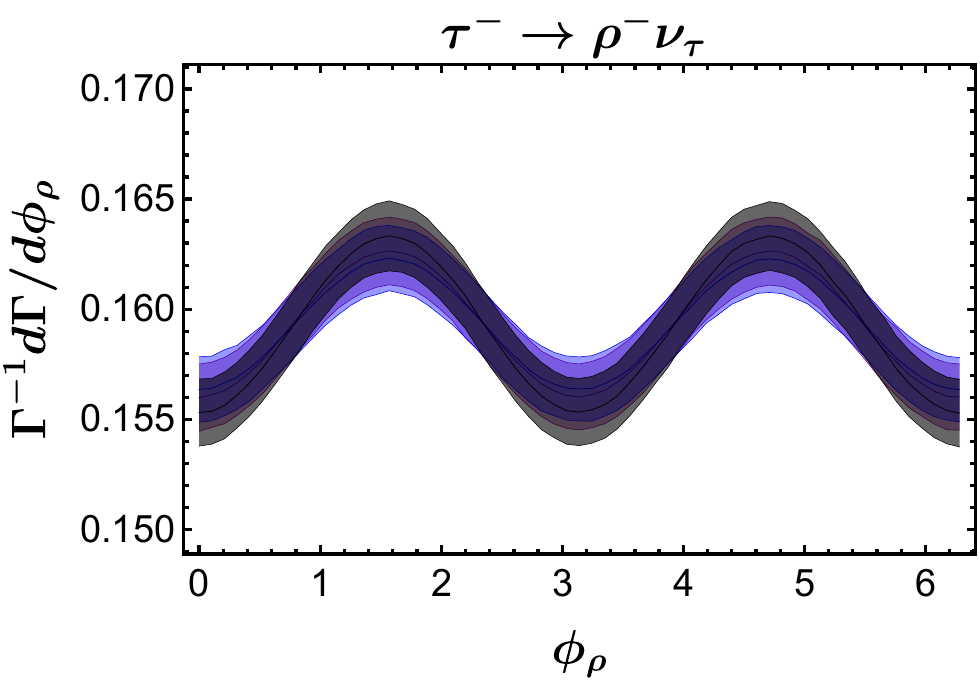}\quad
	\includegraphics[width=0.31\textwidth]{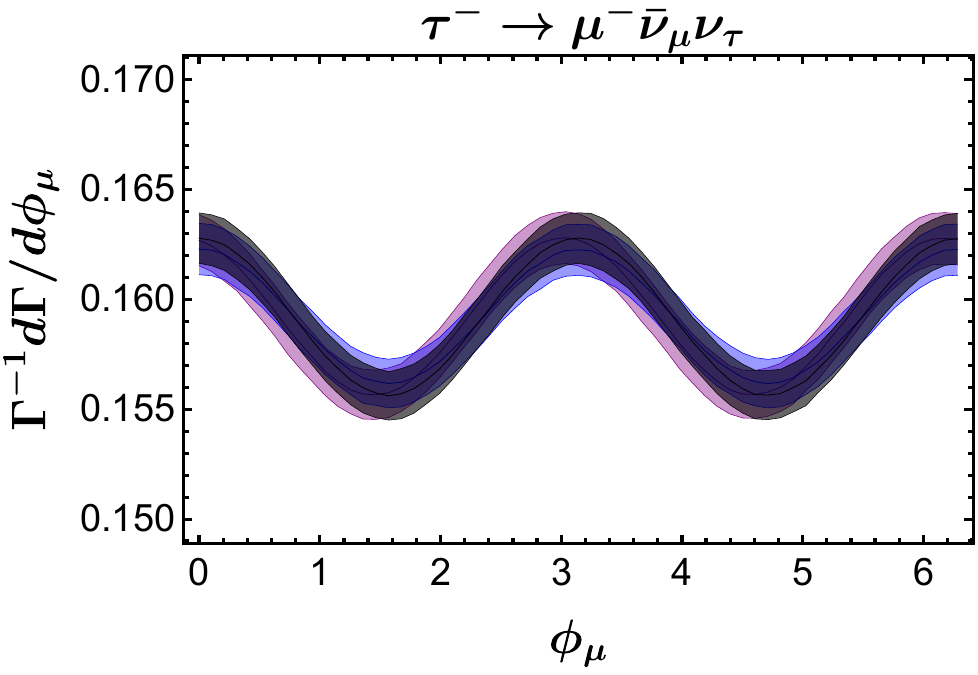}
	\\[4mm]
	\includegraphics[width=0.31\textwidth]{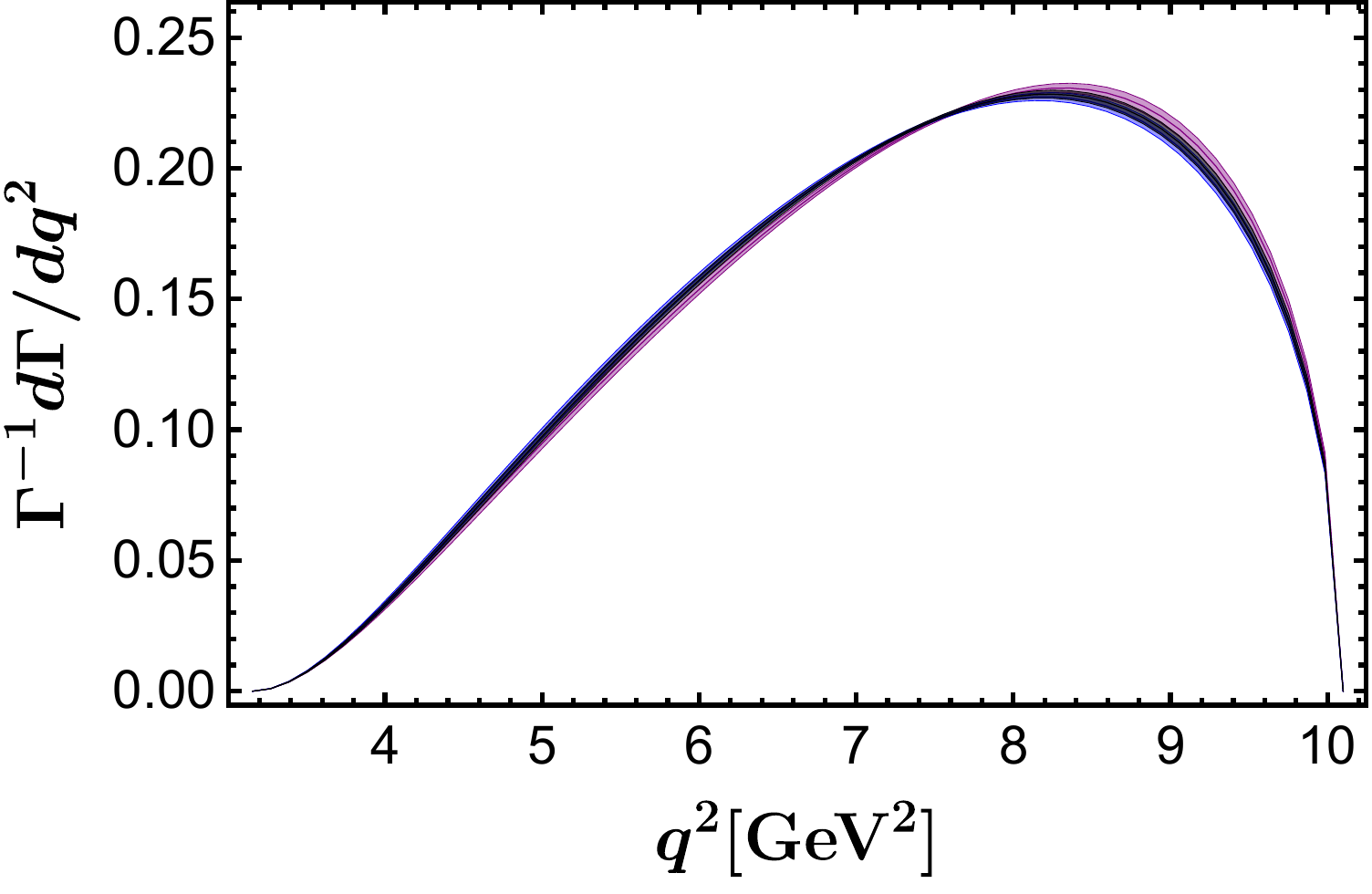}\qquad
	\includegraphics[width=0.31\textwidth]{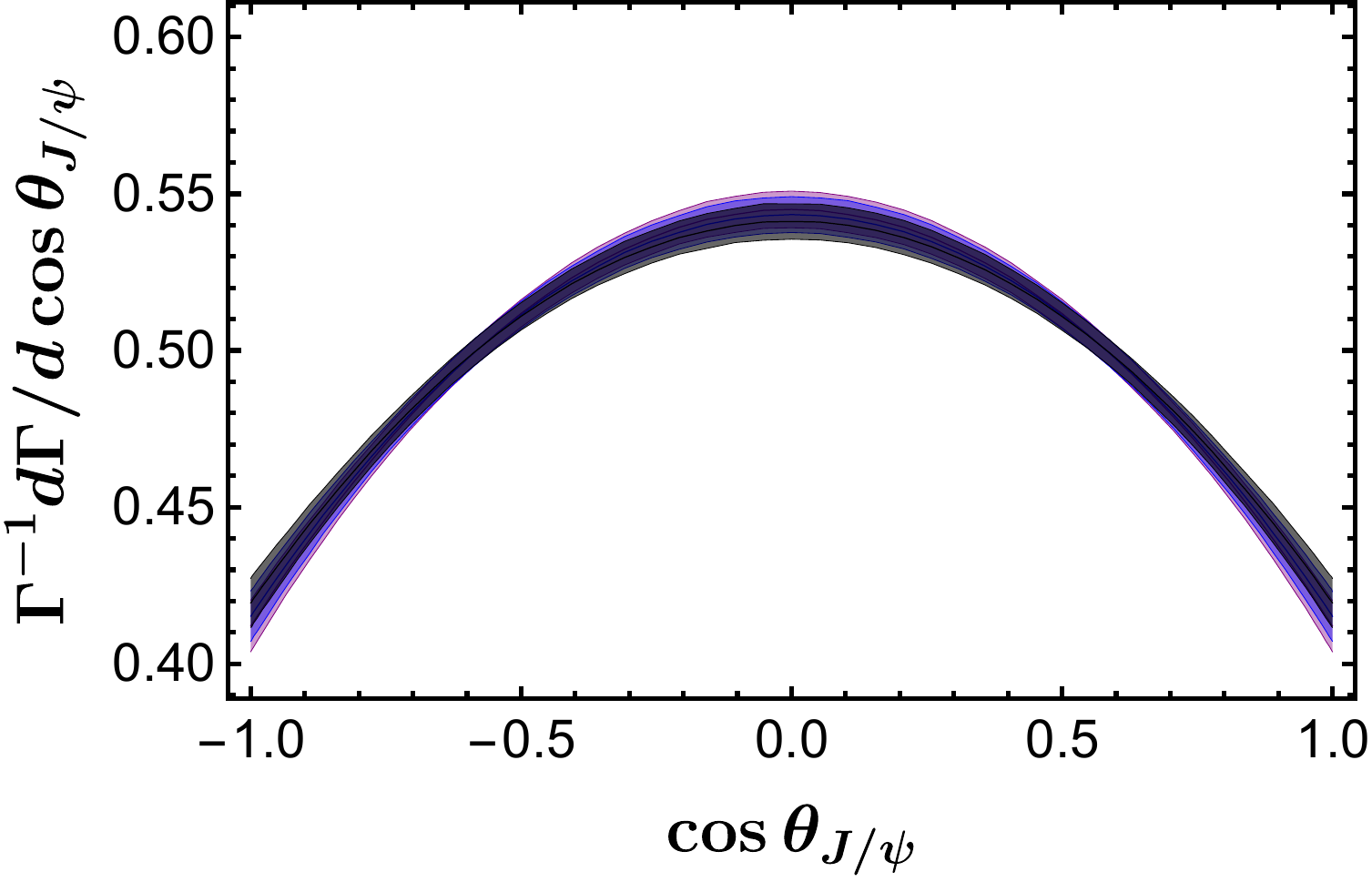}
	\caption{The single differential decay rates with only one kinematic variable left, predicted both within the SM and in the two NP benchmark points, BP4 and BP6, where they contain only the left- and the right-handed neutrinos respectively. The other captions are the same as in figures~\ref{fig:CP-odd} and \ref{fig:CP-even}.} \label{fig:Integrated observables}
\end{figure}

Our numerical results of these integrated observables with only one visible kinematic variable left are illustrated in figure~\ref{fig:Integrated observables}. Since the integration over some kinematic variables will sometimes result in a loss of information on the spin and spin-angular asymmetries, these observables are not as sensitive to the different NP scenarios as are the ones discussed before. Therefore, we only show the sensitivities of these observables to the two different NP benchmark points, BP4 and BP6, where they contain only the left- and the right-handed neutrinos respectively. However, one should notice that, by accumulating events with all allowed values of the kinematic variables but providing the distribution with respect to only one variable, we can largely increase the statistics~\cite{Penalva:2022vxy}. This makes these observables much more promising to be measured with sufficiently high statistics in the future.

\section{Sensitivity of the observables at LHCb}
\label{sec:sensitivity}

In this section, we examine what experimental precision can be achieved when extracting the spin and spin-angular asymmetries from the fully differential distribution given by eq.~\eqref{eq:differential decay rate}. Since a detailed simulation of backgrounds and detector effects is beyond the scope of this work, we shall only give the statistical uncertainty under an ideal experiment with unlimited resolution in all the five kinematic variables $q^2$, $E_d$, $\theta_d$, $\phi_d$ and $\theta_{J/\psi}$. 

For brevity, let us denote the 34 normalized observables as $\vec{O}$ with the corresponding coefficient for each observable $O_i$ designated as $f_{O_i}(q^2,\vec{\xi})$, where $\vec{\xi}$ refer to the rest four kinematic variables $E_d$, $\theta_d$, $\phi_d$ and $\theta_{J/\psi}$. Then, the decay distribution can be written as
\begin{equation}
	\label{eq:probability density}
	\frac{d^5\Gamma}{dq^2dE_d d\cos\theta_d d\phi_dd\cos\theta_{J/\psi}}/\frac{d\Gamma}{dq^2}=\sum_{i=1}^{34}f_{O_i}(q^2,\vec{\xi})O_i.
\end{equation}
Here we only consider the SM case, and thus the condition $C^X_{AB}= 0$ is implied in eq.~\eqref{eq:probability density}. Assuming that there were $N(q^2)$ sample events $\{\vec{\xi}_l\}$ for a fixed $q^2$ but with different $\vec{\xi}$, the true values of the observables $\vec{O}$ will maximize the likelihood function $\mathcal{L}\left(\vec{O}\Big|\{\vec{\xi}_l\}\right)$ and are then the solutions of the following equations: 
\begin{equation}
	\frac{\partial}{\partial O_i}\ln\mathcal{L}\left(\vec{O}\Big|\{\vec{\xi}_l\}\right)=\sum_{l=1}^{N(q^2)}\frac{f_{O_i}(q^2,\vec{\xi}_l)}{\sum_{j=1}^{34}f_{O_j}(q^2,\vec{\xi}_l)O_j}=0.
\end{equation}
The covariance matrix is then given by
\begin{equation}
	\mathrm{cov}^{-1}\left(\hat{\vec{O}}\right)_{ij}=-\frac{\partial^2}{\partial O_i\partial O_j}\ln\mathcal{L}\left(\vec{O}\Big|\{\vec{\xi}_l\}\right) \Big|_{\vec{O}=\hat{\vec{O}}},
\end{equation}
where $\hat{\vec{O}}$ denote the true values of the observables $\vec{O}$, which are taken to be the central values predicted with the lattice QCD input for the $B_c\to J/\psi$ transition form factors~\cite{Harrison:2020gvo}. In this work, for simplicity, we are not going to discuss the correlations among the different observables, and estimate the experimental sensitivity of each observable individually. The statistical uncertainty of an observable $O_i$ can then be estimated as~\cite{Davier:1992nw,Tanaka:2010se,Alonso:2017ktd}
\begin{equation}
	\delta O_i=\frac{1}{\sqrt{N(q^2)}\,S_{O_i}(q^2)},
\end{equation}
with the sensitivity $S_{O_i}(q^2)$ given by~\cite{Davier:1992nw,Tanaka:2010se,Alonso:2017ktd}
\begin{equation}
	S^2_{O_i}(q^2)=\left\langle \left(\frac{f_{O_i}(q^2,\vec{\xi})}{\sum_{j=1}^{34}f_{O_j}(q^2,\vec{\xi})\hat{O}_j}\right)^2\right\rangle=\int d^4\vec{\xi}\,\frac{f^2_{O_i}(q^2,\vec{\xi})}{\sum_{j=1}^{34}f_{O_j}(q^2,\vec{\xi})\hat{O}_j}.
\end{equation}
Similarly, the statistical uncertainties of the binned observables defined by eq.~\eqref{eq:binned observable} can be estimated as
\begin{equation}
	\delta \overline{O_i}^j=\frac{1}{\sqrt{N^j}\,\overline{S_{O_i}}^j},
\end{equation}
with
\begin{equation}	     
    (\overline{S_{O_i}}^j)^{-2}=\frac{1}{\Gamma}\int_{(q^2_-)^j}^{(q^2_+)^j} dq^2\frac{d\Gamma}{dq^2}S^{-2}_{O_i}(q^2).
\end{equation}
Here $N^j$ is the number of events in the full sample for the $q^2$ interval $\{(q^2_-)^j,(q^2_+)^j\}$, and $\overline{S_{O_i}}^j$ the corresponding averaged sensitivity. In the following, we consider only the averaged observables integrated over the full $q^2$ range, which constitutes a single bin for the analysis, and denote these averaged observables simply as $\overline{O_i}$.

The expected number of events $N$ for $\tau^-\to\mu^-\bar{\nu}_\mu\nu_\tau$ decay at LHCb is estimated as
\begin{equation}
	N=\sum_{i,j}\mathcal{L}\times\sigma_{b\bar{b}}(\eta^i)\times f_c(p^j_T)\mathcal{B}(B_c^- \to J/\psi\mu^-\bar{\nu}_\mu)\times R(J/\psi)\times\mathcal{B}_\tau \times\mathcal{B}_{J/\psi}\times\epsilon_\mu(\eta^i,p^j_T)\times\frac{\epsilon_\tau}{\epsilon_\mu},
\end{equation}
where $\mathcal{L}\approx300\,\mathrm{fb}^{-1}$ is the expected integrated luminosity of the LHCb experiment until 2035~\cite{LHCb:2018roe}, and $\sigma_{b\bar{b}}(\eta^i)$ is the $b\bar{b}$ cross section as a function of  the pseudo-rapidity bin $\eta^i$, which sums up to be about $144\,\mathrm{\mu b}$ in the covered $\eta$ range~\cite{LHCb:2016qpe}.\footnote{Here, as a good approximation, the $b\bar{b}$ cross section measured in $pp$ collisions at $13\,\mathrm{TeV}$ center-of-mass energy is adopted.} Product of the hadronization factor $f_c(p^j_T)$ and the branching fraction $\mathcal{B}(B_c^- \to J/\psi\mu^- \bar{\nu}_\mu)$, which is given in different bins of the pseudo-rapidity $\eta$ and the transverse momentum $p_T$ of the $B_c^-$ meson, together with the corresponding signal efficiency $\epsilon_\mu(\eta^i,p^j_T)$, can be found in ref.~\cite{LHCb:2019tea}. Notice that $f_c(p^j_T)$ depends only marginally on $\eta$ and hence can be regarded as a single function of $p_T$. Then, we can use the measured $R(J/\psi)$ and the ratio of the signal efficiencies $\epsilon_\tau/\epsilon_\mu$~\cite{Aaij:2017tyk} to connect the number of events between $B_c^- \to J/\psi\tau^- \bar{\nu}_\tau$ and $B_c^- \to J/\psi\mu^- \bar{\nu}_\mu$ decays. Since the LHCb measurement~\cite{Aaij:2017tyk} uses the decay channel $\tau^-\to\mu^-\bar{\nu}_\mu\nu_\tau$ to identify the $\tau$ lepton, we can estimate the number of events for this channel to be about $2.7\times10^5$. On the other hand, there exists no LHCb measurement of these factors by using a single hadronic $\tau$ decay channel yet, but the number of events for $B \to D\tau \bar{\nu}_\tau$ at Belle is roughly the same for the three channels $\tau^-\to \pi^-\nu_\tau$, $\tau^-\to \rho^-\nu_\tau$ and $\tau^-\to\ell^-\bar{\nu}_\ell\nu_\tau$~\cite{Aushev:2010bq}. Therefore, as an approximation, we shall assume that a similar circumstance is also applied to the LHCb experiment and roughly set the number of events for $\tau^-\to \pi^-(\rho^-)\nu_\tau$ channels to be twice as that of $\tau^-\to\mu^-\bar{\nu}_\mu\nu_\tau$ channel. In addition, we shall not consider the electronic channel $\tau^-\to e^-\bar{\nu}_e\nu_\tau$ due to the poor reconstruction efficiency at LHCb, which results from the high bremsstrahlung rate for electrons~\cite{LHCb:2014vgu}.

\begin{figure}[t]
	\centering
	\includegraphics[width=0.31\textwidth]{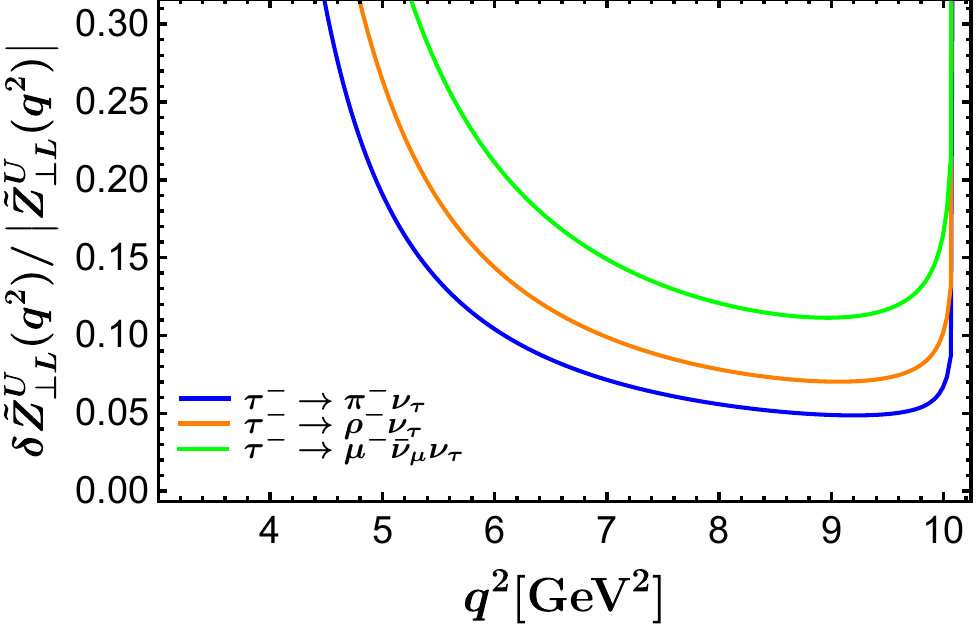}\quad
	\includegraphics[width=0.31\textwidth]{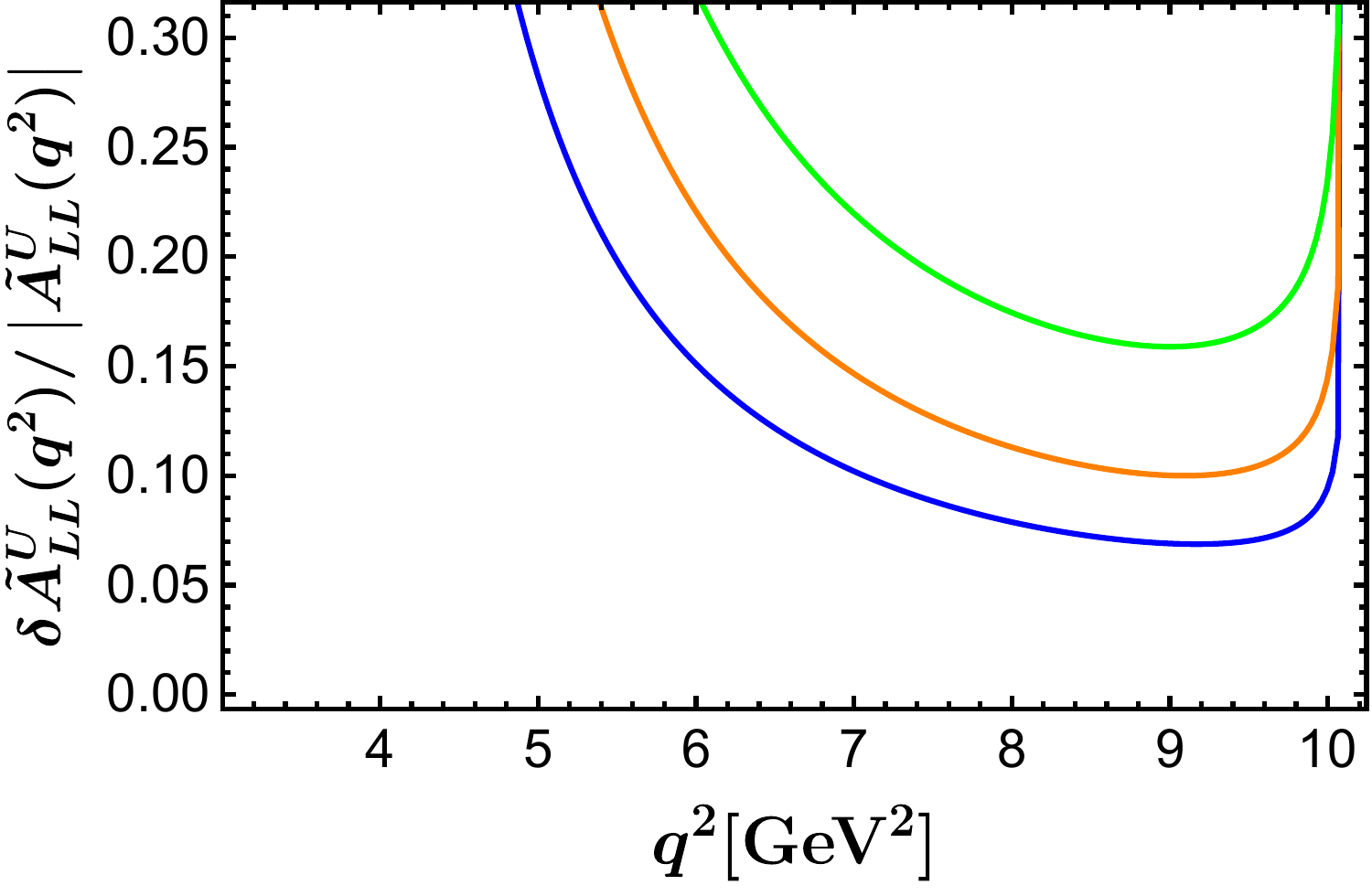}\quad
	\includegraphics[width=0.31\textwidth]{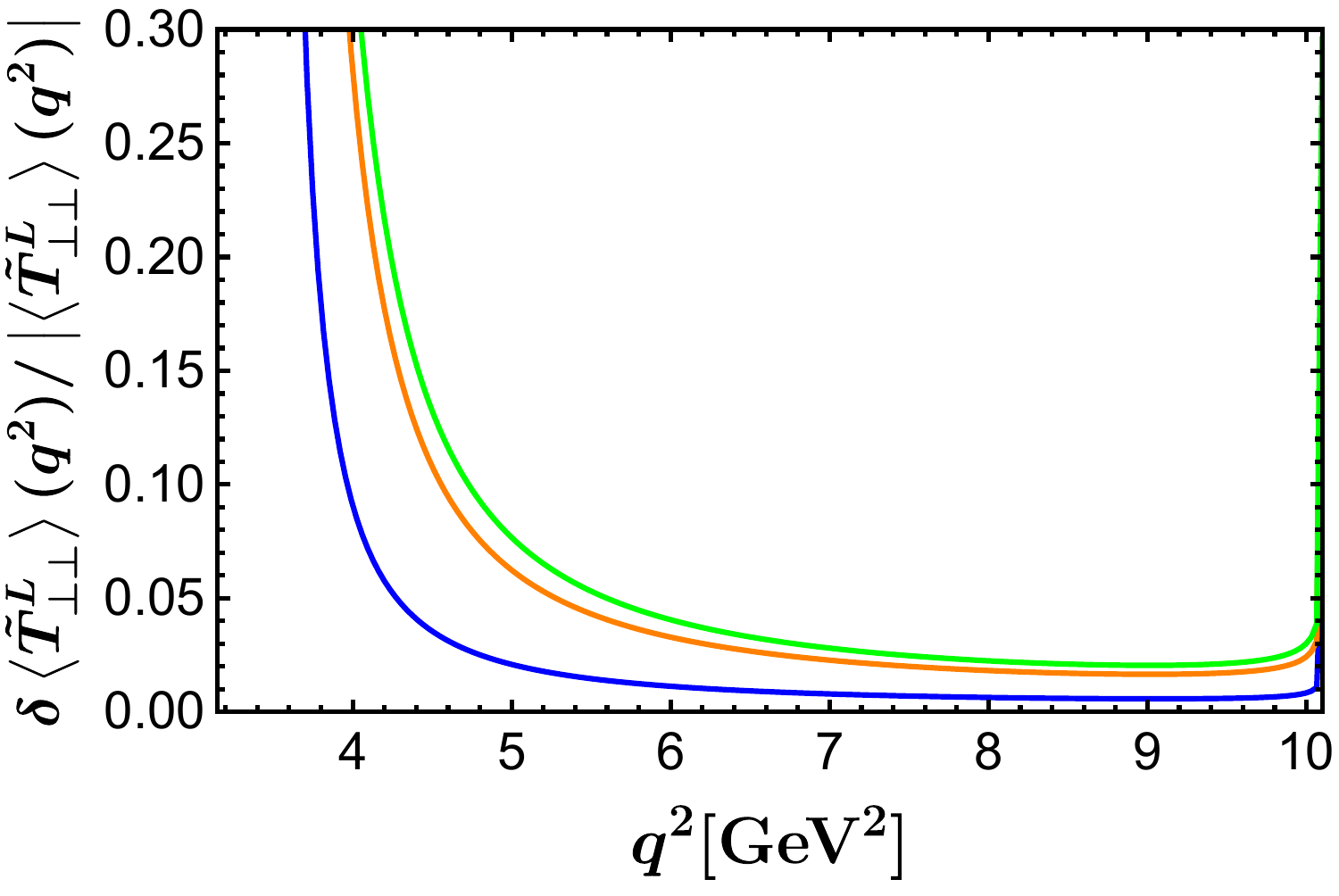}
	\\[4mm]
	\includegraphics[width=0.31\textwidth]{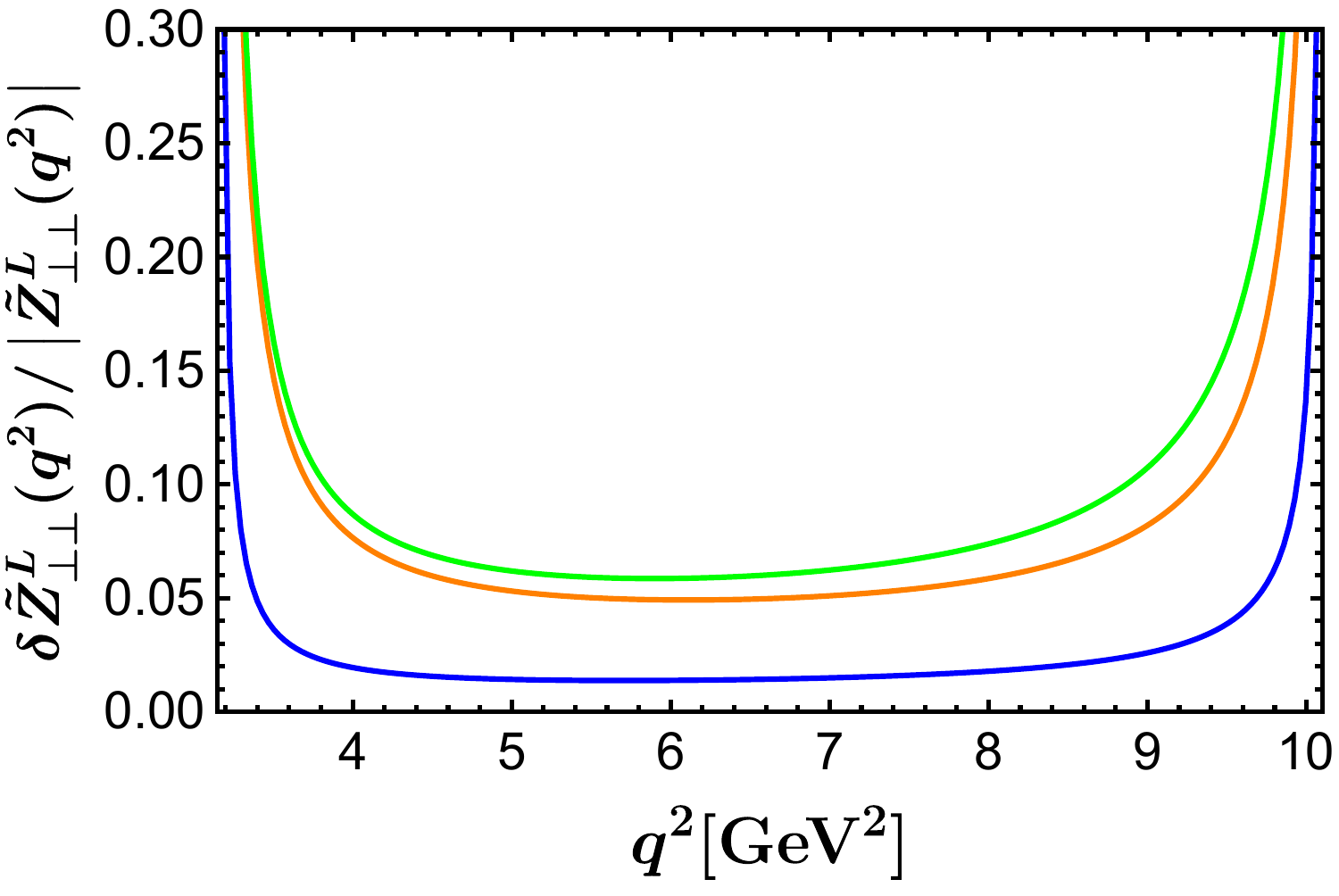}\quad
	\includegraphics[width=0.31\textwidth]{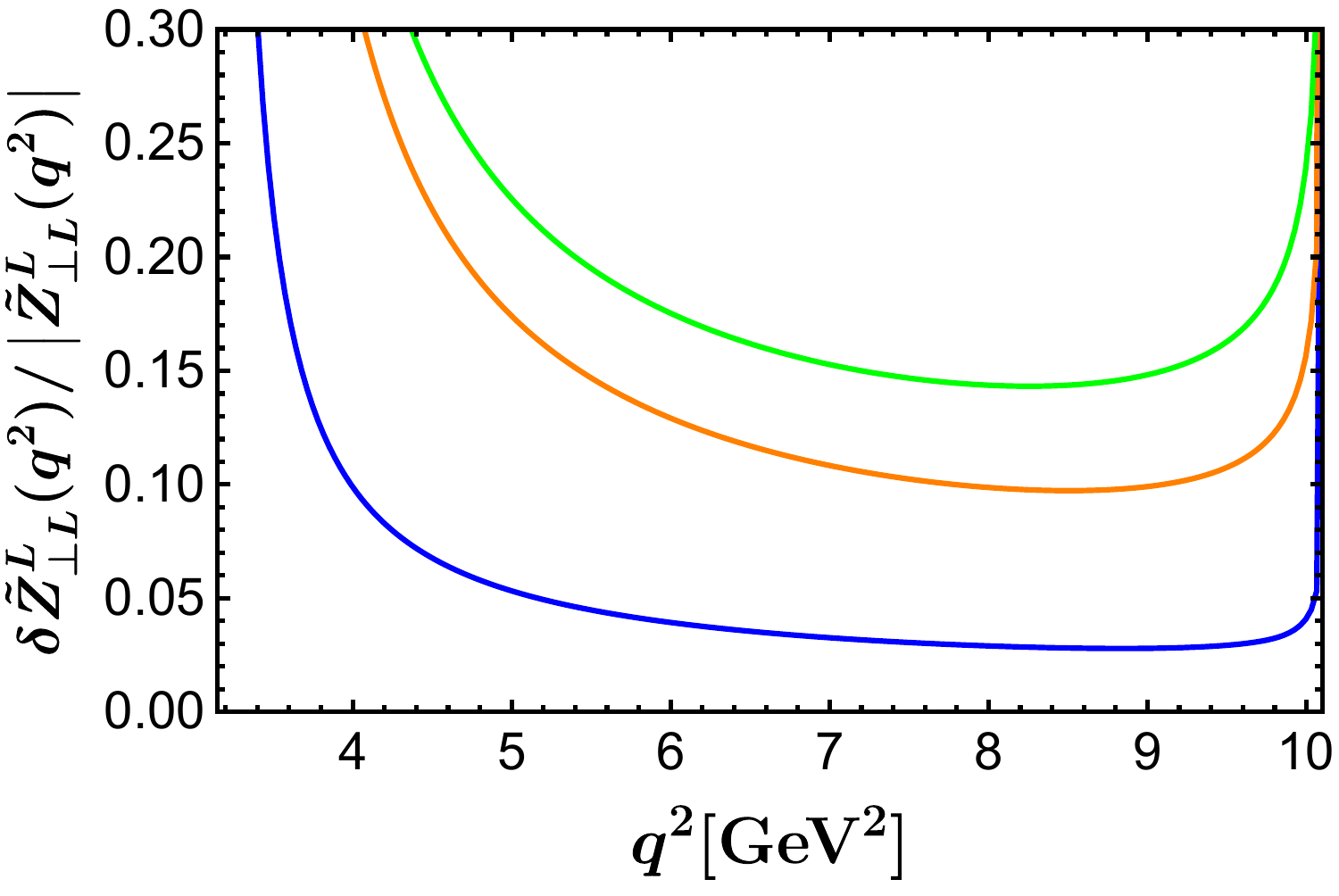}\quad
	\includegraphics[width=0.31\textwidth]{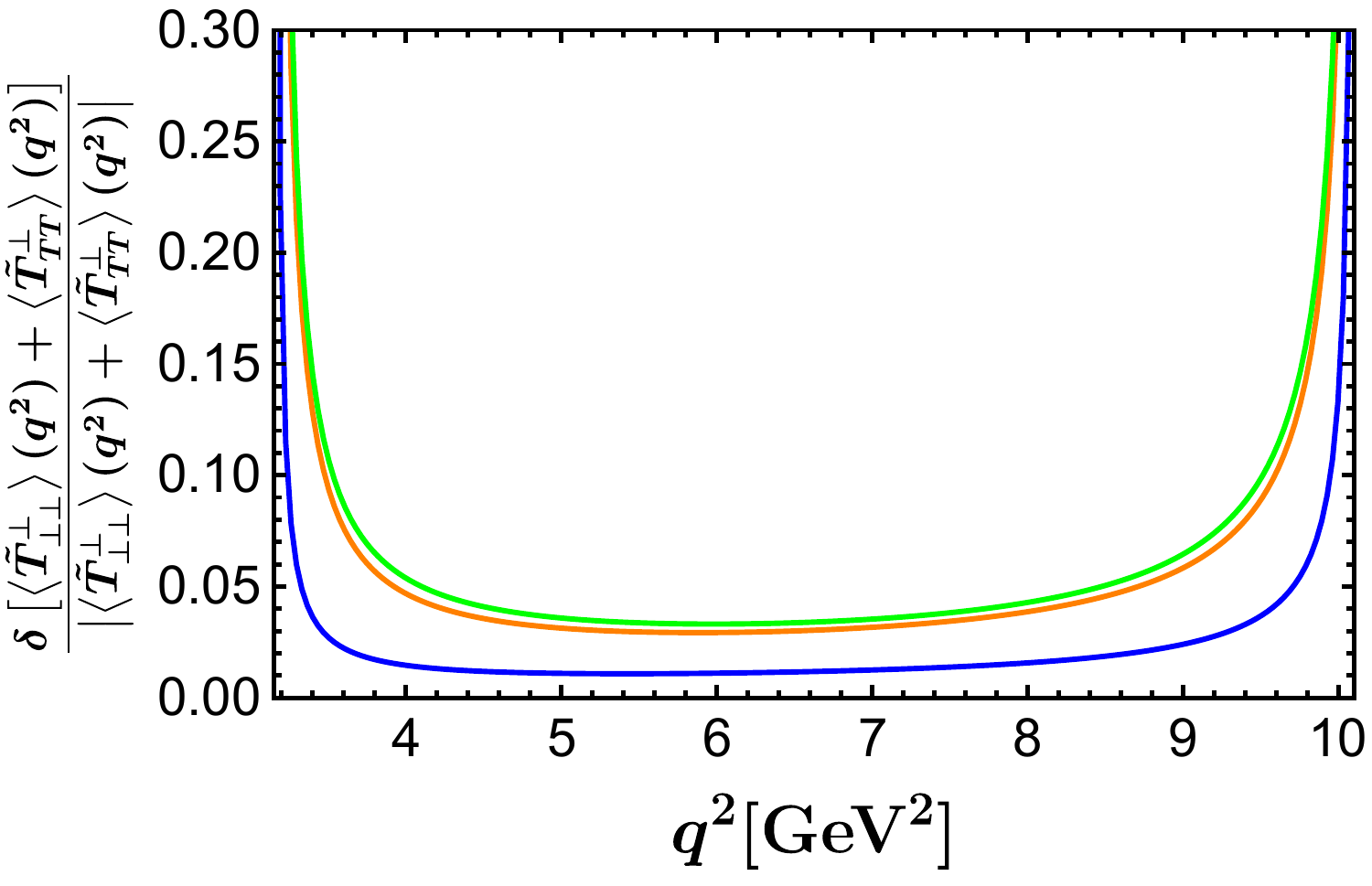}
	\\[4mm]
	\includegraphics[width=0.31\textwidth]{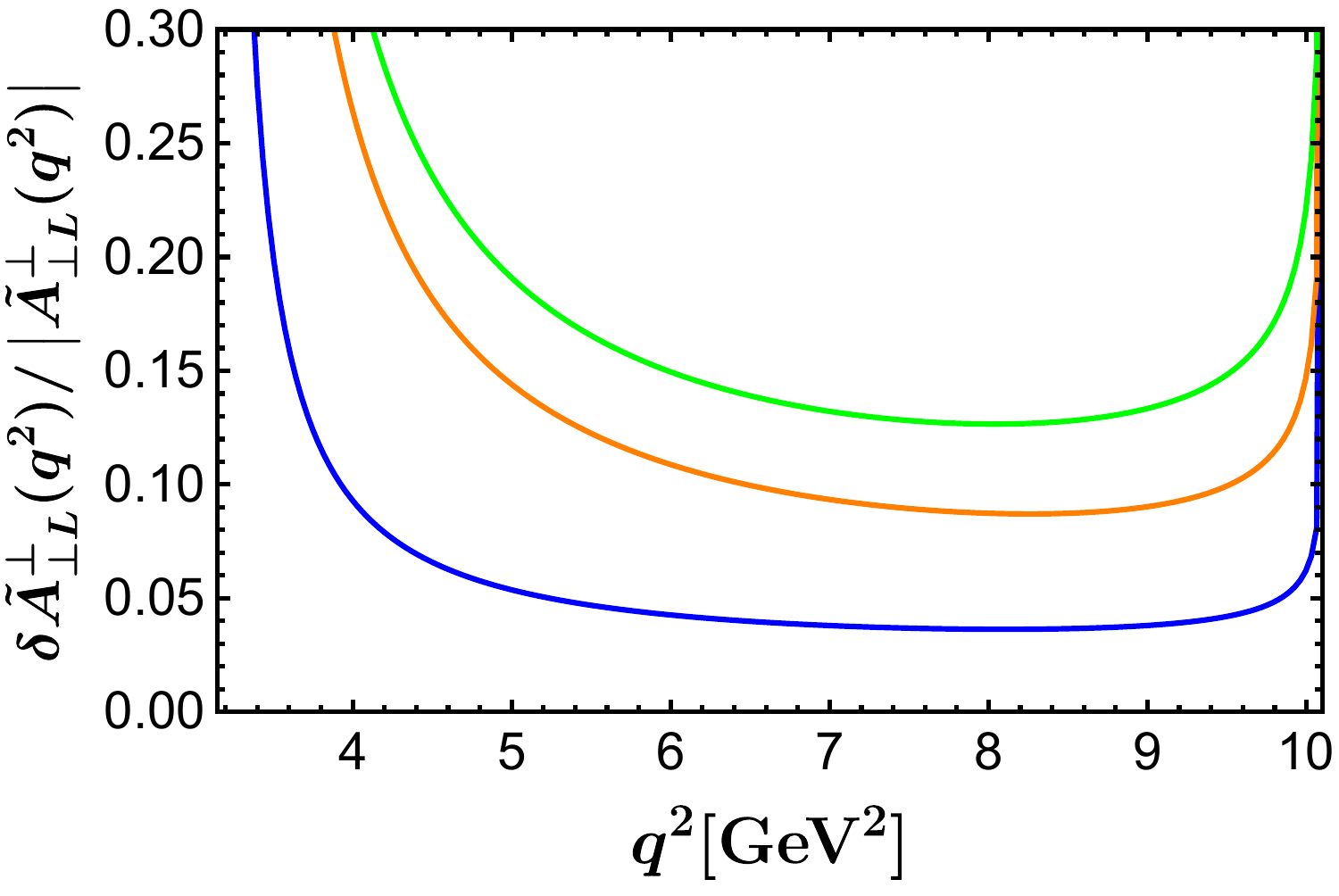}\quad
	\includegraphics[width=0.31\textwidth]{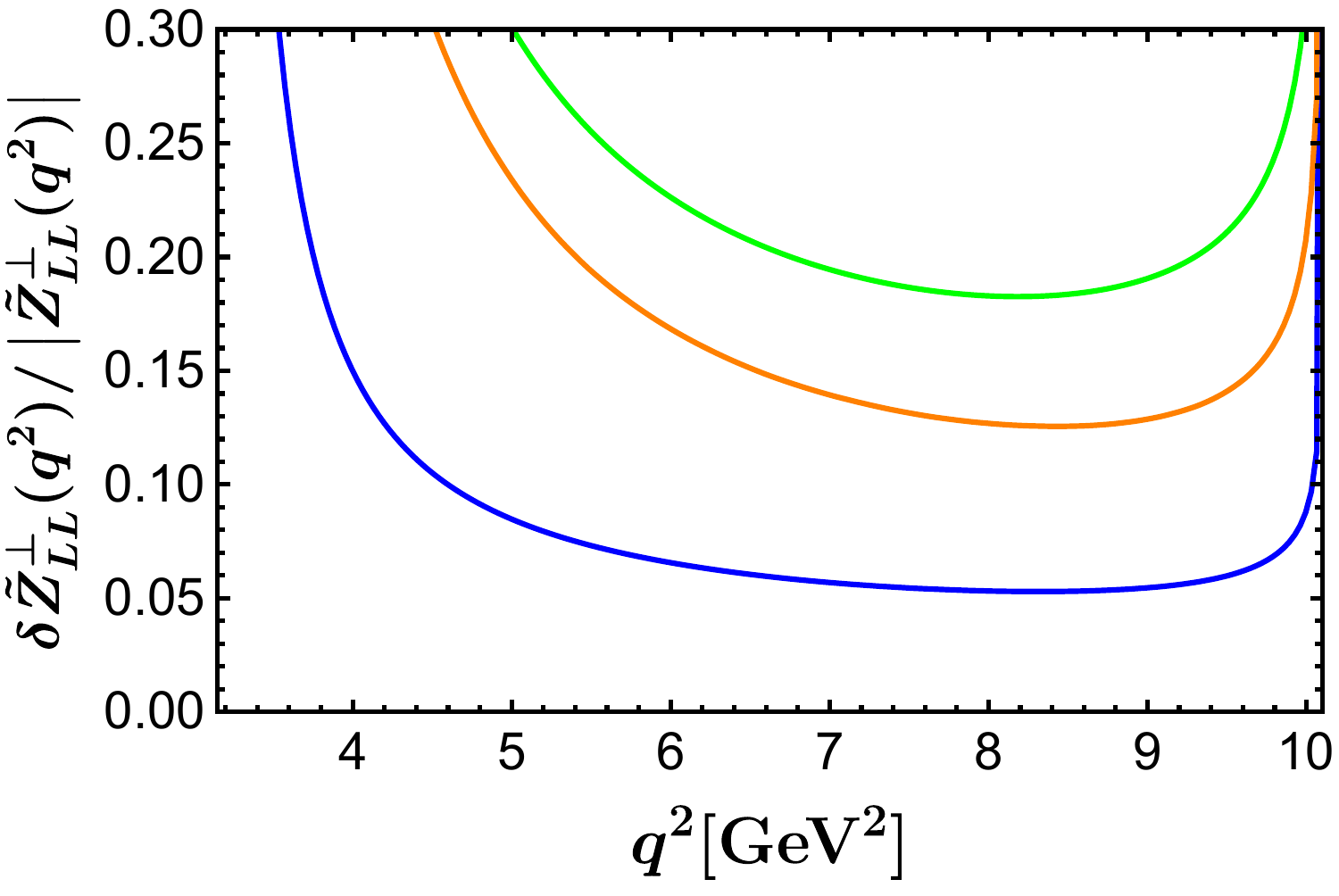}\quad
	\includegraphics[width=0.31\textwidth]{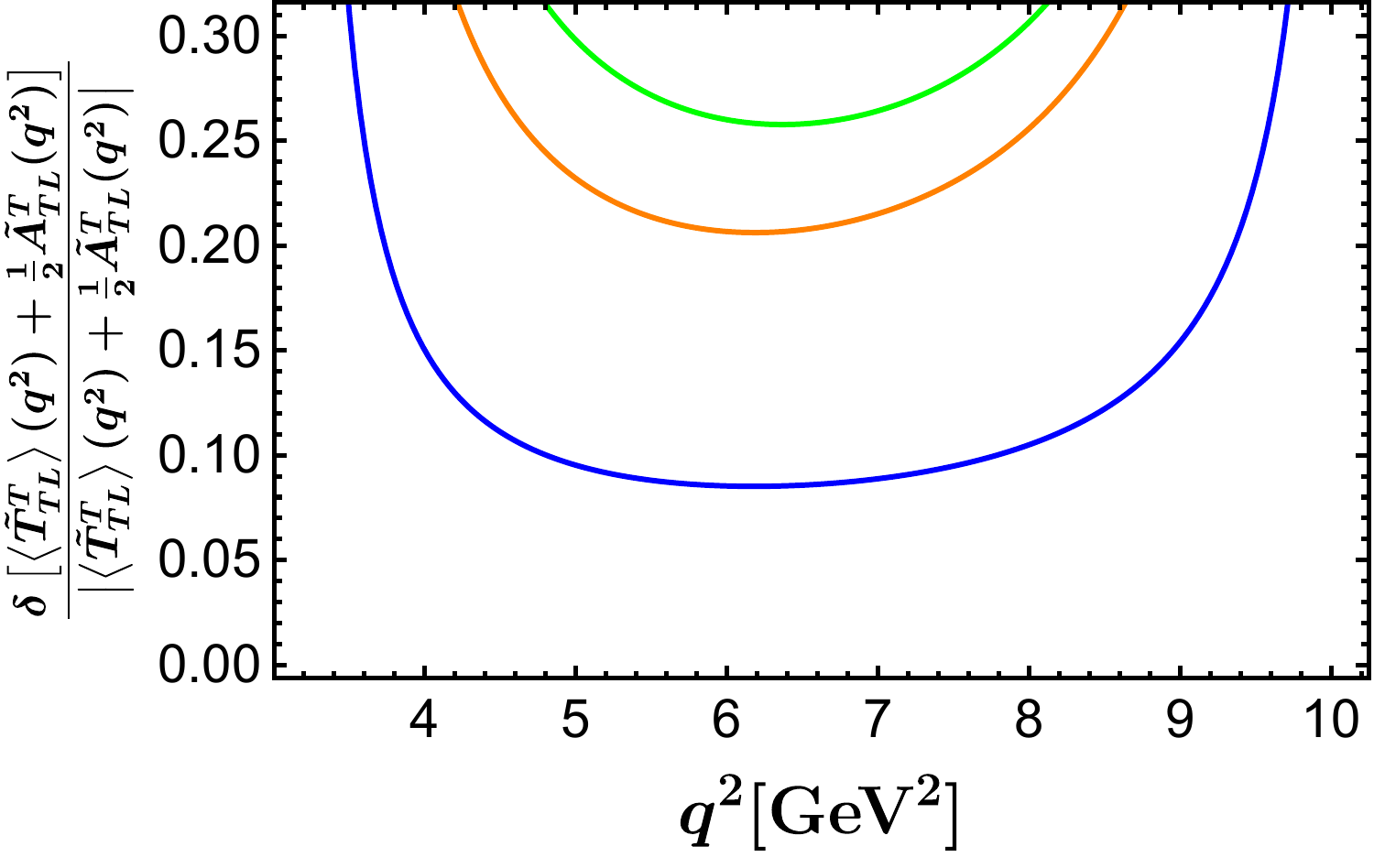}
	\caption{The relative statistical uncertainties of the nine CP-conserving observables at LHCb, estimated for the three $\tau$ decay channels $\tau^-\to \pi^-\nu_\tau$, $\tau^-\to \rho^-\nu_\tau$ and $\tau^-\to\mu^-\bar{\nu}_\mu\nu_\tau$, which are illustrated by the blue, orange and green curves, respectively.} \label{fig:statistical uncertainty}
\end{figure}

\begin{table}[t]
	\begin{center}	
		\tabcolsep 0.165in
		\renewcommand\arraystretch{1.5}
		\begin{tabular}{ccccc}
			\hline \hline  & $\tau^-\to \pi^-\nu_\tau$ & $\tau^-\to \rho^-\nu_\tau$ & $\tau^-\to\mu^-\bar{\nu}_\mu\nu_\tau$ & theoretical\\
			\hline
			$\delta\overline{\tilde {Z}^U_{\perp L}}/\left|\overline{\tilde {Z}^U_{\perp L}}\right|$ &0.030&0.044&0.071 &0.008\\[2mm]
			$\delta \overline{\tilde {A}^U_{L L}}/\left|\overline{\tilde{A}^U_{L L}}\right|$ &0.043&0.063&0.095 &0.016\\[2mm]
			$\delta \overline{\left\langle \tilde {T}^L_{\perp \perp}\right\rangle}/\left|\overline{\left\langle \tilde {T}^L_{\perp \perp}\right\rangle}\right|$ &0.003&0.010&0.013 &0.017\\[2mm]
			$\delta\overline{\tilde{Z}^L_{\perp \perp}}/\left|\overline{\tilde {Z}^L_{\perp \perp}}\right|$ &0.008&0.025&0.038 &0.029\\[2mm]
			$\delta \overline{\tilde {Z}^L_{\perp L}}/\left|\overline{\tilde {Z}^L_{\perp L}}\right|$ &0.014&0.056&0.073 &0.007\\[2mm]
			$\frac{\delta \left[\overline{\left\langle \tilde {T}^\perp_{\perp\perp}\right\rangle}+\overline{\left\langle \tilde{T}^\perp_{TT}\right\rangle}\right]}{\left|\overline{\left\langle \tilde{T}^\perp_{\perp\perp}\right\rangle}+\overline{\left\langle \tilde{T}^\perp_{TT}\right\rangle}\right|}$ &0.006&0.016&0.019 &0.027\\[2mm]
			$\delta\overline{\tilde{A}^\perp_{\perp L}}/\left|\overline{\tilde{A}^\perp_{\perp L}}\right|$ &0.017&0.044&0.066 &0.007\\[2mm]
			$\delta \overline{\tilde {Z}^\perp_{L L}}/\left|\overline{\tilde{Z}^\perp_{LL}}\right|$ &0.025&0.061&0.092& 0.017\\[2mm]
			$\frac{\delta \left[\overline{\left\langle \tilde{T}^T_{TL}\right\rangle}+\frac{1}{2}\overline{\tilde{A}^T_{TL}}\right]}{\left|\overline{\left\langle \tilde{T}^T_{TL}\right\rangle}+\frac{1}{2}\overline{\tilde{A}^T_{TL}}\right|}$ &0.046&0.113&0.156 &0.088\\
			\hline \hline
		\end{tabular}
		\caption{The relative uncertainties of the averaged observables $\overline{O_i}$. The first three columns refer to the relative statistical uncertainties under an ideal experiment, estimated for the three $\tau$ decay channels $\tau^-\to \pi^-\nu_\tau$, $\tau^-\to \rho^-\nu_\tau$ and $\tau^-\to\mu^-\bar{\nu}_\mu\nu_\tau$, respectively. The last column refers to the relative theoretical uncertainties for the SM predictions, which come from the $B_c\to J/\psi$ transition form factors~\cite{Harrison:2020gvo}.} \label{tab:statistical uncertainty}
	\end{center}
\end{table}	

In such an ideal circumstance, we take as an example the nine CP-conserving observables discussed in section~\ref{sec:CP-conserving observables}, and illustrate in figure~\ref{fig:statistical uncertainty} the estimated relative statistical uncertainties of these observables at LHCb. Furthermore, the relative statistical uncertainties of the corresponding averaged observables are given in table~\ref{tab:statistical uncertainty}. We can see that $\tau^-\to \pi^-\nu_\tau$ has the highest analyzing power among the three $\tau$ decay channels.

\section{Conclusions}
\label{sec:conclusions}

The observed $R(D^{(*)})$ anomalies may indicate possible NP in $b \to c \tau^- \bar{\nu}_\tau$ transitions. In this context, the $B_c^- \to J/\psi \tau^-\bar{\nu}_\tau$ decay, which is also induced by the same quark-level transition, provides an ideal and clean mode to search for these possible NP effects. However, since the $\tau$ lepton is very short-lived and its decay products contain at least one undetected neutrino, the $\tau$ three-momentum in the decay cannot be determined precisely. Therefore, in this paper, we have proposed to extract the maximum information from the visible kinematic distributions of the cascade decays $B_c^- \to J/\psi (\to \mu^+ \mu^-) \tau^- (\to \pi^- \nu_\tau, \rho^- \nu_\tau, \ell^-\bar{\nu}_\ell\nu_\tau)\bar{\nu}_\tau$ by considering the polarizations of both $\tau$ and $J/\psi$ at the same time. We found that there are in total 34 normalized observables that can be extracted from the fully differential decay rate given by eq.~\eqref{eq:differential decay rate}. Starting with the most general dimension-six effective Hamiltonian including both the left- and right-handed neutrinos, we can express these normalized observables in terms of 14 independent transversity amplitudes. 

To illustrate the sensitivities of these observables to the different NP scenarios, we have calculated their numerical results by considering six different NP benchmark points, which include four scenarios with purely left-handed neutrinos and two ones with purely right-handed neutrinos. We have also used the latest lattice results for the $B_c\to J/\psi$ (axial-)vector and the lattice+NRQCD results for the tensor form factors. The observables considered can be divided into two parts: the CP-conserving and the CP-violating ones. It is found that the SM contributions to the CP-violating observables vanish to a very good approximation. Therefore, any non-zero measurements of them would be a smoking-gun signature of NP. Although the CP-conserving observables are non-zero within SM, they can also serve to distinguish the different NP scenarios by comparing the experimental measurements from the SM predictions. As an illustration, we have picked up nine of these kinds of observables to show their potential role in distinguishing the different NP scenarios. Finally, we found that some combinations of the observables, which are defined by $R_n$ in eq.~\eqref{eq:Rn}, can only be attributed to the right-handed neutrinos. 

On the other hand, considering the low statistics of the fully differential distribution, we have also studied the integrated observables with only one kinematic variable left. Due to the largely increased statistics, these observables are much more promising to be measured in the future with certain precision. In addition, assuming an ideal circumstance, we have estimated the statistical uncertainties of the nine CP-conserving observables at LHCb, and found that, among the three $\tau$ decay channels, $\tau^-\to \pi^-\nu_\tau$ is the most sensitive one to measure the $\tau$ polarizations.

As a final comment, we would like to emphasize again that, in order to confirm the presence of NP effects and to further distinguish the different NP scenarios, it is essential to go beyond the purely total decay rate measurements. The high statistics required to extract the whole energy and angular distributions may be achieved in the future high-luminosity LHCb~\cite{LHCb:2018roe} and Belle II~\cite{Belle-II:2018jsg} experiments, and then provide a definite answer to the currently observed $R(D^{(*)})$ anomalies.

\acknowledgments
This work is supported by the National Natural Science Foundation of China under Grant Nos.~12135006 and 12075097, as well as by the Fundamental Research Funds for the Central Universities under Grant Nos.~CCNU20TS007, CCNU19TD012 and CCNU22LJ004.

\appendix

\section{Calculation of the spin density matrices}
\label{app:Spin density matrices}

\begin{figure}[t]
	\centering
	\includegraphics[width=0.75\textwidth]{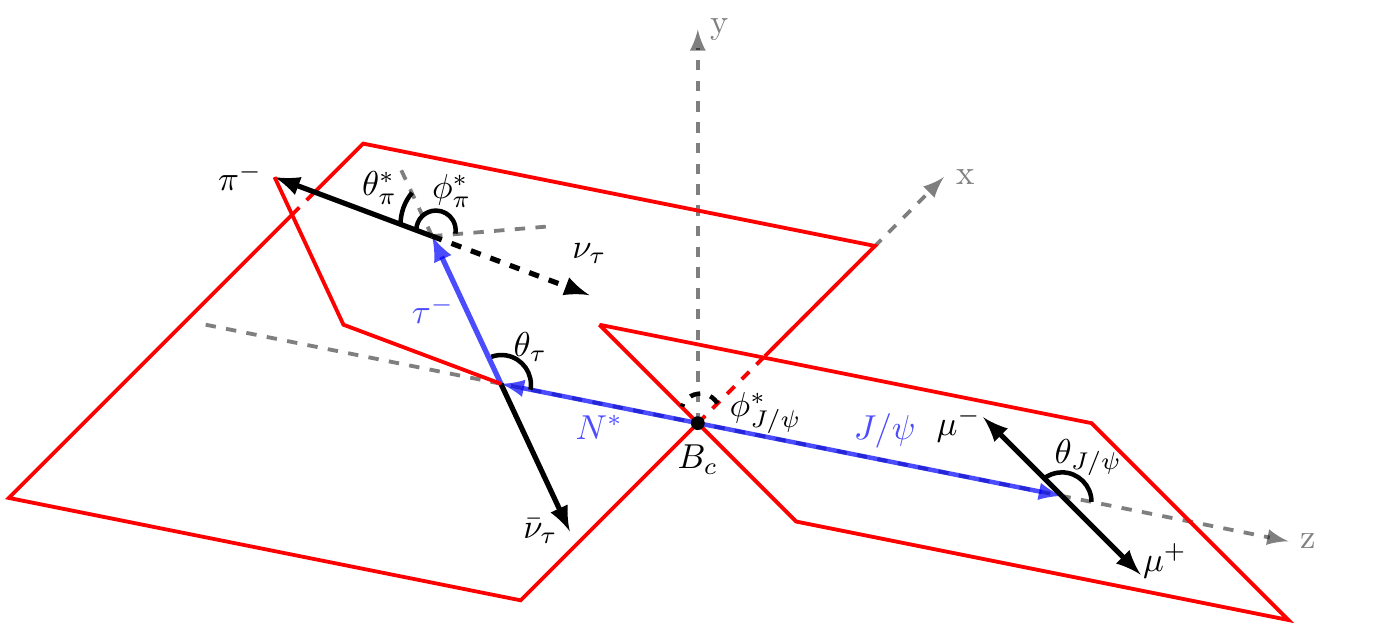}
	\caption{Definitions of the angles in the $B_c^- \to J/\psi (\to \mu^+ \mu^-)\tau^- (\to \pi^- \nu_\tau)\bar{\nu}_\tau$ decay. Here all four successive decays are analyzed in the rest frames of the corresponding parent particles.} \label{fig:angle2}
\end{figure}

In this appendix, we detail the derivations of the three spin density matrices appearing in eq.~\eqref{eq:decay rate}. Taking the hadronic decay $\tau^- \to \pi^- \nu_\tau$ as an example, we illustrate in figure~\ref{fig:angle2} the frames of reference chosen in this appendix. Note that they are different from that adopted in section~\ref{sec:Visible kinematics}, and we use the superscript ``*" to characterize the angles in these specific reference frames.

\subsection{$\tau$ decay density matrix}
\label{app:rho_tau}

The $\tau$ decay density matrix $\rho^\tau(s,s^\prime)$ is calculated in the $\tau$ rest frame with the z-axis chosen to be the direction of $\tau$. We denote the solid angle of a particle $p$ relative to the z-axis as $(\theta^*_{p},\phi^*_{p})$. With these notations, we can write the spinors of the decaying $\tau$ and its decay product $\nu_\tau$, respectively, as~\cite{Auvil:1966eao,Haber:1994pe}
\begin{equation}
	\label{eq:tauspinors}
	\begin{aligned}
		u_{\tau}\left(\frac{1}{2}\right)&=\sqrt{2m_{\tau}}\left(1, \ 0,\ 0, \ 0\right)^T,\quad
		u_{\tau}\left(-\frac{1}{2}\right)=\sqrt{2m_{\tau}}\left(0,\ 1,\ 0,\ 0\right)^T,
		\\[2mm]
		u_{\nu_\tau}\left(\frac{1}{2}\right)&=\sqrt{\left|\vec{p}_{\nu_\tau}\right|} \left(\cos \frac{\theta^*_{\nu_\tau}}{2}, \ \mathrm{e}^{i\phi^*_{\nu_\tau}}\sin \frac{\theta^*_{\nu_\tau}}{2},\ \cos \frac{\theta^*_{\nu_\tau}}{2}, \ \mathrm{e}^{i\phi^*_{\nu_\tau}}\sin \frac{\theta^*_{\nu_\tau}}{2}\right)^T,
		\\[2mm]
		u_{\nu_\tau}\left(-\frac{1}{2}\right)&=\sqrt{\left|\vec{p}_{\nu_\tau}\right|} \left(-\mathrm{e}^{-i\phi^*_{\nu_\tau}}\sin \frac{\theta^*_{\nu_\tau}}{2},\ \cos \frac{\theta^*_{\nu_\tau}}{2},\ \mathrm{e}^{-i\phi^*_{\nu_\tau}}\sin \frac{\theta^*_{\nu_\tau}}{2},\ -\cos \frac{\theta^*_{\nu_\tau}}{2}\right)^T.
	\end{aligned}
\end{equation}
The helicity amplitudes for the $\tau^-\to \pi^- \nu_\tau$ decay can be written as
\begin{equation}
	{\cal M}_{\lambda_\tau}(\tau^-\to \pi^- \nu_\tau) = i\sqrt{2}G_F V_{ud}^\ast f_{\pi} p_{\pi}^{\mu}\, \bar{u}_{\nu_\tau} \gamma_\mu P_L u_\tau(\lambda_\tau),
\end{equation}
where $f_{\pi}$ is the pion decay constant. Similarly, for the $\tau^-\to \rho^- \nu_\tau$ decay, we have
\begin{equation}
	{\cal M}^{\lambda_\rho}_{\lambda_\tau}(\tau^-\to \rho^- \nu_\tau) = i\sqrt{2}G_F V_{ud}^\ast f_\rho m_\rho \tilde{\varepsilon}_{\rho}^{\mu*}(\lambda_\rho)\,\bar{u}_{\nu_\tau} \gamma_\mu P_L u_\tau(\lambda_\tau),
\end{equation}
where $f_{\rho}$ and $\tilde{\varepsilon}_{\rho}^\mu(\lambda_\rho)$ are the decay constant and polarization vectors of the $\rho$ meson. With our notations, we can write $\tilde{\varepsilon}_{\rho}^\mu\left( \lambda_\rho \right)$ explicitly as~\cite{Auvil:1966eao,Haber:1994pe}
\begin{equation} \label{eq:rhopolarization}
	\begin{aligned}
		\tilde{\varepsilon}_{\rho}^\mu(\pm1)&=\frac{1}{\sqrt{2}}\mathrm{e}^{\pm i\phi^*_\rho}\left(0,\mp\cos\theta^*_\rho\cos\phi^*_\rho+i\sin\phi^*_\rho,-i\cos\phi^*_\rho\mp\cos\theta^*_\rho\sin\phi^*_\rho,\pm\sin\theta^*_\rho\right),\\[2mm]
		\tilde{\varepsilon}_{\rho}^\mu(0)&=\left(\frac{|\vec{p}_\rho|}{m_\rho},\frac{p_\rho^0}{m_\rho}\sin\theta^*_\rho\cos\phi^*_\rho,\frac{p_\rho^0}{m_\rho}\sin\theta^*_\rho\sin\phi^*_\rho,\frac{p_\rho^0}{m_\rho}\cos\theta^*_\rho\right).
	\end{aligned}
\end{equation}
The normalized $\tau$ decay density matrices for these two channels can be written as~\cite{Bourrely:1980mr,Boudjema:2009fz}
\begin{equation} \label{eq:rhotH}
	\frac{\rho^\tau(s,s^\prime)}{\mathrm{Tr}\left[\rho^\tau\right]}=\begin{pmatrix}
		\frac{1+\alpha_\tau\cos\theta^*_d}{2} &\frac{\alpha_\tau\sin\theta^*_d}{2}\mathrm{e}^{i\phi^*_d}\\[2mm] \frac{\alpha_\tau\sin\theta^*_d}{2}\mathrm{e}^{-i\phi^*_d} &\frac{ 1-\alpha_\tau\cos\theta^*_d}{2} 
	\end{pmatrix},
\end{equation}
where $d=\{\pi,\rho\}$ and $\alpha_\tau=\{1,(m_\tau^2-2m_\rho^2)/(m_\tau^2+2m_\rho^2)\}$ are the polarization asymmetries for the two channels $\{\tau^-\to\pi^-\nu_\tau, \tau^-\to\rho^-\nu_\tau\}$. We can then easily get the decay rates for these two channels, which read respectively as
\begin{equation}
	\begin{aligned}
		\Gamma(\tau^-\to\pi^-\nu_\tau)&=\frac{1}{2}\frac{1}{2 m_{\tau}}\frac{\left|\vec{p}_{\pi}\right|}{4\pi m_{\tau}}\mathrm{Tr}\left[\rho^{\tau}\right]=\frac{G_F^2f_\pi^2|V_{ud}|^2} {16\pi m_{\tau}}\left(m_\tau^2-m_\pi^2\right)^2,\\[2mm]
		\Gamma(\tau^-\to\rho^-\nu_\tau)&=\frac{1}{2}\frac{1}{2 m_{\tau}}\frac{\left|\vec{p}_{\rho}\right|}{4\pi m_{\tau}}\mathrm{Tr}\left[\rho^{\tau}\right]=\frac{G_F^2f_\rho^2|V_{ud}|^2} {16\pi m_{\tau}^3}\left(m_\tau^2-m_\rho^2\right)^2\left(m_\tau^2+2m_\rho^2\right).
	\end{aligned}
\end{equation}

The helicity amplitudes for the leptonic decay $\tau^-\to\ell^-\bar{\nu}_\ell\nu_\tau$ are given by
\begin{equation}
	{\cal M}^{\lambda_\ell}_{\lambda_\tau}(\tau^-\to\ell^-\bar{\nu}_\ell\nu_\tau) = i2\sqrt{2}G_F\,\bar{u}_{\nu_\tau} \gamma^\mu P_L u_\tau(\lambda_\tau)\,
	\bar{u}_{\ell}(\lambda_\ell) \gamma_\mu P_L v_{\bar{\nu}_\ell},
\end{equation}
from which we can then write the corresponding normalized decay density matrix as~\cite{Bourrely:1980mr,Leader:2011vwq}
\begin{equation}
	\label{eq:rhotL}
	\frac{\rho^\tau(s,s^\prime)}{\mathrm{Tr}\left[\rho^\tau\right]}=\begin{pmatrix}
		\frac{1+\cos\theta^*_{\bar{\nu}_\ell}}{2} &\frac{\sin\theta^*_{\bar{\nu}_\ell}}{2}\mathrm{e}^{i\phi^*_{\bar{\nu}_\ell}}\\[2mm] \frac{\sin\theta^*_{\bar{\nu}_\ell}}{2}\mathrm{e}^{-i\phi^*_{\bar{\nu}_\ell}} &\frac{ 1-\cos\theta^*_{\bar{\nu}_\ell}}{2} 
	\end{pmatrix},
\end{equation}
with the normalization given by
\begin{equation}
	\mathrm{Tr}\left[\rho^\tau\right]=128G_F^2\left(p_{\tau}\cdot p_{\bar{\nu}_\ell}\right)\left(p_{\ell}\cdot p_{\nu_\tau}\right).
\end{equation}
After performing the phase-space integration, we can get the corresponding decay rate as
\begin{equation}
	\Gamma(\tau^-\to\ell^-\bar{\nu}_\ell\nu_\tau)=\frac{G_F^2m_\tau^5}{192\pi^3}\left(1-8y^2+8y^6-y^8-24y^4\ln y\right),
\end{equation}
where $y=m_\ell/m_\tau$.

\subsection{$J/\psi$ decay density matrix}
\label{app:rho_J}

The $J/\psi$ decay density matrix $\rho^{J/\psi}(\lambda,\lambda^\prime)$ is calculated in the $J/\psi$ rest frame with the z-axis chosen to be the direction of $J/\psi$. We denote the solid angles of the $\mu^-$ lepton relative to the z-axis by $(\theta_{J/\psi},\phi^*_{J/\psi})$. In this reference frame, the polarization vectors of the $J/\psi$ meson and the spinors of the massless $\mu^-$ and $\mu^+$ leptons can be written as~\cite{Auvil:1966eao,Haber:1994pe,Hu:2021emb} 
\begin{equation} \label{eq:Jpsipolarizationrestframe}
	\varepsilon_{J/\psi}^\mu\left( 0 \right) = \left(0, \, 0, \, 0,\,1  \right) ,\quad\varepsilon_{J/\psi}^\mu\left(\pm 1 \right) = \left(0,\, \mp 1, \, -i, \, 0 \right)/\sqrt{2},
\end{equation}
and
\begin{equation}\label{eq:muspinors}
	\begin{aligned}
	u_{\mu^-}\left(\frac{1}{2}\right)&=\sqrt{\frac{m_{J/\psi}}{2}}\left(\cos \frac{\theta_{J/\psi}}{2}, \ \mathrm{e}^{i\phi^*_{J/\psi}}\sin \frac{\theta_{J/\psi}}{2},\ \cos \frac{\theta_{J/\psi}}{2}, \ \mathrm{e}^{i\phi^*_{J/\psi}}\sin \frac{\theta_{J/\psi}}{2}\right)^T,
	\\[2mm]
	u_{\mu^-}\left(-\frac{1}{2}\right)&=\sqrt{\frac{m_{J/\psi}}{2}}\left(-\mathrm{e}^{-i\phi^*_{J/\psi}}\sin \frac{\theta_{J/\psi}}{2},\ \cos \frac{\theta_{J/\psi}}{2},\ \mathrm{e}^{-i\phi^*_{J/\psi}}\sin \frac{\theta_{J/\psi}}{2},\ -\cos \frac{\theta_{J/\psi}}{2}\right)^T,
	\\[2mm]
	v_{\mu^+}\left(\frac{1}{2}\right)&=\sqrt{\frac{m_{J/\psi}}{2}} \left(\cos \frac{\theta_{J/\psi}}{2}, \ \mathrm{e}^{i\phi^*_{J/\psi}}\sin \frac{\theta_{J/\psi}}{2},\ -\cos \frac{\theta_{J/\psi}}{2}, \ -\mathrm{e}^{i\phi^*_{J/\psi}}\sin \frac{\theta_{J/\psi}}{2}\right)^T,
	\\[2mm]
	v_{\mu^+}\left(-\frac{1}{2}\right)&=\sqrt{\frac{m_{J/\psi}}{2}} \left(-\mathrm{e}^{-i\phi^*_{J/\psi}}\sin \frac{\theta_{J/\psi}}{2},\ \cos \frac{\theta_{J/\psi}}{2},\ -\mathrm{e}^{-i\phi^*_{J/\psi}}\sin \frac{\theta_{J/\psi}}{2},\ \cos \frac{\theta_{J/\psi}}{2}\right)^T,
	\end{aligned}
\end{equation}
respectively. Here we have used the Jacob-Wick second particle convention~\cite{Jacob:1959at}, which defines the helicity states of a particle moving in the negative z-direction. 

As the $J/\psi \to \mu^+\mu^-$ decay is an electromagnetic process, we can write its helicity amplitudes as
\begin{align}
	{\cal M}^{\lambda_{J/\psi}}_{\lambda_{\mu^-},\lambda_{\mu^+}}\left(J/\psi \to \mu^+\mu^-\right) = \frac{-8i\pi \alpha_{\rm EM} f_{J/\psi}}{3 m_{J/\psi}} \varepsilon_{J/\psi}^{\mu} \left(\lambda_{J/\psi}\right) \bar{u}_{\mu^-}\left(\lambda_{\mu^-}\right) \gamma_\mu v_{\mu^+}\left(\lambda_{\mu^+}\right),
\end{align}
where $f_{J/\psi}$ is the decay constant of the $J/\psi$ meson, and $\alpha_{\rm EM}=e^2/(4\pi)$ is the fine-structure constant. The normalized $J/\psi$ decay density matrix can then be written as
\begin{equation} \label{eq:rhoJ}
	\frac{\rho^{J/\psi}(\lambda,\lambda^\prime)}{\mathrm{Tr}\left[\rho^{J/\psi}\right]}=\begin{pmatrix}
	\frac{1+\cos^2 \theta_{J/\psi}}{4} &\frac{\cos \theta_{J/\psi}\sin \theta_{J/\psi}}{2\sqrt{2}} \mathrm{e}^{i\phi^*_{J/\psi}}&&\frac{\sin^2 \theta_{J/\psi}}{4}\mathrm{e}^{2i\phi^*_{J/\psi}}\\[2mm] 
	\frac{\cos \theta_{J/\psi}\sin \theta_{J/\psi}}{2\sqrt{2}} \mathrm{e}^{-i\phi^*_{J/\psi}} & \frac{\sin^2 \theta_{J/\psi}}{2}&&-\frac{\cos \theta_{J/\psi}\sin \theta_{J/\psi}}{2\sqrt{2}} \mathrm{e}^{i\phi^*_{J/\psi}}\\[2mm]
	\frac{\sin^2 \theta_{J/\psi}}{4}\mathrm{e}^{-2i\phi^*_{J/\psi}}&-\frac{\cos \theta_{J/\psi}\sin \theta_{J/\psi}}{2\sqrt{2}} \mathrm{e}^{-i\phi^*_{J/\psi}}&&\frac{1+\cos^2 \theta_{J/\psi}}{4}
	\end{pmatrix},
\end{equation}
and the total decay rate is given by
\begin{equation}
	\Gamma(J/\psi \to \mu^+\mu^-)=\frac{1}{3}\frac{1}{2 m_{J/\psi}}\frac{\left|\vec{p}_{J/\psi}\right|}{4\pi m_{J/\psi}}\mathrm{Tr}\left[\rho^{J/\psi}\right]=\frac{16\pi \alpha^2_{\rm EM} f^2_{J/\psi}}{27 m_{J/\psi}}.
\end{equation}

\subsection{$B_c^- \to J/\psi\tau^- \bar{\nu}_\tau$ spin density matrix}
\label{app:rho_Bc}

In order to discuss the semitauonic $B_c^- \to J/\psi\tau^- \bar{\nu}_\tau$ decay, we can divide it into the two successive processes $B_c^- \to J/\psi N^{*-}$ and $N^{*-} \to \tau^- \bar{\nu}_\tau$, where $N^*$ is a virtual intermediate state that refers to the $W$ boson within the SM. They can be discussed most conveniently in the $B_c$ and $N^*$ rest frames respectively, with the z-axes  both chosen to be the direction of the $J/\psi$ meson. In the $B_c$ rest frame, the polarization vectors of $J/\psi$ and $N^*$ can be written as~\cite{Auvil:1966eao,Haber:1994pe} 
\begin{equation} \label{eq:Jpsipolarizationmovingframe}
	\tilde{\varepsilon}_{J/\psi}^\mu\left( 0 \right) = \left(\left|\vec{p}_{J/\psi} \right|, \, 0, \, 0,\, E_{J/\psi}  \right)/m_{J/\psi}, \quad \tilde{\varepsilon}_{J/\psi}^\mu\left(\pm 1 \right) = \left(0,\, \mp 1, \, -i, \, 0 \right)/\sqrt{2},
\end{equation}
and
\begin{equation} \label{eq:N*movingframe}
	\tilde{\varepsilon}_{N^*}^\mu\left(t \right)  = q^\mu/\sqrt{q^2},\quad\tilde{\varepsilon}_{N^*}^\mu\left(0 \right)  = \left(\left|\vec{q} \right|,\, 0,\, 0,\, -q^0  \right) /\sqrt{q^2}, \quad
	\tilde{\varepsilon}_{N^*}^\mu\left(\pm 1 \right)  = \left(0,\, \pm1,\, -i,\, 0 \right)/\sqrt{2},
\end{equation}
respectively. Throughout this paper, all the polarization vectors with the symbol `` $\tilde{\;}$ " are defined in a moving reference frame, while the ones without `` $\tilde{\;}$ " are given in the rest frame of a vector particle; this explains our notations for the $\rho$, $J/\psi$ and $N^*$ polarization vectors, as given by eqs.~\eqref{eq:rhopolarization}, \eqref{eq:Jpsipolarizationrestframe}~(\eqref{eq:Jpsipolarizationmovingframe}) and \eqref{eq:N*movingframe}~(\eqref{eq:N*restframe}), respectively.

The helicity amplitudes for the hadronic part with different Lorentz structures are defined, respectively, as~\cite{Hu:2021emb}
\begin{equation}
   \begin{aligned}
   H^{\{L,R\}}_{\lambda_{J/\psi}} & = \left\langle J/\psi(\lambda_{J/\psi}) \left|\left(C^S_{R\{L,R\}}+C^S_{L\{L,R\}}\right) {\bar c} b +\left(C^S_{R\{L,R\}}-C^S_{L\{L,R\}}\right) {\bar c} \gamma_5 b \right| B_c^- \right\rangle,
  \\[2mm]
  H^{\{L,R\}}_{\lambda_{J/\psi},\lambda} & = \tilde{\varepsilon}_{N^*}^{\mu*}(\lambda) \Big\langle J/\psi(\lambda_{J/\psi}) \Big|\left(\{1,0\}+C^V_{L\{L,R\}}+C^V_{R\{L,R\}}\right) {\bar c}\gamma_\mu b\\[1mm]
  &\hspace{5.0cm} -\left(\{1,0\}+C^V_{L\{L,R\}}-C^V_{R\{L,R\}}\right){\bar c}\gamma_\mu \gamma_5 b  \Big| B_c^- \Big \rangle,
  \\[2mm]
  H^{\{L,R\}}_{\lambda_{J/\psi},\lambda,\lambda^\prime} & =C^T_{\{LL,RR\}}\tilde{\varepsilon}_{N^*}^{\mu*}(\lambda) \tilde{\varepsilon}_{N^*}^{\nu*}(\lambda^\prime) \left\langle J/\psi(\lambda_{J/\psi}) \left| {\bar c} i \sigma_{\mu\nu}\left(1\mp\gamma_5\right) b \right| B_c^- \right\rangle, 
  \end{aligned}
\end{equation}
where the entry ``1" refers to the SM contribution. For the parametrization of the hadronic matrix elements in terms of the $B_c\to J/\psi$ transition form factors, we use the same definitions as in our previous work~\cite{Hu:2021emb}. For the scalar and pseudo-scalar operators, there are only two non-zero hadronic helicity amplitudes,
\begin{equation}
	H^{\{L,R\}}_0 = {\cal A}_{\{L,R\},t}^{SP},
\end{equation}
while for the vector and axial-vector operators, we have eight non-zero hadronic helicity amplitudes,
\begin{equation}
	\begin{aligned}
	H^{\{L,R\}}_{0,t} &= {\cal A}_{\{L,R\},t}^{VA}, &
	H^{\{L,R\}}_{0,0} &= {\cal A}_{\{L,R\},0}, 
    \\[2mm]
	H^{\{L,R\}}_{1,1} &= \left({\cal A}_{\{L,R\},\perp} + {\cal A}_{\{L,R\},\parallel} \right)/\sqrt{2}, &
	H^{\{L,R\}}_{-1,-1} &= \left({\cal A}_{\{L,R\},\perp} - {\cal A}_{\{L,R\},\parallel} \right)/\sqrt{2}. 
    \end{aligned}
\end{equation}
For the tensor operators, on the other hand, there are in total twenty-four non-zero hadronic helicity amplitudes,
\begin{equation}
	\begin{aligned}
	H^{\{L,R\}}_{0,t,0} &=\pm H^{\{L,R\}}_{0,-1,1} = -H^{\{L,R\}}_{0,0,t} = \mp H^{\{L,R\}}_{0,1,-1} = {\cal A}_{\{L,R\},0}^T, 
    \\[2mm]
	H^{\{L,R\}}_{1,t,1} &= \pm H^{\{L,R\}}_{1,0,1} = -H^{\{L,R\}}_{1,1,t} =\mp H^{\{L,R\}}_{1,1,0}  = \left({\cal A}_{\{L,R\},\parallel}^T + {\cal A}_{\{L,R\},\perp}^T \right)/\sqrt{2},
    \\[2mm]
	 H^{\{L,R\}}_{-1,-1,t}&=\pm H^{\{L,R\}}_{-1,0,-1} = - H^{\{L,R\}}_{-1,t,-1} =\mp H^{\{L,R\}}_{-1,-1,0}  = \left({\cal A}_{\{L,R\},\parallel}^T - {\cal A}_{\{L,R\},\perp}^T \right)/\sqrt{2}.
    \end{aligned}
\end{equation}
Here the non-zero hadronic helicity amplitudes are all expressed in terms of the transversity amplitudes, with the latter given explicitly as
\begin{align}
	\mathcal{A}^{SP}_{\{L,R\},t}&=-\left(C^S_{R\{L,R\}}-C^S_{L\{L,R\}}\right)A_0(q^2)\frac{\sqrt{Q_+Q_-}}{m_b+m_c},\nonumber\\[2mm]
	\mathcal{A}^{VA}_{\{L,R\},t}&=-\left(\{1,0\}+C^V_{L\{L,R\}}-C^V_{R\{L,R\}}\right)A_0(q^2)\frac{\sqrt{Q_+Q_-}}{\sqrt{q^2}},\nonumber\\[2mm]
	\mathcal{A}_{\{L,R\},0}&=-\left(\{1,0\}+C^V_{L\{L,R\}}-C^V_{R\{L,R\}}\right)\frac{m_{B_c}+ m_{J/\psi}}{2m_{J/\psi}\sqrt{q^2}}\Bigg[A_1(q^2)(m_{B_c}^2- m_{J/\psi}^2-q^2)\nonumber\\
	&\hspace{8.5cm} -A_2(q^2)\frac{Q_+Q_-}{(m_{B_c}+ m_{J/\psi})^2}\Bigg], \nonumber\\[2mm]
	\mathcal{A}_{\{L,R\},\perp}&=-\left(\{1,0\}+C^V_{L\{L,R\}}-C^V_{R\{L,R\}}\right)\sqrt{2}A_1(q^2)(m_{B_c}+ m_{J/\psi}), \nonumber\\[2mm]
	\mathcal{A}_{\{L,R\},\parallel}&=\left(\{1,0\}+C^V_{L\{L,R\}}+C^V_{R\{L,R\}}\right)\sqrt{2}V(q^2)\frac{\sqrt{Q_+Q_-}}{m_{B_c}+ m_{J/\psi}},\nonumber\\[2mm]
	\mathcal{A}^T_{\{L,R\},0}&=\pm C^T_{\{LL,RR\}}\frac{1}{2m_{J/\psi}}\left[T_2(q^2)(m_{B_c}^2+3 m_{J/\psi}^2-q^2)-T_3(q^2)\frac{Q_+Q_-}{ m_{B_c}^2-m_{J/\psi}^2}\right],
	\nonumber\\[2mm]
	\mathcal{A}^T_{\{L,R\},\perp}&=\pm C^T_{\{LL,RR\}}\sqrt{2}T_2(q^2)\frac{m_{B_c}^2- m_{J/\psi}^2}{\sqrt{q^2}},
	\nonumber\\[2mm]
	\mathcal{A}^T_{\{L,R\},\parallel}&=C^T_{\{LL,RR\}}\sqrt{2}T_1(q^2)\frac{\sqrt{Q_+Q_-}}{\sqrt{q^2}},
\end{align}
where $m_b$ and $m_c$ are the current quark masses evaluated at the scale $\mu= m_b$, and  $Q_\pm \equiv (m_{B_c}\pm  m_{J/\psi})^2 - q^2$, with $q^2$ being the dilepton invariant mass squared. 

The process $N^{*-} \to \tau^- \bar{\nu}_\tau$ is most conveniently described in the $N^*$ rest frame. In this reference frame, the polarization vectors of the intermediate $N^*$ boson are now given by~\cite{Auvil:1966eao,Haber:1994pe} 
\begin{equation} \label{eq:N*restframe}
	\varepsilon_{N^*}^\mu\left(t \right)  = \left(1,\, 0,\, 0,\, 0  \right),\quad\varepsilon_{N^*}^\mu\left(0 \right)  = \left(0,\, 0,\, 0,\, -1  \right), \quad
	\varepsilon_{N^*}^\mu\left(\pm 1 \right)  = \left(0,\, \pm1,\, -i,\, 0 \right)/\sqrt{2}.
\end{equation}
Denoting the polar angle of the $\tau$ lepton relative to the z-axis by $\theta_\tau$, we can write the spinors of $\tau$ and $\bar{\nu}_\tau$ explicitly as~\cite{Auvil:1966eao,Haber:1994pe} 
\begin{align}
		u_{\tau}\left(\frac{1}{2}\right)&=\left(\beta_+\cos \frac{\theta_\tau}{2}, \ \beta_+\sin \frac{\theta_\tau}{2},\ \beta_-\cos \frac{\theta_\tau}{2}, \ \beta_-\sin \frac{\theta_\tau}{2}\right)^T,
		\nonumber\\[2mm]
		u_{\tau}\left(-\frac{1}{2}\right)&=\left(-\beta_+\sin \frac{\theta_\tau}{2}, \ \beta_+\cos \frac{\theta_\tau}{2},\ \beta_-\sin \frac{\theta_\tau}{2}, \ -\beta_-\cos \frac{\theta_\tau}{2}\right)^T,
		\nonumber\\[2mm]
		v_{\bar{\nu}_\tau}\left(\frac{1}{2}\right)&=\sqrt{\left|\vec{p}_\tau\right|} \left(\cos \frac{\theta_\tau}{2}, \ \sin \frac{\theta_\tau}{2},\ -\cos \frac{\theta_\tau}{2}, \ -\sin \frac{\theta_\tau}{2}\right)^T,
		\nonumber\\[2mm]
		v_{\bar{\nu}_\tau}\left(-\frac{1}{2}\right)&=\sqrt{\left|\vec{p}_\tau\right|} \left(-\sin \frac{\theta_\tau}{2}, \ \cos \frac{\theta_\tau}{2},\ -\sin \frac{\theta_\tau}{2}, \ \cos \frac{\theta_\tau}{2}\right)^T,
\end{align}
where, as in eq.~\eqref{eq:muspinors}, we have also used the Jacob-Wick second particle convention~\cite{Jacob:1959at}, and $\beta_{\pm}=\sqrt{E_\tau\pm m_\tau}$ with $E_\tau=(q^2+m_\tau^2)/(2\sqrt{q^2})$. The helicity amplitudes for the leptonic part with different Lorentz structures are defined, respectively, as~\cite{Hu:2021emb}
\begin{equation}
	\begin{aligned}
		L^{\{L,R\}}_{\lambda_{\tau}} &\equiv\left\langle\tau^{-}(\lambda_{\tau}) \bar{\nu}\left|\bar{\tau} P_{\{L,R\}} \nu\right| 0\right\rangle,\\[2mm]
		L_{\lambda_{\tau},\lambda}^{\{L,R\}} &\equiv \varepsilon_{N^*}^{\mu}\left(\lambda\right)\left\langle\tau^{-}(\lambda_{\tau}) \bar{\nu}\left|\bar{\tau} \gamma_{\mu} P_{\{L,R\}} \nu\right| 0\right\rangle ,\\[2mm]
		L_{\lambda_{\tau},\lambda,\lambda^\prime}^{\{L,R\}} &\equiv-i \varepsilon_{N^*}^{\mu}\left(\lambda\right) \varepsilon_{N^*}^{\nu}\left(\lambda^\prime\right)\left\langle\tau^{-}(\lambda_{\tau}) \bar{\nu}\left|\bar{\tau} \sigma_{\mu \nu} P_{\{L,R\}} \nu\right| 0\right\rangle.
	\end{aligned}
\end{equation}

Combining the helicity amplitudes for both the hardronic and leptonic parts, we can obtain the spin density matrix $\rho^{B_c}(s,s^\prime,\lambda,\lambda^\prime)$ introduced in eq.~\eqref{eq:decay rate}, and then work out the explicit forms of the parameters defined in eq.~\eqref{eq:rhoBc}. However, as mentioned already in section~\ref{sec:Observables}, the parity-conserving $J/\psi \to \mu^+\mu^-$ decay cannot be used to extract the vector polarizations of $J/\psi$, we list therefore the remaining parameters redefined in eq.~\eqref{eq:redefine} as
\begin{equation}
	\begin{aligned}
		\tilde{T}^{\{U,L\}}_{ii}&=\frac{1}{2}\left\langle \tilde{T}^{\{U,L\}}_{ii}\right\rangle+\tilde{Z}^{\{U,L\}}_{ii}P^0_1\left(\cos\theta_\tau\right)+\tilde{A}^{\{U,L\}}_{ii}P^0_2\left(\cos\theta_\tau\right),\\[2mm]
		\tilde{T}^{\{U,L\}}_{\{\perp L,TL\}}&=-\frac{2}{\pi}\left\langle \tilde{T}^{\{U,L\}}_{\{\perp L,TL\}}\right\rangle P^1_1\left(\cos\theta_\tau\right)-\frac{1}{2}\tilde{Z}^{\{U,L\}}_{\{\perp L,TL\}}P^1_2\left(\cos\theta_\tau\right),\\[2mm]
		\tilde{T}^{\{U,L\}}_{\perp T}&=\frac{1}{4}\left\langle \tilde{T}^{\{U,L\}}_{\perp T}\right\rangle P^2_2\left(\cos\theta_\tau\right),\\[2mm]
		\tilde{T}^{\{\perp,T\}}_{\{\perp L,TL\}}&=\frac{1}{2}\left\langle \tilde{T}^{\{\perp,T\}}_{\{\perp L,TL\}}\right\rangle+\tilde{Z}^{\{\perp,T\}}_{\{\perp L,TL\}}P^0_1\left(\cos\theta_\tau\right)+\tilde{A}^{\{\perp,T\}}_{\{\perp L,TL\}}P^0_2\left(\cos\theta_\tau\right),\\[2mm]
		\tilde{T}^{\{\perp,T\}}_{\{ii,\perp T\}}&=-\frac{2}{\pi}\left\langle \tilde{T}^{\{\perp,T\}}_{\{ii,\perp T\}}\right\rangle P^1_1\left(\cos\theta_\tau\right)-\frac{1}{2}\tilde{Z}^{\{\perp,T\}}_{\{ii,\perp T\}}P^1_2\left(\cos\theta_\tau\right),
	\end{aligned}
\end{equation}
where $P^m_l\left(\cos\theta_\tau\right)$ are the associated Legendre functions. Meanwhile, the integration over the unmeasurable kinematic variables will also cause a loss of some information. Thus, we list only the observables that can be extracted from the five-fold differential distribution of eq.~\eqref{eq:differential decay rate}. Explicitly, we have 
\begin{align}
		\frac{d\Gamma}{dq^2}=&\frac{2\mathcal{N}}{3}\bigg\{\bigg[m_\tau^2\left( \left| \mathcal{A}_{L,0}^+\right|
		^2+\left| \mathcal{A}_{L,\parallel}^+\right|
		^2+\left| \mathcal{A}_{L,\perp}^+\right|
		^2+3\left| \mathcal{A}_{L,t}\right|
		^2\right)\nonumber\\
		&\hspace{3.5cm} +2q^2\left(\left| \mathcal{A}_{L,0}^-\right|
		^2+\left| \mathcal{A}_{L,\parallel}^-\right|
		^2+\left| \mathcal{A}_{L,\perp}^-\right|
		^2\right)\bigg]+\left(L\rightarrow R\right)\bigg\},\nonumber\\[2mm]
		\frac{d\Gamma}{dq^2}\left\langle \tilde{T}^U_{\perp\perp}\right\rangle=&\frac{\mathcal{N}}{\sqrt{6}}\bigg\{\bigg[2m_\tau^2\left( \left| \mathcal{A}_{L,0}^+\right|
		^2+\left| \mathcal{A}_{L,\parallel}^+\right|
		^2+3\left| \mathcal{A}_{L,t}\right|
		^2\right)\nonumber\\
		&\hspace{3.5cm}+q^2\left(4\left| \mathcal{A}_{L,0}^-\right|
		^2+\left| \mathcal{A}_{L,\parallel}^-\right|
		^2+3\left| \mathcal{A}_{L,\perp}^-\right|
		^2\right)\bigg]+\left(L\rightarrow R\right)\bigg\},\nonumber\\[2mm]
		\frac{d\Gamma}{dq^2}\tilde{Z}^U_{\perp\perp}=&\frac{3\mathcal{N}}{\sqrt{6}}\left\{2m_\tau^2\left({\rm Re}\left[\mathcal{A}_{L,t} \mathcal{A}_{L,0}^{+*} \right]+\left(L\to R\right)\right)+q^2\left({\rm Re}\left[\mathcal{A}_{L,\parallel}^- \mathcal{A}_{L,\perp}^{-*} \right]-\left(L\to R\right)\right)\right\},\nonumber\\[2mm]
		\frac{d\Gamma}{dq^2}\tilde{A}^U_{\perp\perp}=&\frac{\mathcal{N}}{\sqrt{6}}\left\{\left[m_\tau^2\left( 2\left| \mathcal{A}_{L,0}^+\right|
		^2-\left| \mathcal{A}_{L,\parallel}^+\right|
		^2\right)-q^2\left(2\left| \mathcal{A}_{L,0}^-\right|
		^2-\left| \mathcal{A}_{L,\parallel}^-\right|
		^2\right)\right]+\left(L\rightarrow R\right)\right\},\nonumber\\[2mm]
		\frac{d\Gamma}{dq^2}\tilde{A}^U_{TT}=&\frac{\mathcal{N}}{\sqrt{6}}\left\{\left[m_\tau^2\left( 2\left| \mathcal{A}_{L,0}^+\right|
		^2-\left| \mathcal{A}_{L,\perp}^+\right|
		^2\right)-q^2\left(2\left| \mathcal{A}_{L,0}^-\right|
		^2-\left| \mathcal{A}_{L,\perp}^-\right|
		^2\right)\right]+\left(L\rightarrow R\right)\right\},\nonumber\\[2mm]
		\frac{d\Gamma}{dq^2}\left\langle \tilde{T}^U_{\perp T}\right\rangle=&\frac{2\mathcal{N}}{\sqrt{6}}\left\{\left(m_\tau^2{\rm Im}\left[\mathcal{A}_{L,\parallel}^+ \mathcal{A}_{L,\perp}^{+*} \right]-q^2{\rm Im}\left[\mathcal{A}_{L,\parallel}^- \mathcal{A}_{L,\perp}^{-*} \right]\right)+\left(\left(L\rightarrow R\right)\right)\right\},\nonumber\\[2mm]
		\frac{d\Gamma}{dq^2}\left\langle \tilde{T}^U_{\perp L}\right\rangle=&-\pi\frac{\sqrt{3}\mathcal{N}}{4}\left\{m_\tau^2\left({\rm Re}\left[\mathcal{A}_{L,t} \mathcal{A}_{L,\perp}^{+*} \right]+\left(L\to R\right)\right) \right. \nonumber\\
		&\hspace{3.5cm}\left. -q^2\left({\rm Re}\left[\mathcal{A}_{L,0}^- \mathcal{A}_{L,\parallel}^{-*} \right]-\left(L\to R\right)\right)\right\},\nonumber\\[2mm]
		\frac{d\Gamma}{dq^2}\tilde{Z}^U_{\perp L}=&-\frac{\mathcal{N}}{\sqrt{3}}\left\{\left(m_\tau^2{\rm Re}\left[\mathcal{A}_{L,0}^+ \mathcal{A}_{L,\perp}^{+*} \right]-q^2{\rm Re}\left[\mathcal{A}_{L,0}^- \mathcal{A}_{L,\perp}^{-*} \right]\right)+\left(L\to R\right)\right\},\nonumber\\[2mm]
		\frac{d\Gamma}{dq^2}\left\langle \tilde{T}^U_{T L}\right\rangle=&-\pi\frac{\sqrt{3}\mathcal{N}}{4}\left\{m_\tau^2\left({\rm Im}\left[\mathcal{A}_{L,t} \mathcal{A}_{L,\parallel}^{+*} \right]+\left(L\to R\right)\right)\right. \nonumber\\
		&\hspace{3.5cm}\left. -q^2\left({\rm Im}\left[\mathcal{A}_{L,0}^- \mathcal{A}_{L,\perp}^{-*} \right]-\left(L\to R\right)\right)\right\},\nonumber\\[2mm]
		\frac{d\Gamma}{dq^2}\tilde{Z}^U_{T L}=&-\frac{\mathcal{N}}{\sqrt{3}}\left\{\left(m_\tau^2{\rm Im}\left[\mathcal{A}_{L,0}^+ \mathcal{A}_{L,\parallel}^{+*} \right]-q^2{\rm Im}\left[\mathcal{A}_{L,0}^- \mathcal{A}_{L,\parallel}^{-*} \right]\right)+\left(L\to R\right)\right\},\nonumber\\[2mm]
		\frac{d\Gamma}{dq^2}\left\langle \tilde{T}^U_{LL}\right\rangle=&\frac{2\mathcal{N}}{\sqrt{6}}\left\{\left[m_\tau^2\left( \left| \mathcal{A}_{L,\parallel}^+\right|
		^2+\left| \mathcal{A}_{L,\perp}^+\right|
		^2\right)+2q^2\left(\left| \mathcal{A}_{L,\parallel}^-\right|
		^2+\left| \mathcal{A}_{L,\perp}^-\right|
		^2\right)\right]+\left(L\rightarrow R\right)\right\},\nonumber\\[2mm]
		\frac{d\Gamma}{dq^2}\tilde{Z}^U_{L L}=&\sqrt{6}\mathcal{N}q^2\left\{{\rm Re}\left[\mathcal{A}_{L,\parallel}^- \mathcal{A}_{L,\perp}^{-*} \right]-\left(L\to R\right)\right\},\nonumber\\[2mm]
		\frac{d\Gamma}{dq^2}\tilde{A}^U_{LL}=&-\frac{\mathcal{N}}{\sqrt{6}}\left\{\left[m_\tau^2\left( \left| \mathcal{A}_{L,\parallel}^+\right|
		^2+\left| \mathcal{A}_{L,\perp}^+\right|
		^2\right)-q^2\left(\left| \mathcal{A}_{L,\parallel}^-\right|
		^2+\left| \mathcal{A}_{L,\perp}^-\right|
		^2\right)\right]+\left(L\rightarrow R\right)\right\},\nonumber\\[2mm]
		\frac{d\Gamma}{dq^2}\left\langle \tilde{T}^L_{\perp\perp}\right\rangle=&\frac{\mathcal{N}}{\sqrt{6}}\bigg\{\bigg[2m_\tau^2\left( \left| \mathcal{A}_{L,0}^+\right|
		^2+\left| \mathcal{A}_{L,\parallel}^+\right|
		^2+3\left| \mathcal{A}_{L,t}\right|
		^2\right)\nonumber\\
		&\qquad\qquad\qquad\qquad-q^2\left(4\left| \mathcal{A}_{L,0}^-\right|
		^2+\left| \mathcal{A}_{L,\parallel}^-\right|
		^2+3\left| \mathcal{A}_{L,\perp}^-\right|
		^2\right)\bigg]-\left(L\rightarrow R\right)\bigg\},\nonumber\\[2mm]
		\frac{d\Gamma}{dq^2}\tilde{Z}^L_{\perp\perp}=&\frac{3\mathcal{N}}{\sqrt{6}}\left\{2m_\tau^2\left({\rm Re}\left[\mathcal{A}_{L,t} \mathcal{A}_{L,0}^{+*} \right]-\left(L\to R\right)\right)-q^2\left({\rm Re}\left[\mathcal{A}_{L,\parallel}^- \mathcal{A}_{L,\perp}^{-*} \right]+\left(L\to R\right)\right)\right\},\nonumber\\[2mm]
		\frac{d\Gamma}{dq^2}\tilde{A}^L_{\perp\perp}=&\frac{\mathcal{N}}{\sqrt{6}}\left\{\left[m_\tau^2\left( 2\left| \mathcal{A}_{L,0}^+\right|
		^2-\left| \mathcal{A}_{L,\parallel}^+\right|
		^2\right)+q^2\left(2\left| \mathcal{A}_{L,0}^-\right|
		^2-\left| \mathcal{A}_{L,\parallel}^-\right|
		^2\right)\right]-\left(L\rightarrow R\right)\right\},\nonumber\\[2mm]
		\frac{d\Gamma}{dq^2}\tilde{A}^L_{TT}=&\frac{\mathcal{N}}{\sqrt{6}}\left\{\left[m_\tau^2\left( 2\left| \mathcal{A}_{L,0}^+\right|
		^2-\left| \mathcal{A}_{L,\perp}^+\right|
		^2\right)+q^2\left(2\left| \mathcal{A}_{L,0}^-\right|
		^2-\left| \mathcal{A}_{L,\perp}^-\right|
		^2\right)\right]-\left(L\rightarrow R\right)\right\},\nonumber\\[2mm]
		\frac{d\Gamma}{dq^2}\left\langle \tilde{T}^L_{\perp T}\right\rangle=&\frac{2\mathcal{N}}{\sqrt{6}}\left\{\left(m_\tau^2{\rm Im}\left[\mathcal{A}_{L,\parallel}^+ \mathcal{A}_{L,\perp}^{+*} \right]+q^2{\rm Im}\left[\mathcal{A}_{L,\parallel}^- \mathcal{A}_{L,\perp}^{-*} \right]\right)-\left(L\to R\right)\right\},\nonumber\\[2mm]
		\frac{d\Gamma}{dq^2}\left\langle \tilde{T}^L_{\perp L}\right\rangle=&-\pi\frac{\sqrt{3}\mathcal{N}}{4}\left\{m_\tau^2\left({\rm Re}\left[\mathcal{A}_{L,t} \mathcal{A}_{L,\perp}^{+*} \right]-\left(L\to R\right)\right)\right. \nonumber\\
		&\hspace{3.5cm}\left. +q^2\left({\rm Re}\left[\mathcal{A}_{L,0}^- \mathcal{A}_{L,\parallel}^{-*} \right]+\left(L\to R\right)\right)\right\},\nonumber\\[2mm]
		\frac{d\Gamma}{dq^2}\tilde{Z}^L_{\perp L}=&-\frac{\mathcal{N}}{\sqrt{3}}\left\{\left(m_\tau^2{\rm Re}\left[\mathcal{A}_{L,0}^+ \mathcal{A}_{L,\perp}^{+*} \right]+q^2{\rm Re}\left[\mathcal{A}_{L,0}^- \mathcal{A}_{L,\perp}^{-*} \right]\right)-\left(L\to R\right)\right\},\nonumber\\[2mm]
		\frac{d\Gamma}{dq^2}\left\langle \tilde{T}^L_{T L}\right\rangle=&-\pi\frac{\sqrt{3}\mathcal{N}}{4}\left\{m_\tau^2\left({\rm Im}\left[\mathcal{A}_{L,t} \mathcal{A}_{L,\parallel}^{+*} \right]-\left(L\to R\right)\right)\right. \nonumber\\
		&\hspace{3.5cm}\left. +q^2\left({\rm Im}\left[\mathcal{A}_{L,0}^- \mathcal{A}_{L,\perp}^{-*} \right]+\left(L\to R\right)\right)\right\},\nonumber\\[2mm]
		\frac{d\Gamma}{dq^2}\tilde{Z}^L_{T L}=&-\frac{\mathcal{N}}{\sqrt{3}}\left\{\left(m_\tau^2{\rm Im}\left[\mathcal{A}_{L,0}^+ \mathcal{A}_{L,\parallel}^{+*} \right]+q^2{\rm Im}\left[\mathcal{A}_{L,0}^- \mathcal{A}_{L,\parallel}^{-*} \right]\right)-\left(L\to R\right)\right\},\nonumber\\[2mm]
		\frac{d\Gamma}{dq^2}\left\langle \tilde{T}^L_{LL}\right\rangle=&\frac{2\mathcal{N}}{\sqrt{6}}\left\{\left[m_\tau^2\left( \left| \mathcal{A}_{L,\parallel}^+\right|
		^2+\left| \mathcal{A}_{L,\perp}^+\right|
		^2\right)-2q^2\left(\left| \mathcal{A}_{L,\parallel}^-\right|
		^2+\left| \mathcal{A}_{L,\perp}^-\right|
		^2\right)\right]-\left(L\rightarrow R\right)\right\},\nonumber\\[2mm]
		\frac{d\Gamma}{dq^2}\tilde{Z}^L_{L L}=&-\sqrt{6}\mathcal{N}q^2\left\{{\rm Re}\left[\mathcal{A}_{L,\parallel}^- \mathcal{A}_{L,\perp}^{-*} \right]+\left(L\to R\right)\right\},\nonumber\\[2mm]
		\frac{d\Gamma}{dq^2}\tilde{A}^L_{LL}=&-\frac{\mathcal{N}}{\sqrt{6}}\left\{\left[m_\tau^2\left( \left| \mathcal{A}_{L,\parallel}^+\right|
		^2+\left| \mathcal{A}_{L,\perp}^+\right|
		^2\right)+q^2\left(\left| \mathcal{A}_{L,\parallel}^-\right|
		^2+\left| \mathcal{A}_{L,\perp}^-\right|
		^2\right)\right]-\left(L\rightarrow R\right)\right\},\nonumber\\[2mm]
		\frac{d\Gamma}{dq^2}\left(\left\langle \tilde{T}^\perp_{\perp\perp}\right\rangle\right.&\left.+\left\langle \tilde{T}^\perp_{TT}\right\rangle\right)=-\pi\frac{\sqrt{6}\mathcal{N}}{4}m_\tau\sqrt{q^2}\bigg\{4\left({\rm Re}\left[\mathcal{A}_{L,t} \mathcal{A}_{L,0}^{-*} \right]-\left(L\to R\right)\right)\nonumber\\
		&\hspace{3.5cm}-\left({\rm Re}\left[\mathcal{A}_{L,\parallel}^+ \mathcal{A}_{L,\perp}^{-*} \right]+{\rm Re}\left[\mathcal{A}_{L,\perp}^+ \mathcal{A}_{L,\parallel}^{-*} \right]+\left(L\to R\right)\right)\bigg\},\nonumber\\[2mm]
		\frac{d\Gamma}{dq^2}\tilde{Z}^\perp_{\perp\perp}=&-\frac{2\mathcal{N}}{\sqrt{6}}m_\tau\sqrt{q^2}\left\{\left(2{\rm Re}\left[\mathcal{A}_{L,0}^+ \mathcal{A}_{L,0}^{-*} \right]-{\rm Re}\left[\mathcal{A}_{L,\parallel}^+ \mathcal{A}_{L,\parallel}^{-*} \right]\right)-\left(L\to R\right)\right\},\nonumber\\[2mm]
		\frac{d\Gamma}{dq^2}\tilde{Z}^\perp_{\perp T}=&\frac{\mathcal{N}}{\sqrt{6}}m_\tau\sqrt{q^2}\left\{\left({\rm Im}\left[\mathcal{A}_{L,\parallel}^+ \mathcal{A}_{L,\perp}^{-*} \right]-{\rm Im}\left[\mathcal{A}_{L,\perp}^+ \mathcal{A}_{L,\parallel}^{-*} \right]\right)-\left(L\to R\right)\right\},\nonumber\\[2mm]
		\frac{d\Gamma}{dq^2}\tilde{A}^\perp_{\perp L}=&-\frac{\mathcal{N}}{\sqrt{3}}m_\tau\sqrt{q^2}\left\{\left({\rm Re}\left[\mathcal{A}_{L,0}^+ \mathcal{A}_{L,\perp}^{-*} \right]+{\rm Re}\left[\mathcal{A}_{L,\perp}^+ \mathcal{A}_{L,0}^{-*} \right]\right)-\left(L\to R\right)\right\},\nonumber\\[2mm]
		\frac{d\Gamma}{dq^2}\tilde{Z}^\perp_{TT}=&-\frac{2\mathcal{N}}{\sqrt{6}}m_\tau\sqrt{q^2}\left\{\left(2{\rm Re}\left[\mathcal{A}_{L,0}^+ \mathcal{A}_{L,0}^{-*} \right]-{\rm Re}\left[\mathcal{A}_{L,\perp}^+ \mathcal{A}_{L,\perp}^{-*} \right]\right)-\left(L\to R\right)\right\},\nonumber\\[2mm]
		\frac{d\Gamma}{dq^2}\tilde{A}^\perp_{TL}=&-\frac{\mathcal{N}}{\sqrt{3}}m_\tau\sqrt{q^2}\left\{\left({\rm Im}\left[\mathcal{A}_{L,0}^+ \mathcal{A}_{L,\parallel}^{-*} \right]-{\rm Im}\left[\mathcal{A}_{L,\parallel}^+ \mathcal{A}_{L,0}^{-*} \right]\right)-\left(L\to R\right)\right\},\nonumber\\[2mm]
		\frac{d\Gamma}{dq^2}\left\langle \tilde{T}^\perp_{LL}\right\rangle=&\pi\frac{\sqrt{6}\mathcal{N}}{4}m_\tau\sqrt{q^2}\left\{\left({\rm Re}\left[\mathcal{A}_{L,\parallel}^+ \mathcal{A}_{L,\perp}^{-*} \right]+{\rm Re}\left[\mathcal{A}_{L,\perp}^+ \mathcal{A}_{L,\parallel}^{-*} \right]\right)+\left(L\to R\right)\right\},\nonumber\\[2mm]
		\frac{d\Gamma}{dq^2}\tilde{Z}^\perp_{LL}=&\frac{2\mathcal{N}}{\sqrt{6}}m_\tau\sqrt{q^2}\left\{\left({\rm Re}\left[\mathcal{A}_{L,\parallel}^+ \mathcal{A}_{L,\parallel}^{-*} \right]+{\rm Re}\left[\mathcal{A}_{L,\perp}^+ \mathcal{A}_{L,\perp}^{-*} \right]\right)-\left(L\to R\right)\right\},\nonumber\\[2mm]
		\frac{d\Gamma}{dq^2}\left(\left\langle \tilde{T}^T_{\perp L}\right\rangle\right.&\left.+\frac{1}{2}\tilde{A}^T_{\perp L}\right)=\frac{\sqrt{3}\mathcal{N}}{2}m_\tau\sqrt{q^2}\bigg\{2\left({\rm Im}\left[\mathcal{A}_{L,t} \mathcal{A}_{L,\parallel}^{-*} \right]-\left(L\to R\right)\right)\nonumber\\
		&\hspace{3.5cm}+\left[\left({\rm Im}\left[\mathcal{A}_{L,0}^+ \mathcal{A}_{L,\perp}^{-*} \right]-{\rm Im}\left[\mathcal{A}_{L,\perp}^+ \mathcal{A}_{L,0}^{-*} \right]\right)+\left(L\to R\right)\right]\bigg\},\nonumber\\[2mm]
		\frac{d\Gamma}{dq^2}\left(\left\langle \tilde{T}^T_{TL}\right\rangle\right.&\left.+\frac{1}{2}\tilde{A}^T_{TL}\right)=-\frac{\sqrt{3}\mathcal{N}}{2}m_\tau\sqrt{q^2}\bigg\{2\left({\rm Re}\left[\mathcal{A}_{L,t} \mathcal{A}_{L,\perp}^{-*} \right]-\left(L\to R\right)\right)\nonumber\\
		&\hspace{2.5cm}+\left[\left({\rm Re}\left[\mathcal{A}_{L,0}^+ \mathcal{A}_{L,\parallel}^{-*} \right]+{\rm Re}\left[\mathcal{A}_{L,\parallel}^+ \mathcal{A}_{L,0}^{-*} \right]\right)+\left(L\to R\right)\right]\bigg\}.
	\end{align}
Here the abbreviation $\mathcal{N}$ is defined as
\begin{equation}\label{eq:abbreviation_N}
	\mathcal{N}\equiv\frac{G_{F}^{2}\left|V_{c b}\right|^{2}\left|\vec{p}_{J/\psi}\right|}{128 \pi^{3} m_{B_c}^{2}}\left(1-\frac{m_\tau^2}{q^2}\right)^2 \mathcal{B}_\tau \mathcal{B}_{J/\psi},
\end{equation}
with $\mathcal{B}_\tau$ and $\mathcal{B}_{J/\psi}$ being the branching fractions of the decay channels of $\tau$ and $J/\psi$ respectively. Here the combinations of the transversity amplitudes are given by~\cite{Alguero:2020ukk}
\begin{equation}
	\begin{aligned}
		\mathcal{A}_{\{L,R\},i}^+&=\mathcal{A}_{\{L,R\},i}+4\frac{\sqrt{q^2}}{m_\tau}\mathcal{A}_{\{L,R\},i}^T,\quad\mathcal{A}_{\{L,R\},i}^-=\mathcal{A}_{\{L,R\},i}+4\frac{m_\tau}{\sqrt{q^2}}\mathcal{A}_{\{L,R\},i}^T,\\[2mm]
		\mathcal{A}_{\{L,R\},t}&=\frac{\sqrt{q^2}}{m_\tau}\mathcal{A}_{\{L,R\},t}^{SP}+\mathcal{A}_{\{L,R\},t}^{VA}
	\end{aligned}
\end{equation}

\section{Phase-space integrations}
\label{app:phase space}

In this appendix, we detail the phase-space integrations in eq.~\eqref{eq:decay rate}. Firstly, for a generic two-body phase-space integration, we have 
\begin{equation} \label{eq:two-body}
  \int d\Pi_2(p;k_1,k_2) = \int \frac{d^3\vec{k}_1}{(2\pi)^3 2 E_1} \frac{d^3\vec{k}_2}{(2\pi)^3 2 E_2} (2\pi)^4 \delta^{(4)}(p-k_1-k_2) =\frac{\lambda^{1/2}(p^2,k_1^2,k_2^2)}{8\pi p^2},
\end{equation}
where the standard K\"allen function is defined by
\begin{equation}
  \lambda(a,b,c)=a^2+b^2+c^2-2 a b-2 a c-2 b c.
\end{equation}
With the help of eq.~\eqref{eq:two-body}, we can easily evaluate the integrations $\int d\Pi_2(p_{B_c};p_{J/\psi},q)$ and $\int d\Pi_2(p_{J/\psi};p_{\mu^-},p_{\mu^+})$ in eq.~\eqref{eq:decay rate}. 

For the phase-space integrations $\int d\Pi_2(q;p_{\tau},p_{\bar{\nu}_{\tau}})$ and $\int d\Pi_{2(3)}(p_\tau;p_d(,p_{\bar{\nu}_\ell}),p_{\nu_\tau})$, on the other hand, it is convenient to perform the evaluation simultaneously in the $\tau\bar{\nu}_\tau$ centre-of-mass frame. For the hadronic decays $\tau^- \to \pi^-\nu_\tau$ and $\tau^- \to \rho^-\nu_\tau$, we have~\cite{Bhattacharya:2020lfm,Hu:2020axt,Hu:2021emb}
\begin{equation}
	\begin{aligned}
		&\int d \Pi_{2}(q; p_{\tau}, p_{\bar{\nu}_\tau}) d \Pi_{2}(p_{\tau} ; p_{d}, p_{\nu_\tau}) = \frac{1}{(4\pi)^{4}} \int \frac{d\left|\vec{p}_{\tau}\right|}{\sqrt{q^{2}}} d \cos \theta_{\tau d} d \phi_{\tau d} d E_{d} d \cos \theta_{d}d\phi_d \\
		&\hspace{4.2cm} \times\delta\left(\left|\vec{p}_{\tau}\right|-\frac{q^{2}-m_{\tau}^{2}}{2 \sqrt{q^{2}}}\right) \delta\left(\cos \theta_{\tau d}-\frac{2 E_{\tau} E_{d}-m_{\tau}^{2}-m_{d}^{2}}{2\left|\vec{p}_{\tau}\right|\left|\vec{p}_{d}\right|}\right),
	\end{aligned}
\end{equation}
where the kinematic variables $q^2$ and $E_{d=\pi,\rho}$ are restricted, respectively, within the ranges
\begin{equation}
	m_{\tau}^{2} \leq q^{2} \leq\left(m_{B_c}-m_{J/\psi}\right)^{2}, \quad \frac{m_{\tau}^{4}+m_d^{2} q^{2}}{2 m_{\tau}^{2} \sqrt{q^{2}}} \leq E_d \leq \frac{m_d^{2}+q^{2}}{2 \sqrt{q^{2}}}.
\end{equation}
For the leptonic decay $\tau^- \to \ell^-\bar{\nu}_\ell\nu_\tau$, we can firstly integrate over the momenta of the two neutrinos by using the formula specific for massless particles~\cite{Penalva:2021wye}
\begin{equation}
	\int\frac{d^3\vec{p}_{\nu_\tau}}{2\left|\vec{p}_{\nu_\tau}\right|}\int\frac{d^3\vec{p}_{\bar{\nu}_\ell}}{2\left|\vec{p}_{\bar{\nu}_\ell}\right|}\delta^{(4)}(Q-p_{\nu_\tau}-p_{\bar{\nu}_\ell})p_{\nu_\tau}^\alpha p_{\bar{\nu}_\ell}^\beta=\frac{\pi Q^2}{24}\left(g^{\alpha\beta}+2\frac{Q^\alpha Q^\beta}{Q^2}\right)\theta\left(Q^2\right),
\end{equation}
where $Q=p_\tau-p_\ell$ and $\theta\left(Q^2\right)$ is the step function. We can then get
\begin{equation}
	\begin{aligned}
		& \int\frac{\mathrm{Tr}\left[\rho^{B_c}(s,s^\prime,\lambda,\lambda^\prime)\left(\rho^\tau(s,s^\prime)\otimes\rho^{J/\psi}(\lambda,\lambda^\prime)\right)^T\right]}{m_\tau\Gamma\left(\tau\to\ell\bar{\nu}_\ell\nu_\tau\right)\mathrm{Tr}\left[\rho^{B_c}\right]\mathrm{Tr}\left[\rho^{J/\psi}\right]}\,d\Pi_3(p_\tau;p_d,p_{\bar{\nu}_\ell},p_{\nu_\tau})\\[2mm]
		&\quad=\frac{8}{3\pi m_\tau^3}\int\frac{d^3\vec{p}_d}{2E_d}\frac{\theta(1+y^2-x)}{1-8y^2+8y^6-y^8-24y^4\ln y}\\[2mm]
		&\quad\times\Bigg\{m_\tau\left[x\left(3-2x\right)-y^2\left(4-3x\right)\right]\left[1+\sqrt{\frac{3}{2}}T^U_{ij}(\vec{n}_i\cdot\hat{p}_{\mu^-})(\vec{n}_j\cdot\hat{p}_{\mu^-})\right]\\
		&\qquad \quad+2\left(2x-3y^2-1\right)\left[ P_U^{i^\prime}(N^{i^\prime}\cdot p_d)+\sqrt{\frac{3}{2}} T^{i^\prime}_{ij}(N^{i^\prime}\cdot p_d)(\vec{n}_i\cdot\hat{p}_{\mu^-})(\vec{n}_j\cdot\hat{p}_{\mu^-})\right]\Bigg\},
	\end{aligned}
\end{equation}
and
\begin{equation} \label{eq:leftphasespaceintegration}
  \int d \Pi_{2}(q; p_{\tau}, p_{\bar{\nu}_\tau}) \frac{d^3 \vec{p}_d}{2E_d} =\frac{m_\tau^2}{(8\pi)^2} \int \frac{d\left|\vec{p}_{\tau}\right|}{\sqrt{q^{2}}} dx d \phi_{\tau d} d E_{d} d \cos\theta_{d}d\phi_d\,\delta\left(\left|\vec{p}_{\tau}\right|-\frac{q^{2}-m_{\tau}^{2}}{2 \sqrt{q^{2}}}\right).
\end{equation}
The available range of $q^2$ in this channel is the same as in the other two channels, but the angle $\theta_{\tau d}$ is now a free variable and varies from 0 to $\pi$. In terms of the parameter $x$ introduced in eq.~\eqref{eq:x-costaud}, we have explicitly
\begin{equation}
 x_-\leq x\leq x_+, \quad x_\pm=\frac{2}{m^2_\tau}\left(E_\tau    E_d\pm\left|\vec{p}_\tau\right|\left|\vec{p}_d\right|\right).
\end{equation}
Moreover, due to the simultaneous presence of two neutrinos in the leptonic $\tau$ decay, the visible product $\ell$ can be at rest in the $\tau\bar{\nu}_\tau$ centre-of-mass frame. Taking account of the extra constrain $x\leq1+y^2$ from the step function $\theta(1+y^2-x)$, it is convenient to split the region of integration in eq.~\eqref{eq:leftphasespaceintegration} into the following two parts:
\begin{equation}
	\begin{aligned}
		&\mbox{Part I:} \quad m_d\leq E_d \leq \frac{m_{\tau}^{4}+m_d^{2} q^{2}}{2 m_{\tau}^{2} \sqrt{q^{2}}}, \qquad x_-\leq x\leq x_+,\\[2mm]
		&\mbox{Part II:} \quad \frac{m_{\tau}^{4}+m_d^{2} q^{2}}{2 m_{\tau}^{2} \sqrt{q^{2}}} \leq E_d \leq \frac{m_d^{2}+q^{2}}{2 \sqrt{q^{2}}}, \qquad x_-\leq x\leq1+y^2.
	\end{aligned}
\end{equation}

\bibliographystyle{JHEP}
\bibliography{ref}

\end{document}